\documentstyle[preprint,aps,tabularx]{revtex}
\begin{document}
\preprint{\vbox{\hbox{IFP-780-UNC}
\hbox{VAND-TH-000-09}
\hbox{hep-th/0011186}
\hbox{November 2000}}}

\draft
\title{Classification of Conformality Models Based on Nonabelian Orbifolds}
\author{{\bf Paul H. Frampton$^{(a)}$ and Thomas W. Kephart$^{(b)}$}}
\address{(a)Department of Physics and Astronomy,\\
University of North Carolina, Chapel Hill, NC 27599.}
\address{(b)Department of Physics and Astronomy,\\
Vanderbilt University, Nashville, TN 37325.}
\maketitle

\begin{abstract}
A systematic analysis is presented of compactifications of the IIB
superstring on $AdS_5 \times S^5/\Gamma$ where $\Gamma$ is a non-abelian
discrete group. Every possible $\Gamma$ with order $g \leq 31$ is
considered. There exist 45 such groups but a majority cannot yield chiral
fermions due to a certain theorem that is proved. The lowest order to embrace
the nonSUSY standard $SU(3) \times SU(2) \times U(1)$ model with three chiral
families is $\Gamma = D_4 \times Z_3$, with $g=24$; this is the only 
successful model found in the search. The 
consequent uniqueness of the successful model arises primarily
from the scalar sector, prescribed by the construction, being
sufficient to allow the correct symmetry
breakdown. 
\end{abstract}


\bigskip

\newpage

\section{Introduction}

In particle phenomenology, the impressive success of the standard theory
based on $SU(3) \times SU(2) \times U(1)$ has naturally led to the question
of how to extend the theory to higher energies? One is necessarily led by
weaknesses and incompleteness in the standard theory. If one extrapolates
the standard theory as it stands one finds (approximate) unification of the
gauge couplings at $\sim 10^{16}$ GeV. But then there is the {\it hierarchy}
problem of how to explain the occurrence of the tiny dimensionless ratio $%
\sim 10^{-14}$ of the weak scale to the unification scale. Inclusion of
gravity leads to a {\it super-hierarchy} problem of the ratio of the weak
scale to the Planck scale, $\sim 10^{19}$ GeV, an even tinier $\sim 10^{-17}$ dimensionless ratios. 
Although this is obviously a very important problem about which
conformality by itself is not informative, we shall discuss first the
hierarchy rather than the super-hierarchy.

\bigskip

There are four well-defined approaches to the hierarchy problem:

\begin{itemize}
\item  1. Supersymmetry

\item  2. Technicolor.

\item  3. Extra dimensions.

\item  4. Conformality.
\end{itemize}

\noindent {\it Supersymmetry} has the advantage of rendering the hierarchy
technically natural, that once the hierarchy is put in to the lagrangian it
need not be retuned in perturbation theory. Supersymmetry predicts
superpartners of all the known particles and these are predicted to be at or
below a TeV scale if supersymmetry is related to the electroweak breaking.
Inclusion of such hypothetical states improves the gauge coupling
unification. On the negative side, supersymmetry does not explain the origin
of the hierarchy.

\bigskip

\noindent {\it Technicolor} postulates that the Higgs boson is a composite
of fermion-antifermion bound by a new (technicolor) strong dynamics at or
below the TeV scale. This obviates the hierarchy problem. On the minus side,
no convincing simple model of technicolor has been found.

\bigskip

\noindent {\it Extra dimensions} can have a range as large as $1 ({\rm TeV}%
)^{-1}$ and the gauge coupling unification can happen quite differently than
in only four spacetime dimensions. This replaces the hierarchy problem with
a different fine-tuning question of why the extra dimension is restricted to
a distance corresponding to the weak interaction scale.
There is also a potentially serious problem with the proton lifetime.

\bigskip

\noindent {\it Conformality} is inspired by superstring duality and assumes
that the particle spectrum of the standard model is enriched such that there
is a conformal fixed point of the renormalization group at the TeV scale.
Above this scale the coupling do not run so the hierarchy is nullified.

\bigskip

Conformality is the approach followed in this paper. We shall systematicaly
analyse the compactification of the IIB superstring on $AdS_5 \times
S^5/\Gamma$ where $\Gamma$ is a discrete non-abelian group.

The duality between weak and strong coupling field theories and then between
all the different superstring theories has led to a revolution in our
understanding of strings. Equally profound, is the AdS/CFT duality which is
the subject of the present article. This AdS/CFT duality is between string
theory compactified on Anti-de-Sitter space and Conformal Field Theory.

Until very recently, the possibility of testing string theory seemed at best
remote. The advent of $AdS/CFT$s and large-scale string compactification
suggest this point of view may be too pessimistic, since both could lead to $%
\sim 100TeV$ evidence for strings. With this thought in mind, we are
encouraged to build $AdS/CFT$ models with realistic fermionic structure, and
reduce to the standard model below $\sim 1TeV$.

Using AdS/CFT duality, one arrives at a class of gauge field theories of
special recent interest. The simplest compactification of a ten-dimensional
superstring on a product of an AdS space with a five-dimensional spherical
manifold leads to an ${\cal N} = 4~SU(N)$ supersymmetric gauge theory, well
known to be conformally invariant\cite{mandelstam}. By replacing the
manifold $S^5$ by an orbifold $S^5/\Gamma$ one arrives at less
supersymmetries corresponding to ${\cal N} = 2,~1 ~{\rm or}~ 0$ depending%
\cite{KS} on whether: (i) $\Gamma \subset SU(2)$, (ii) $\Gamma \subset 
SU(3)$ but $\Gamma \not\subset SU(2)$, 
or (iii)  
$\Gamma \subset SU(4)$ but $\Gamma \not\subset SU(3)$ respectively, 
where $\Gamma$ is in all cases a subgroup of 
$SU(4) \sim SO(6)$, the isometry of the $S^5$ manifold.

It was conjectured in \cite{maldacena} that such $SU(N)$ gauge theories are
conformal in the $N \rightarrow \infty$ limit. In \cite{F1} it was
conjectured that at least a subset of the resultant nonsupersymmetric ${\cal %
N} = 0$ theories are conformal even for finite $N$ and that
one of this subset provides the right extension
of the standard model. Some first steps to
check this idea were made in \cite{WS}. Model-building based on abelian $%
\Gamma$ was studied further in \cite{CV,F2,F3}, arriving in \cite{F3} at an $%
SU(3)^7$ model based on $\Gamma = Z_7$ which has three families of chiral
fermions, a correct value for ${\rm sin}^2 \theta$ and a conformal scale $%
\sim 10$~~TeV.

The case of non-abelian orbifolds bases on non-abelian $\Gamma$ has not
previously been studied, partially due to the fact that it is apparently
somewhat more mathematically sophisticated. However, we shall show here that
it can be handled equally as systematically as the abelian case and leads to
richer structures and interesting results.

In such constructions, the cancellation of chiral anomalies in the
four-dimensional theory, as is necessary in extension
of the standard model ({\it e.g.} \cite{chiral,331}),
follows from the fact that the progenitor ten-dimensional
superstring theory has cancelling hexagon anomaly\cite{hexagon}.

We consider all non-abelian discrete groups of order $g < 32$. These are
described in detail in \cite{books,FK}. There are exactly 45 such
non-abelian groups. Because the gauge group arrived at by this construction%
\cite{CV} is $\otimes_i SU(Nd_i)$ where $d_i$ are the dimensions of the
irreducible representations of $\Gamma$, one can expect to arrive at models
such as the Pati-Salam $SU(4) \times SU(2) \times SU(2)$ type\cite{PS} by
choosing $N = 2$ and combining two singlets and a doublet in the {\bf 4} of $%
SU(4)$. Indeed we shall show that such an accommodation of the standard
model is possible by using a non-abelian $\Gamma$.

The procedures for building a model within such a conformality approach are:
(1) Choose $\Gamma$; (2) Choose a proper embedding $\Gamma \subset SU(4)$ by
assigning the components of the {\bf 4} of $SU(4)$ to irreps of $\Gamma$,
while at the same time ensuring that the {\bf 6} of $SU(4)$ is real; (3)
Choose $N$, in the gauge group $\otimes_i SU(Nd_i)$. (4) Analyse the
patterns of spontaneous symmetry breaking.

In the present study we shall most often choose $N = 2$ and aim at the gauge group $%
SU(4) \times SU(2) \times SU(2)$. To obtain chiral fermions, it is necessary%
\cite{CV} that the {\bf 4} of $SU(4)$ be complex ${\bf 4} \neq {\bf 4}^*$.
Actually this condition is not quite sufficient to ensure chirality in the
present case because of the pseudoreality of $SU(2)$. We must ensure that
the {\bf 4} is not just pseudoreal.

This last condition means that many of our 45 candidates for $\Gamma$ do not
lead to chiral fermions. For example, $\Gamma = Q_{2n} \subset SU(2)$ has
irreps of appropriate dimensionalities for our purpose but with $N=2$ it will not
sustain chiral fermions under $SU(4)\times SU(2) \times SU(2)$ because these
irreps are all, like $SU(2)$, pseudoreal.\footnote{%
Note that were we using $N \geq 3$ then a pseudoreal {\bf 4} would give
chiral fermions.} Applying the rule that {\bf 4} must be neither real nor
pseudoreal leaves a total of only 19 possible non-abelian discrete groups of
order $g \leq 31$. The smallest group which avoids pseudoreality has order $%
g = 16$ but gives only two families. The technical details of our systematic
search will be given in Sections V and VI.
The simplest interesting non-abelian case which has $g = 24$ and gives three
chiral families in a Pati-Salam-type model\cite{PS}.

Before proceeding to details, it is worth
reminding the reader that the Conformal Field Theory (CFT) that it
exemplifies should be free of all divergences, even logarithmic ones, if the
conformality conjecture is correct, and be completely finite. Further the
theory is originating from a superstring theory in a higher-dimension (ten)
and contains gravity\cite{V,RS,GW} by compactification of the
higher-dimensional graviton already contained in that superstring theory. In
the CFT as we derive it, gravity is absent because we have not kept these
graviton modes - of course, their influence on high-energy physics
experiments is generally completely negligible unless the compactification
scale is ``large''\cite{antoniadis}; here we shall neglect the effects of
gravity.

It is worthwhile
noting the degree of constraint imposed on the symmetry and particle content
of a model as the number of irreps $N_{R}$ of the discrete group $\Gamma $
associated with the choice of orbifold changes. The number of gauge groups
grows linearly in $N_{R}$, the number of scalar irreps grows roughly
quadratically with $N_{R}$, and the chiral fermion content is highly $\Gamma 
$ dependent. If we require the minimal $\Gamma $ that is large enough for
the model generated to contain the fermions of the standard model and have
sufficient scalars to break the symmetry to that of the standard model, then 
$\Gamma = Q \times Z_{3}$ appears to be that minimal choice\cite{FK2}.

Although a decade ago the chances of testing string theory seemed at best
remote, recent progress has given us hope that such tests may indeed be
possible in AdS/CFTs. The model provided here demonstrates the standard
model can be accommodated in these theories and suggests the possibility of a
rich spectrum of new physics just around the TeV corner.

\bigskip \bigskip \bigskip

\newpage

\bigskip
\bigskip
\bigskip

\section{Non-Abelian Groups with order $g\leq 31$}

>From any good textbook on finite groups\cite{books} we may find a tabulation
of
the number of finite groups as a function of the order g, the number of
elements in the group. Up to order 31 there is a total of 93 different
finite groups of which slightly over one half (48) are abelian.

Amongst finite groups, the non-abelian examples have the advantage
of non-singlet irreducible representations which can be used to inter-relate
families. Which such group to select is based on simplicity: the minimum
order and most economical use of representations\cite{guts}.

Let us first dispense with the abelian groups. These are all made up from
the basic unit $Z_p$, the order p group formed from the $p^{th}$ roots
of unity. It is important to note that the the product $Z_pZ_q$ is identical
to $Z_{pq}$ if and only if p and q have no common prime factor.

If we write the prime factorization of g as:
\begin{equation}
g = \prod_{i}p_i^{k_i}
\end{equation}
where the product is over primes, it follows that the number
$N_a(g)$ of inequivalent abelian groups of order g is given by:
\begin{equation}
N_a(g) = \prod_{k_i}P(k_i)
\end{equation}
where $P(x)$ is the number of unordered partitions of $x$.
For example, for order $g = 144 = 2^43^2$ the value would be
$N_a(144) = P(4)P(2) = 5\times2 = 10$. For $g\leq31$ it is simple
to evaluate $N_a(g)$ by inspection. $N_a(g) = 1$ unless g contains
a nontrivial power ($k_i\geq2$) of a prime. These exceptions are:
$N_a(g = 4,9,12,18,20,25,28) = 2; N_a(8,24,27) = 3$; and $N_a(16) = 5$.
This confirms that:
\begin{equation}
\sum_{g = 1}^{31}N_a(g) = 48
\end{equation}
We do not consider the abelian cases further in this paper.\\

Of the nonabelian finite groups, the best known are perhaps the
permutation groups $S_N$ (with $N \geq 3$) of order $N!$
The smallest non-abelian finite group is $S_3$ ($\equiv D_3$),
the symmetry of an equilateral triangle with respect to all
rotations in a three dimensional sense. This group initiates two
infinite series, the $S_N$ and the $D_N$. Both have elementary
geometrical significance since the symmetric permutation group
$S_N$ is the symmetry of the N-plex in N dimensions while the dihedral group
$D_N$ is the symmetry of the planar N-agon in 3 dimensions.
As a family symmetry, the $S_N$ series becomes uninteresting rapidly
as the order and the dimensions of the representions increase. Only $S_3$
and $S_4$ are of any interest as symmetries associated with the particle
spectrum\cite{Pak}, also the order (number of elements) of the $S_N$ groups
grow factorially with N. The order of the dihedral groups increase only
linearly with N and their irreducible representations are all one- and
two- dimensional. This is reminiscent of the representations of the
electroweak $SU(2)_L$ used in Nature.

Each $D_N$ is a subgroup of $O(3)$ and has a counterpart double dihedral (also known as dicyclic)
group $Q_{2N}$, of order $4N$, which is a subgroup of the double covering
$SU(2)$ of $O(3)$.

With only the use of $D_N$, $Q_{2N}$, $S_N$ and the tetrahedral group T ( of
order
12, the even permutations subgroup of $S_4$ ) we find 32 of the 45
nonabelian groups up to order 31, either as simple groups or as
products of simple nonabelian groups with abelian groups:
(Note that $D_6 \simeq Z_2 \times D_3, D_{10} \simeq Z_2 \times D_5$ and $
D_{14} \simeq Z_2 \times D_7$ ) Some of these groups are firmiliar from crystalography and chemistry, but the 
nonabelian groups that do not embed in in $SU(2)$ are less to have seen wide usage.

\begin{center}

\begin{tabular}{||c||c||}   \hline
g & \\    \hline
$6$  & $D_3 \equiv S_3$\\  \hline
$8$ & $ D_4 , Q = Q_4 $\\    \hline
$10$& $D_5$\\   \hline
$12$&  $D_6, Q_6, T$ \\ \hline
$14$& $D_7$\\  \hline
$16$& $D_8, Q_8, Z_2 \times D_4, Z_2 \times Q$\\  \hline
$18$& $D_9, Z_3 \times D_3$\\  \hline
$20$& $D_{10}, Q_{10}$ \\  \hline
$22$& $D_{11}$\\  \hline
$24$& $D_{12}, Q_{12}, Z_2 \times D_6, Z_2 \times Q_6, Z_2 \times T$,\\  \hline
 & $Z_3 \times D_4, Z_3 \times Q, Z_4 \times D_3, S_4$\\  \hline
$26$& $D_{13}$\\  \hline
$28$& $D_{14}, Q_{14}$ \\  \hline
$30$& $D_{15}, D_5 \times Z_3, D_3 \times Z_5$\\  \hline
\end{tabular}

\end{center}

\bigskip
\bigskip

There remain thirteen others formed by twisted products of abelian factors.
Only certain such twistings are permissable, namely (completing all $g \leq 31$
)

$$\begin{tabular}{||c||c||}   \hline
g & \\    \hline
$16$  & $Z_2 \tilde{\times} Z_8$ (two, excluding $D_8$), $Z_4 \tilde{\times}
Z_4, Z_2 \tilde{\times}(Z_2 \times Z_4)$
(two)\\  \hline
$18$ & $Z_2 \tilde{\times} (Z_3 \times Z_3)$\\    \hline
$20$&  $Z_4 \tilde{\times} Z_5$ \\   \hline
$21$&  $Z_3 \tilde{\times} Z_7$ \\    \hline
$24$&  $Z_3 \tilde{\times} Q, Z_3 \tilde{\times} Z_8, Z_3 \tilde{\times} D_4$
\\  \hline
$27$&  $ Z_9 \tilde{\times} Z_3, Z_3 \tilde{\times} (Z_3 \times Z_3)$ \\
\hline
\end{tabular}$$

It can be shown that these thirteen exhaust the classification of {\it all}
inequivalent finite groups up to order thirty-one\cite{books}.

Of the 45 nonabelian groups, the dihedrals ($D_N$) and double dihedrals
($Q_{2N}$), of order 2N and 4N respectively,
form the simplest sequences. In particular, they fall into subgroups of
$O(3)$ and $SU(2)$ respectively,
the two simplest nonabelian continuous groups.

For the $D_N$ and $Q_{2N}$, the multiplication tables, as derivable from the
character tables,
are simple to express in general. $D_N$, for odd N, has two singlet
representations $1,1^{'}$ and $m = (N-1)/2$
doublets $2_{(j)}$ ($1 \leq j \leq m$). The multiplication rules are:

\begin{equation}
1^{'}\times 1^{'} = 1 ; ~~~1^{'}\times 2_{(j)} = 2_{(j)}
\end{equation}
\begin{equation}
2_{(i)}\times 2_{(j)} = \delta_{ij} (1 + 1^{'}) + 2_{(min[i+j,N-i-j])}
+ (1 - \delta_{ij}) 2_{(|i - j|)}
\end{equation}
\noindent

For even N, $D_N$ has four singlets $1, 1^{'},1^{''},1^{'''}$ and $(m - 1)$
doublets
$2_{(j)}$ ($ 1 \leq j \leq m - 1$)where $m = N/2$ with multiplication rules:

\begin{equation}
1^{'}\times 1^{'} = 1^{''} \times 1^{''} = 1^{'''} \times 1^{'''} = 1
\end{equation}
\begin{equation}
1^{'} \times 1^{''} = 1^{'''}; 1^{''} \times 1^{'''} = 1^{'}; 1^{'''} \times
1^{'} = 1^{''}
\end{equation}
\begin{equation}
1^{'}\times 2_{(j)} = 2_{(j)}
\end{equation}
\begin{equation}
1^{''}\times 2_{(j)} = 1^{'''} \times 2_{(j)} = 2_{(m-j)}
\end{equation}
\begin{equation}
2_{(j)} \times 2_{(k)} = 2_{|j-k|} + 2_{(min[j+k,N-j-k])}
\end{equation}

\noindent
(if $k \neq j, (m - j)$)

\begin{equation}
2_{(j)} \times 2_{(j)} = 2 _{(min[2j,N-2j])} + 1 + 1^{'}
\end{equation}

\noindent
(if $j \neq m/2$ )

\begin{equation}
2_{(j)} \times 2_{(m - j)} = 2_{|m - 2j|} + 1^{''} + 1^{'''}
\end{equation}

\noindent
(if $j \neq m/2 $)

\begin{equation}
2_{m/2} \times 2_{m/2} = 1 + 1^{'} + 1^{''} + 1^{'''}
\end{equation}

\noindent
This last is possible only if m is even and hence if N is divisible by {\it
four}.\\

For $Q_{2N}$, there are four singlets $1$, $1^{'}$ ,$1^{''}$, $1^{'''}$ and
$(N - 1)$ doublets $2_{(j)}$ ($ 1 \leq j \leq (N-1) $).
The singlets have the multiplication rules:

\begin{equation}
1 \times 1 = 1^{'} \times 1^{'} = 1
\end{equation}
\begin{equation}
1^{''} \times 1^{''} = 1^{'''} \times 1^{'''} = 1^{'}
\end{equation}
\begin{equation}
 1^{'} \times 1^{''} = 1^{'''} ; 1^{'''} \times 1^{'} = 1^{''}
\end{equation}

\noindent
for $N = (2k + 1)$ but are identical to those for $D_N$ when N = 2k.

The products involving the $2_{(j)}$ are identical to those given
for $D_N$ (N even) above.

This completes the multiplication rules for 19 of the 45 groups. 
As they are not available in the literature, and somewhat tedious to work out, we have 
provided the complete multiplication tables for all the nonabelian groups
with order $g \leq 31$ in the Appendix.

\bigskip

\newpage

\bigskip
\bigskip
\bigskip

\section{Mathematical Theorem}

\underline{Theorem: A Pseudoreal $4$ of $SU(4)$ Cannot Yield Chiral Fermions.}

\bigskip

In \cite{CV} it was proved that if the embedding in $SU(4)$ is such that
the ${\bf 4}${\bf \ }is real: ${\bf 4}={\bf 4}^{{\bf *}}$, then the
resultant fermions are always non-chiral. It was implied there that the
converse holds, that if ${\bf 4}$ is complex, ${\bf 4}={\bf 4}^{{\bf *}}$ ,
then the resulting fermions are necessarily chiral. Actually for $\Gamma $ $%
\subset $ $SU(2)$ one encounters the intermediate possibility that the {\bf 4%
} is {\it pseudoreal}. In the present section we shall show that if ${\bf 4}$
is pseudoreal then the resultant fermions are necessarily non-chiral. The
converse now holds: if the ${\bf 4}$ is neither real nor pseudoreal then the
resutant fermions are chiral.\bigskip

For $\Gamma \subset SU(2)$ it is important that the embedding be consistent
with the chain $\Gamma \subset SU(2)\subset SU(4)$ otherwise the embedding
is not a consistent one. One way to see the inconsistency is to check the
reality of the ${\bf 6}=({\bf 4}\otimes {\bf 4)}_{antisymmetric}$. If ${\bf 6%
}\neq {\bf 6}^{{\bf *}}${\bf \ }then the embedding is clearly improper. To
avoid this inconsistency it is sufficient to include in the ${\bf 4}$ of $%
SU(4)$ only complete irreducible representations of $SU(2)$.\bigskip

An explicit example will best illustrate this propriety constraint on
embeddings. Let us consider $\Gamma =Q_{6}$, the dicyclic group of order $%
g=12$. This group has six inequivalent irreducible representations: $%
1,1^{\prime },1^{\prime \prime },1^{\prime \prime \prime },2_{1},2_{2}$. The
1, $1^{\prime }$, 2$_{1}$ are real. The $1^{\prime \prime }$ and $1^{\prime
\prime \prime }$ are a complex conjugate pair, The $2_{2}$ is pseudoreal. To
embed $\Gamma =Q_{6}\subset SU(4)$ we must choose from the special
combinations which are complete irreducible representations of $SU(2)$
namely 1, $2=2_{2}$, $3=1^{\prime }+2_{1}$ and $4=1^{\prime \prime
}+1^{\prime \prime \prime }+2_{2}$. In this way the embedding either makes
the ${\bf 4}$ of $SU(4)$ real {\it e.g}. $4=1+1^{\prime }+2_{1}\ $and the
theorem of \cite{CV} applies, and non-chirality results, or the ${\bf 4}$
is pseudoreal {\it e.g}. $4=2_{2}+2_{2}$. In this case one can check that
the embedding is consistent because $({\bf 4}\otimes {\bf 4)}%
_{antisymmetric} $ is real. But it is equally easy to check that the product
of this pseudoreal ${\bf 4}$ with the complete set of irreducible
representations of $Q_{6}$ is again real and that the resultant fermions are
non-chiral.

The lesson is:

{\it To obtain chiral fermions from compactification on }${\it AdS}_{{\it 5}%
}\times S_{5}/\Gamma ${\it , the embedding of }${\it \Gamma }${\it \ in }$%
SU(4)${\it \ must be such that the 4 of }$SU(4)${\it \ is neither real nor
pseudoreal}.\bigskip

\bigskip

\newpage

\bigskip
\bigskip
\bigskip

\section{Chiral Fermions for all nonabelian $g\leq 31$}

Looking at the full list of non-abelian discrete groups
of order $g \leq 31$ as given explicitly in \cite{FK} we see that
of the 45 such groups 32 are simple groups or semi-direct products thereof; these
32 are listed in the Table on page 4691 of \cite{FK}, and  reproduced in section II above.
The remaining 13 are formed as semi-direct product groups (SDPGs)
and are listed in the Table on page 4692 of \cite{FK} and in section II. We shall
follow closely this classification.

\bigskip

Using the pseudoreality considerations of the previous section, we can pare
down the full list of 45 to only 19 which include 13 SDPGs. The
lowest order nonabelian group $\Gamma$ which can lead to
chiral fermions is $g = 16$. The only possible orders
$g \leq 31$ are the seven values:
$g = 16(5[5 SDPGs]),~~ 18(2[1 SDPG]),~~ 20(1[1 SDPG]),~~ 21(1[1 SDPG]),$

\noindent $24(6[3 SDPGs]),~~ 27(2[2 SDPGs]), ~~{\rm and} ~~ 30(2[0 SDPG])$. 

\noindent In parenthesis are the number of groups
at order $g$ and the number of these that are SDPGS is in square brackets;
they add to (19[13 SDPGs]). We shall proceed with the analysis 
systematically, in progressively increasing magnitude of $g$.

\bigskip
\bigskip

\newpage

\bigskip
\bigskip

\underline{{\bf g = 16.}}

\bigskip

\noindent The non-pseudoreal groups number five, and all are SDPGs. In the
notation of Thomas and Wood\cite{books}, which we shall follow for
definiteness both here and in Appendix A, they are:
$16/8,9,10,11,13$. So we now treat these in the order they are
enumerated by Thomas and Wood. Again, the relevant multiplication tables are collected in Appendix A.

\bigskip

\noindent \underline{Group 16/8; also designated $(Z_4 \times Z_2) \tilde{\times} Z_2$}.

\bigskip

\noindent This group has eight singlets $1_1, 1_2, ......, 1_8$ and two doublets
$2_1$ and $2_2$. In the embedding of 16/8 in $SU(4)$ we must avoid the singlet
$1_1$ otherwise there will be a residual supersymmetry with ${\cal N} \geq 1$.
Consider the embedding defined by ${\bf 4} = (2_1, 2_1)$. To find the 
surviving chiral fermions we need to product the ${\bf 4}$ with all ten of the irreps
of 16/8. This results in the table:

\bigskip

\begin{tabular}{||c||c|c|c|c|c|c|c|c||c|c||}
\hline
 & $1_1$ & $1_2$ & $1_3$ & $1_4$ & $1_5$ & $1_6$ & $1_7$ & $1_8$ & $2_1$ & $2_2$ \\
\hline\hline
$1_1$&&&&&&&&&$\times\times$& \\
\hline
$1_2$&&&&&&&&&$\times\times$& \\
\hline
$1_3$&&&&&&&&&$\times\times$& \\
\hline
$1_4$&&&&&&&&&$\times\times$& \\
\hline
$1_5$&&&&&&&&&&$\times\times$ \\
\hline
$1_6$&&&&&&&&&&$\times\times$ \\
\hline
$1_7$&&&&&&&&&&$\times\times$ \\
\hline
$1_8$&&&&&&&&&&$\times\times$ \\
\hline
\hline
$2_1$&&&&&$\times\times$&$\times\times$&$\times\times$&$\times\times$&& \\
\hline
$2_2$&$\times\times$&$\times\times$&$\times\times$&$\times\times$&&&&&& \\
\hline
\hline
\end{tabular}

\bigskip

If we choose $N = 2$, the gauge group is $SU(2)^8 \times SU(4)^2$, and the entries in the table correspond to bifundamental representations of this group (e.g., the entry nearest
the top right corner at the position ($1_1$, $2_1$) is the representation $2(2,1,1,1,1,1,1,1;\bar{4},1))$. If
we identify the diagonal subgroup of the first four SU(2)s as $SU(2)_L$,
of the second four as $SU(2)_R$ and of the two SU(4) as color SU(4)
the result is non-chiral due to the symmetry about the main diagonal of the above table.

On the other hand, if we identify ${\bf 4_1}$ with ${\bf \bar{4}_2}$ there
are potentially eight chiral families:

\begin{equation}
8[(2, 1, 4) + (1, 2, \bar{4})]
\end{equation}
under $SU(2)_L \times SU(2)_R \times SU(4)$. This is the maximum total chirality for this orbifold, 
but as we will see in section IV, the allowed chiral at any stage is as usual determined by 
spontaneous symmetry breaking (SSB) generated in the scalar sector.
In this section we give the maximum chirality for each orbifold, in the next section we
study SSB
for those models with sufficient chirality too accomodate at least three families.

\bigskip

Because $2_1=2_2^*$ form a complex conjugate pair, the embedding ${\bf 4} = (2_1, 2_2)$
is pseudoreal ${\bf 4} \equiv {\bf 4^*}$ and the fermions are non-chiral as
easily confirmed by the resultant table:

\bigskip

\begin{tabular}{||c||c|c|c|c|c|c|c|c||c|c||}
\hline
 & $1_1$ & $1_2$ & $1_3$ & $1_4$ & $1_5$ & $1_6$ & $1_7$ & $1_8$ & $2_1$ & $2_2$ \\
\hline\hline
$1_1$&&&&&&&&&$\times$&$\times$ \\
\hline
$1_2$&&&&&&&&&$\times$&$\times$ \\
\hline
$1_3$&&&&&&&&&$\times$&$\times$ \\
\hline
$1_4$&&&&&&&&&$\times$&$\times$ \\
\hline
$1_5$&&&&&&&&&$\times$&$\times$ \\
\hline
$1_6$&&&&&&&&&$\times$&$\times$ \\
\hline
$1_7$&&&&&&&&&$\times$&$\times$ \\
\hline
$1_8$&&&&&&&&&$\times$&$\times$ \\
\hline
\hline
$2_1$&$\times$&$\times$&$\times$&$\times$&$\times$&$\times$&$\times$&$\times$&& \\
\hline
$2_2$&$\times$&$\times$&$\times$&$\times$&$\times$&$\times$&$\times$&$\times$& & \\
\hline
\hline
\end{tabular}

\bigskip

\noindent For this embedding, the result is non-chiral for either of the
cases ${\bf 4_1} \equiv {\bf 4_2}$ or
${\bf 4_1} \equiv {\bf \bar{4}_2}$.
(In the future, we shall not even consider such trivially real non-chiral embeddings).

\bigskip

\noindent Finally, for 16/8, consider the embedding
${\bf 4} = (1_2,1_5, 2_1)$. (In general there will be many equivalent embeddings. We will give
one member of each equivalence class. Cases that are obviously nonchiral (vectorlike)
will, in general, be ignored, except for a few instructive examples as order 16 and 18.) The table is now:

\bigskip

\begin{tabular}{||c||c|c|c|c|c|c|c|c||c|c||}
\hline
 & $1_1$ & $1_2$ & $1_3$ & $1_4$ & $1_5$ & $1_6$ & $1_7$ & $1_8$ & $2_1$ & $2_2$ \\
\hline\hline
$1_1$&&$\times$&&&$\times$&&&&$\times$& \\
\hline
$1_2$&$\times$&&&&&$\times$&&&$\times$& \\
\hline
$1_3$&&&&$\times$&&&$\times$&&$\times$& \\
\hline
$1_4$&&&$\times$&&&&&$\times$&$\times$& \\
\hline
$1_5$&$\times$&&&&&&$\times$&&&$\times$ \\
\hline
$1_6$&&$\times$&&&&$\times$&&&&$\times$ \\
\hline
$1_7$&&&$\times$&&&&&$\times$&&$\times$ \\
\hline
$1_8$&&&&$\times$&&&$\times$&&&$\times$ \\
\hline
\hline
$2_1$&&&&&$\times$&$\times$&$\times$&$\times$&$\times$&$\times$ \\
\hline
$2_2$&$\times$&$\times$&$\times$&$\times$&&&&&$\times$&$\times$ \\
\hline
\hline
\end{tabular}

\bigskip

\noindent which is chiral.

\bigskip
\bigskip

\noindent These examples of embeddings for $\Gamma = 16/8$ show clearly how the number of
chiral families depends critically on the choice of embedding $\Gamma \subset SU(4)$. To
actual achieve a model with a viable phenomenologically, we must study
the possible routes through SSB for each chiral model. This 
we postpone until we have found all models of potential 
interest.

\bigskip

\noindent \underline{Group 16/9; also designated $[(Z_4 \times Z_2) \tilde{\times} Z_2]^{'}$}

\bigskip

\noindent This group has irreps which comprise eight singlets $1_1,..., 1_8$
and two doublets $2_1, 2_2$. With the embedding ${\bf 4} = (2_1, 2_2)$
and using the multiplication table from Appendix A we arrive at the table of fermion bilinears:

\bigskip

\begin{tabular}{||c||c|c|c|c|c|c|c|c||c|c||}
\hline
 & $1_1$ & $1_2$ & $1_3$ & $1_4$ & $1_5$ & $1_6$ & $1_7$ & $1_8$ & $2_1$ & $2_2$ \\
\hline\hline
$1_1$&&&&&&&&&$\times\times$& \\
\hline
$1_2$&&&&&&&&&&$\times\times$ \\
\hline
$1_3$&&&&&&&&&$\times\times$& \\
\hline
$1_4$&&&&&&&&&&$\times\times$ \\
\hline
$1_5$&&&&&&&&&$\times\times$& \\
\hline
$1_6$&&&&&&&&&&$\times\times$ \\
\hline
$1_7$&&&&&&&&&$\times\times$& \\
\hline
$1_8$&&&&&&&&&&$\times\times$ \\
\hline
\hline
$2_1$&$\times\times$&&$\times\times$&&$\times\times$&&$\times\times$&&& \\
\hline
$2_2$&&$\times\times$&&$\times\times$&&$\times\times$&&$\times\times$&& \\
\hline
\hline
\end{tabular}

\bigskip
\bigskip

This is non-chiral and has no families. This was the only potentially chiral embedding. In what follows,
nonchiral models will not be displayed, however, as the unification scale can be rather low in AdS/CFT
models, it would also be interesting to investigate vectorlike models of this class.

\bigskip

\noindent \underline{Group 16/10; also designated $Z_4 \tilde{\times} Z_4$}

\bigskip

The multiplication table is identical to that for 16/9, as mentioned in 
Appendix A; thus the model building for 16/10 is also identical to 16/9
and merits no additional discussion.

\bigskip

\noindent \underline{Group 16/11; also designated $Z_8 \tilde{\times} Z_2$}

\bigskip

Again there are eight singlets and two doublets. 
The singlets $1_{1,3,5,7}$ are real while the other
singlets fall into two conjugate pairs: $1_2 = 1_4^*$
and $1_6 = 1_8^*$. The doublets are complex: $2_1 = 2_2^*$.

The multiplication table in the Appendix includes the products:
$1_{1,3,5,7} \times 2_{1,2} = 2_{1,2}$ 
and
$1_{2,4,6,8} \times 2_{1,2} = 2_{2,1}$.
Also $2_1 \times 2_1 = 2_2 \times 2_2 = 1_2 + 1_4 + 1_6 + 1_8$,
while $2_1 \times 2_2 = 1_1 + 1_3 + 1_5 + 1_7$. 

This means that there are no interesting (legitimate and chiral)
embeddings of the type 1+1+2 or 2+2. 

The most chiral possibility is the embedding ${\bf 4} = (1_2, 1_2, 1_2, 1_2)$
which leads to the fermions in the following table. In this table,
$(\times)^6 \equiv (\times\times\times\times\times\times)$.

\bigskip

\begin{tabular}{||c||c|c|c|c|c|c|c|c||c|c||}
\hline
 & $1_1$ & $1_2$ & $1_3$ & $1_4$ & $1_5$ & $1_6$ & $1_7$ & $1_8$ & $2_1$ & $2_2$ \\
\hline\hline
$1_1$&&$(\times)^6$&&&&&&&& \\
\hline
$1_2$&&&$(\times)^6$&&&&&&& \\
\hline
$1_3$&&&&$(\times)^6$&&&&&& \\
\hline
$1_4$&$(\times)^6$&&&&&&&&& \\
\hline
$1_5$&&&&&&$(\times)^6$&&&& \\
\hline
$1_6$&&&&&&&$(\times)^6$&&& \\
\hline
$1_7$&&&&&&&&$(\times)^6$&& \\
\hline
$1_8$&&&&&$(\times)^6$&&&&& \\
\hline
\hline
$2_1$&&&&&&&&&&$(\times)^6$ \\
\hline
$2_2$&&&&&&&&&$(\times)^6$& \\
\hline
\hline
\end{tabular}

\bigskip
\bigskip

\noindent This gives rise to 
twelve chiral families if we identify $N=3$
and $3_1=3_4=3_5=3_8$, $3_2=3_6$ and $3_3=3_7$.
Under $SU(3)^3$ the chiral fermions are:

\begin{equation}
12[(3, \bar{3}, 1) + (1, 3, \bar{3}) + (\bar{3}, 3, 1)]
\end{equation}

\noindent together with real non-chiral representations. In section VI where we discuss spontaneous 
symmetry breaking, we will see if this type of unification is possible.

\noindent With the different embedding ${\bf 4} = (1_2, 1_2, 1_2, 1_4)$
the model changes to a less chiral but still interesting fermion
configuration:

\bigskip

\begin{tabular}{||c||c|c|c|c|c|c|c|c||c|c||}
\hline
 & $1_1$ & $1_2$ & $1_3$ & $1_4$ & $1_5$ & $1_6$ & $1_7$ & $1_8$ & $2_1$ & $2_2$ \\
\hline\hline
$1_1$&&$\times\times\times$&&$\times$&&&&&& \\
\hline
$1_2$&$\times$&&$\times\times\times$&&&&&&& \\
\hline
$1_3$&&$\times$&&$\times\times\times$&&&&&& \\
\hline
$1_4$&$\times\times\times$&&$\times$&&&&&&& \\
\hline
$1_5$&&&&&&$\times\times\times$&&$\times$&& \\
\hline
$1_6$&&&&&$\times$&&$\times\times\times$&&& \\
\hline
$1_7$&&&&&&$\times$&&$\times\times\times$&& \\
\hline
$1_8$&&&&&$\times\times\times$&&$\times$&&& \\
\hline
\hline
$2_1$&&&&&&&&&&$\times\times\times\times$ \\
\hline
$2_2$&&&&&&&&&$\times\times\times\times$& \\
\hline
\hline
\end{tabular}

\bigskip
\bigskip

\noindent If we can identify
$SU(3)'s$ as $3_1 \equiv 3_4 \equiv 3_5 \equiv 3_8$,
$3_2 \equiv 3_6$ and $3_3 \equiv 3_7$ this embedding give just four chiral families:

\begin{equation}
4[(3, \bar{3}, 1) + (1, 3, \bar{3}) + (\bar{3}, 3, 1)]
\end{equation}
under $SU(3)^3$ together with real representations.

\bigskip

\noindent To check consistency, we have verified that real and legitimate
embeddings for 16/11 like: ${\bf 4} = (1_3, 1_3, 1_3, 1_3)$ and
${\bf 4} = (2_1, 2_2)$ give no chiral fermions.

\bigskip
\bigskip

\newpage

\bigskip

\noindent \underline{Group 16/13; also designated $[Z_8 \tilde{\times} Z_2]^{''}$}

\bigskip
\bigskip

\noindent Of the five non-pseudoreal $g =16$ nonabelian $\Gamma$, 16/13
is unique in having only four inequivalent singlets $1_1, 1_2, 1_3, 1_4$
but three doublets $2_1, 2_2, 2_3$.

All the four singlet are real $1_i = 1_i^*$. The three doublets comprise
a conjugate complex pair $2_1= 2_3^* \neq 2_1^*$ and the real $2_2 = 2_2^*$.

With the embedding ${\bf 4} = (1_3, 1_4, 2_1)$ the resultant model has a chiral 
fermion quiver corresponding to the Table:

\bigskip

\begin{tabular}{||c||c|c|c|c||c|c|c||}
\hline
 & $1_1$ & $1_2$ & $1_3$ & $1_4$ & $2_1$ & $2_2$ & $2_3$ \\
\hline\hline
$1_1$&&&$\times$&$\times$&$\times$&& \\
\hline
$1_2$&&&$\times$&$\times$&&&$\times$ \\
\hline
$1_3$&$\times$&$\times$&&&&&$\times$ \\
\hline
$1_4$&$\times$&$\times$&&&$\times$&& \\
\hline
\hline
$2_1$&&$\times$&$\times$&&$\times$&$\times$&$\times$ \\
\hline
$2_2$&&&&&$\times$&$\times\times$&$\times$ \\
\hline
$2_3$&$\times$&&&$\times$&$\times$&$\times$&$\times$ \\
\hline
\end{tabular}

\bigskip
\bigskip

\noindent If we identify $SU(2)_L$ with the diagonal subgroup of the first
and fourth $SU(2)$s, and $SU(2)_R$ with the diagonal subgroup of the
2nd and 3rd, then there are four chiral families if we embed
${\bf 4_1} \equiv {\bf \bar{4}_3}$ and break $SU(4)_2$ completely.

\bigskip

\noindent Next, for $g=16$, consider 16/13 with
${\bf 4} = (2_1, 2_1)$. The table becomes:

\bigskip
\bigskip

\begin{tabular}{||c||c|c|c|c||c|c|c||}
\hline
 & $1_1$ & $1_2$ & $1_3$ & $1_4$ & $2_1$ & $2_2$ & $2_3$ \\
\hline\hline
$1_1$&&&&&$\times\times$&& \\
\hline
$1_2$&&&&&&&$\times\times$ \\
\hline
$1_3$&&&&&&&$\times\times$ \\
\hline
$1_4$&&&&&$\times\times$&& \\
\hline
\hline
$2_1$&&$\times\times$&$\times\times$&&&$\times\times$& \\
\hline
$2_2$&&&&&$\times\times$&&$\times\times$ \\
\hline
$2_3$&$\times\times$&&&$\times\times$&&$\times\times$& \\
\hline
\end{tabular}

\bigskip
\bigskip

\noindent With ${\bf 4_1} \equiv {\bf \bar{4}_3}$ there are eight
chiral families.

\bigskip
\bigskip

A similar result occurs, of course, for ${\bf 4} = (2_3, 2_3)$. But the
embedding ${\bf 4} = (2_1, 2_3)$ is non-chiral, leading to the symmetric
fermion quiver/table:

\bigskip
\bigskip

\begin{tabular}{||c||c|c|c|c||c|c|c||}
\hline
 & $1_1$ & $1_2$ & $1_3$ & $1_4$ & $2_1$ & $2_2$ & $2_3$ \\
\hline\hline
$1_1$&&&&&$\times$&&$\times$ \\
\hline
$1_2$&&&&&$\times$&&$\times$ \\
\hline
$1_3$&&&&&$\times$&&$\times$ \\
\hline
$1_4$&&&&&$\times$&&$\times$ \\
\hline
\hline
$2_1$&$\times$&$\times$&$\times$&$\times$&&$\times\times$& \\
\hline
$2_2$&&&&&$\times\times$&&$\times\times$ \\
\hline
$2_3$&$\times$&$\times$&$\times$&$\times$&&$\times\times$& \\
\hline
\end{tabular}

\bigskip
\bigskip

\noindent This arrangement is manifestly non-chiral because of the
symmetry of the table. Even though $2_1$ and $2_3$ are complex,  
$2_1={2_3^*}$, so $4^*=(2_1,2_3)^*=(2_3,2_1)$. We can rotate this
within $SU(4)$ to $4=(2_1,2_3)$. Therefore, the $4$ is pseudoreal 
and the fermions are vectorlike as expected.

\bigskip
\bigskip

\noindent The embedding ${\bf 4} = (2_2, 2_2)$ in 16/13 gives rise to
the table:

\bigskip
\bigskip

\begin{tabular}{||c||c|c|c|c||c|c|c||}
\hline
 & $1_1$ & $1_2$ & $1_3$ & $1_4$ & $2_1$ & $2_2$ & $2_3$ \\
\hline\hline
$1_1$&&&&&&$\times\times$& \\
\hline
$1_2$&&&&&&$\times\times$& \\
\hline
$1_3$&&&&&&$\times\times$& \\
\hline
$1_4$&&&&&&$\times\times$& \\
\hline
\hline
$2_1$&&&&&$\times\times$&&$\times\times$ \\
\hline
$2_2$&$\times\times$&$\times\times$&$\times\times$&$\times\times$&&& \\
\hline
$2_3$&&&&&$\times\times$&&$\times\times$ \\
\hline
\end{tabular}

\bigskip
\bigskip

\noindent This embedding leads to no chirality and zero families.

\bigskip

\newpage

\bigskip

\noindent Finally, the embedding ${\bf 4} = (2_1, 2_2)$ of 16/13 leads to the intermediate
situation:

\bigskip
\bigskip

\begin{tabular}{||c||c|c|c|c||c|c|c||}
\hline
 & $1_1$ & $1_2$ & $1_3$ & $1_4$ & $2_1$ & $2_2$ & $2_3$ \\
\hline\hline
$1_1$&&&&&$\times$&$\times$& \\
\hline
$1_2$&&&&&&$\times$&$\times$ \\
\hline
$1_3$&&&&&&$\times$&$\times$ \\
\hline
$1_4$&&&&&$\times$&$\times$& \\
\hline
\hline
$2_1$&&$\times$&$\times$&&$\times$&$\times$&$\times$ \\
\hline
$2_2$&$\times$&$\times$&$\times$&$\times$&$\times$&&$\times$ \\
\hline
$2_3$&$\times$&&&$\times$&$\times$&$\times$&$\times$ \\
\hline
\end{tabular}

\bigskip
\bigskip

\noindent This give rise to four chiral families with the identification
${\bf 4_1} \equiv {\bf \bar{4}_3}$.

\bigskip
\bigskip

To summarize the ``double doublet'' embeddings ${\bf 4} = (2_i, 2_j)$ of 16/13:
for the equivalent embeddings 
(i, j) = (1, 1) or (3, 3), there are up to eight chiral families;
for the other mutually equivalent cases
(i, j) = (1, 2), (3, 2), (2, 3), or (2, 1) there are up to
four chiral families and finally for the pseudoreal
cases (i, j) = (1, 3), (3, 1) and the real case (2, 2) there are, because
of the mathematical theorem (and as we have now is easyverified by direct calculation)
no chiral fermions.

\bigskip
\bigskip
\bigskip

\newpage

\underline{{\bf g = 18.}}

\bigskip

\noindent The non-pseudoreal groups number two, and one is an SDPG. In the
notation of Thomas and Wood\cite{books}
they are:
$18/3,5$. So we now treat these in the order they are
enumerated by Thomas and Wood.

\bigskip

\noindent \underline{Group 18/3; also designated $D_3 \times Z_3$}

\bigskip

\noindent This group has irreps which fall into six singlets
$1, 1^{'}, 1\alpha, 1^{'}\alpha, 1\alpha^2, 1^{'}\alpha^2$ and three
doublets $2, 2\alpha, 2\alpha^2$. Using the $D_3$ multiplication
table from appendix A the embedding ${\bf 4} = (1\alpha, 1^{'}, 2\alpha)$
yields the table:

\bigskip
\bigskip

\bigskip
\bigskip

\begin{tabular}{||c||c|c|c||c|c|c||c|c|c||}
\hline
 & $1$ & $1^{'}$ & $2$ & $1\alpha$ & $1^{'}\alpha$ & $2\alpha$ & $1\alpha^2$ & $1^{'}\alpha^2$ & $ 2\alpha^2$ \\
\hline\hline
$1$&&$\times$&&$\times$&&$\times$&&& \\
\hline
$1^{'}$&$\times$&&&&$\times$&$\times$&&& \\
\hline
$2$&&&$\times$&$\times$&$\times$&$\times\times$&&& \\
\hline\hline
$1\alpha$&&&&&$\times$&&$\times$&&$\times$ \\
\hline
$1^{'}\alpha$&&&&$\times$&&&&$\times$&$\times$ \\
\hline
$2\alpha$&&&&&&$\times$&$\times$&$\times$&$\times\times$ \\
\hline\hline
$1\alpha^2$&$\times$&&$\times$&&&&&$\times$& \\
\hline
$1^{'}\alpha^2$&&$\times$&$\times$&&&&$\times$&& \\
\hline
$2\alpha^2$ & $\times$&$\times$&$\times\times$&&&&&&$\times$ \\
\hline
\end{tabular}

\bigskip
\bigskip

\noindent Identifying $SU(2)_{L,R}$ with the diagonal subgroups
of respectively $SU(2)_3 \times SU(2)_4$ and
$SU(2)_5 \times SU(2)_6$ gives rise to two chiral families
when it is assumed that $SU(2)_{1,2}$ and $SU(4)_{1,2}$
are broken.

\bigskip
\bigskip
\bigskip

\newpage

\noindent \underline{Group 18/5; also designated $(Z_3 \times Z_3) \tilde{\times} Z_2$}

\bigskip
\bigskip
\bigskip

\noindent This group has two singlets $1, 1^{'}$ and four doublets
$2_1, 2_2, 2_3, 2_4$. Using the multiplication table from Appendix A
we compute the models corresponding to the three inequivalent embeddings
${\bf 4} = (1^{'}, 1^{'}, 2_1)$,
${\bf 4} = (2_1, 2_1)$
and
${\bf 4} = (2_1, 2_2)$.

\bigskip

\noindent For ${\bf 4} = (1^{'}, 1^{'}, 2_1)$ the table is:

\bigskip
\bigskip
\bigskip

\begin{tabular}{||c||c|c||c|c|c|c||}
\hline
 & $1$ & $1^{'}$ & $2_1$ & $2_2$ & $2_3$ & $2_4$  \\
\hline\hline
$1$&&$\times\times$&$\times$&&& \\
\hline
$1^{'}$&$\times\times$&&$\times$&&& \\
\hline\hline
$2_1$&$\times$&$\times$&$\times\times\times$&&& \\
\hline
$2_2$&&&&$\times\times$&$\times$&$\times$ \\
\hline
$2_3$&&&&$\times$&$\times\times$&$\times$\\
\hline
$2_4$&&&&$\times$&$\times$&$\times\times$ \\
\hline
\end{tabular}

\bigskip
\bigskip

\noindent This model is manifestly non-chiral due to the symmetry
of the table.

\bigskip
\bigskip

\newpage   

\bigskip

\noindent For ${\bf 4} = (2_1, 2_1)$ the table is:

\bigskip
\bigskip
\bigskip

\begin{tabular}{||c||c|c||c|c|c|c||}
\hline
 & $1$ & $1^{'}$ & $2_1$ & $2_2$ & $2_3$ & $2_4$  \\
\hline\hline
$1$&&&$\times\times$&&& \\
\hline
$1^{'}$&&&$\times\times$&&& \\
\hline\hline
$2_1$&$\times\times$&$\times\times$&$\times\times$&&& \\
\hline
$2_2$&&&&&$\times\times$&$\times\times$ \\
\hline
$2_3$&&&&$\times\times$&&$\times\times$\\
\hline
$2_4$&&&&$\times\times$&$\times\times$& \\
\hline
\end{tabular}

\bigskip
\bigskip

\noindent This model is also manifestly non-chiral due to the symmetry
of the table.

\bigskip
\bigskip

\bigskip

\noindent For ${\bf 4} = (2_1, 2_2)$ the table is:

\bigskip
\bigskip
\bigskip

\begin{tabular}{||c||c|c||c|c|c|c||}
\hline
 & $1$ & $1^{'}$ & $2_1$ & $2_2$ & $2_3$ & $2_4$  \\
\hline\hline
$1$&&&$\times$&$\times$&& \\
\hline
$1^{'}$&&&$\times$&$\times$&& \\
\hline\hline
$2_1$&$\times$&$\times$&$\times$&&$\times$&$\times$ \\
\hline
$2_2$&$\times$&$\times$&&$\times$&$\times$&$\times$ \\
\hline
$2_3$&&&$\times$&$\times$&&$\times\times$\\
\hline
$2_4$&&&$\times$&$\times$&$\times\times$& \\
\hline
\end{tabular}

\bigskip
\bigskip

\noindent Again, this model is manifestly non-chiral.
18/5 does not lend itself to chirality.

This is easy to understand when one realizes that all
of the irreducible representations of 18/5
are individually either real or pseudoreal \cite{books} making a complex embedding of {\bf 4} impossible.

\bigskip
\bigskip

\newpage

\underline{{\bf g = 20.}}

\bigskip

\noindent One non-pseudoreal group, an SDPG. In the
notation of Thomas and Wood\cite{books} it is $20/5$.

\bigskip

\noindent \underline{Group 20/5; also designated $Z_5 \tilde{\times} Z_4$}   

\bigskip
\bigskip
\bigskip

\noindent The group has four singlets $1_1, 1_2, 1_3, 1_4$ and a $4$.
The singlets $1_1, 1_3$ are real and the other two form a
complex conjugate pair $1_2 = 1_4^*$.
The {\bf 6} which is the antisymmetric product
{\bf 6} = $(4 \times 4)_a$ must be real for a legitimate embedding.
The two inequivalent choices, bearing in mind the multiplication table
provided in the Appendix are ${\bf 4} = (1_2, 1_2, 1_2, 1_2)$ 
and ${\bf 4} = (1_2, 1_2, 1_2, 1_4)$.

\bigskip

The first ${\bf 4} = (1_2, 1_2, 1_2, 1_2)$ yields the chiral fermions
in the following table:

\bigskip
\bigskip

\begin{tabular}{||c||c|c|c|c||c||}
\hline
 & $1_1$ & $1_2$ & $1_3$ & $1_4$ & $4$  \\
\hline\hline
$1_1$&&$\times\times\times\times$&&& \\
\hline
$1_2$&&&$\times\times\times\times$&& \\
\hline
$1_3$&&&&$\times\times\times\times$& \\
\hline
$1_4$&$\times\times\times\times$&&&& \\
\hline\hline
$4$&&&&&$\times\times\times\times\times$ \\
\hline
\end{tabular}

\bigskip
\bigskip

\noindent Putting $N=3$ this embedding gives four chiral families
when we identify $SU(3)_3 \equiv SU(3)_4$ and drop
real representations, giving:

\begin{equation}
4[ (3, \bar{3}, 1) + (1, 3, \bar{3}) + (\bar{3}, 1, 3)]
\end{equation}

\noindent under $SU(3) \times SU(3) \times SU(3)$. This possibility
for the 20/5 nonabelian orbifold certainly merits further study.

\bigskip

\noindent The symmetry breaking for this model will be investigated in the
subsequent section.

\bigskip
\bigskip

The second inequivalent embedding ${\bf 4} = (1_2, 1_2, 1_2, 1_4)$
gives rise to the table:

\bigskip
\bigskip

\begin{tabular}{||c||c|c|c|c||c||}
\hline
 & $1_1$ & $1_2$ & $1_3$ & $1_4$ & $4$  \\
\hline\hline
$1_1$&&$\times\times\times$&&$\times$& \\
\hline
$1_2$&$\times$&&$\times\times\times$&& \\
\hline
$1_3$&&$\times$&&$\times\times\times$& \\
\hline
$1_4$&$\times\times\times$&&$\times$&& \\
\hline\hline
$4$&&&&&$\times\times\times\times\times$ \\
\hline
\end{tabular}

\bigskip

\bigskip
\bigskip

\noindent Identifying $SU(3)_3 \equiv SU(3)_4$ as before for $N=3$
this is less chiral and gives rise to just two chiral
families.

\begin{equation}
4[ (3, \bar{3}, 1) + (1, 3, \bar{3}) + (\bar{3}, 1, 3)]
\end{equation}

\noindent under $SU(3) \times SU(3) \times SU(3)$.

\bigskip
\bigskip

\newpage

\bigskip
\bigskip

\underline{{\bf g = 21.}}

\bigskip

\noindent One non-pseudoreal group, an SDPG. In the
notation of Thomas and Wood\cite{books} it is: $21/2$.

\bigskip

\noindent \underline{Group 21/2; also designated $Z_7 \tilde{\times} Z_3$}

\bigskip
\bigskip
\bigskip

This group has irreps which comprise three singlets $1_1, 1_2, 1_3$ and
two triplets $3_1, 3_2$. With the embedding
${\bf 4} = (1_2, 3_1)$ (recall that $1_1$ must be avoided to obtain
${\cal N} = 0$), the resultant fermions are given by:

\bigskip
\bigskip

\bigskip

\begin{tabular}{||c||c|c|c||c|c||}
\hline
 & $1_1$ & $1_2$ & $1_3$ & $3_1$ & $3_2$  \\
\hline\hline
$1_1$&&$\times$&&$\times$& \\
\hline
$1_2$&&&$\times$&$\times$& \\
\hline
$1_3$&$\times$&&&$\times$& \\
\hline\hline
$3_1$&&&&$\times\times$&$\times\times$ \\
\hline
$3_2$&$\times$&$\times$&$\times$&$\times$&$\times\times$ \\
\hline
\end{tabular}

\bigskip
\bigskip

Putting $N = 2$, the gauge group is $SU(2)^3 \times SU(6)^2$. 
Clearly the model is chiral as seen in the asymmetry of the table.
For example, put $SU(2)_L \equiv SU(2)_1$,
$SU(2)_R \equiv SU(2)_2$, break $SU(2)_3$ entirely and
${\bf 6_1} \rightarrow {\bf 4}, {\bf 6_2} \rightarrow {\bf \bar{4}}$,
to find two chiral families.

\newpage

\underline{{\bf g = 24.}}

\bigskip

\noindent The non-pseudoreal groups number six, and three are SDPGs. In the
notation of Thomas and Wood\cite{books}
they are: $24/7,8,9,13,14,15$.
So we now treat these in the order they are
enumerated by Thomas and Wood.

\bigskip
\bigskip

\noindent \underline{Group 24/7; also designated $D_4 \times Z_3$}

\bigskip
\bigskip

\noindent This has twelve singlets $1_1\alpha^i, 1_2\alpha^i, 1_3\alpha^i, 1_4\alpha^i$ 
(i = 0 - 2)
and three doublets $2\alpha^i$ (i = 0 - 2); here $\alpha = exp (i\pi/3)$.
The embedding ${\bf 4} = (1_1\alpha, 1_2, 2\alpha)$
was studied in detail in our previous article 
\cite{FK2} where it was shown how it can lead to precisely three
chiral families in the standard model.

\bigskip

For completeness we include the table for the chiral fermions (it was
presented in a different equivalent way in \cite{FK2}):

\bigskip
\bigskip

\begin{tabular}{||c||c|c|c|c|c||c|c|c|c|c||c|c|c|c|c||}
\hline
 & $1_1$ & $1_2$ & $1_3$ & $1_4$ & $2$ &$1_1\alpha$& $1_2\alpha$ & $1_3\alpha$ & $1_4\alpha$ & $2\alpha$ 
& $1_1\alpha^2$ & $1_2\alpha^2$ & $1_3\alpha^2$ & $1_4\alpha^2$& $2\alpha^2$  \\
\hline\hline
$1_1$&&$\times$&&&&$\times$&&&&$\times$&&&&& \\
\hline
$1_2$&$\times$&&&&&&$\times$&&&$\times$&&&&& \\
\hline
$1_3$&&&&$\times$&&&&$\times$&&$\times$&&&&& \\
\hline
$1_4$&&&$\times$&&&&&&$\times$&$\times$&&&&& \\
\hline
$2$&&&&&$\times$&$\times$&$\times$&$\times$&$\times$&$\times$&&&&& \\
\hline\hline
$1_1\alpha$&&&&&&&$\times$&&&&$\times$&&&&$\times$ \\
\hline
$1_2\alpha$&&&&&&$\times$&&&&&&$\times$&&&$\times$  \\
\hline
$1_3\alpha$&&&&&&&&&$\times$&&&&$\times$&&$\times$ \\
\hline
$1_4\alpha$ &&&&&&&&$\times$&&&&&&$\times$&$\times$ \\
\hline
$2\alpha$ &&&&&&&&&&$\times$&$\times$&$\times$&$\times$&$\times$&$\times$ \\
\hline\hline
$1_1\alpha^2$&$\times$&&&&$\times$&&&&&&&$\times$&&&  \\
\hline
$1_2\alpha^2$&&$\times$&&&$\times$&&&&&&$\times$&&&&\\
\hline
$1_3\alpha^2$&&&$\times$&&$\times$&&&&&&&&&$\times$& \\
\hline
$1_4\alpha^2$&&&&$\times$&$\times$&&&&&&&&$\times$&&\\
\hline
$2\alpha^2$&$\times$&$\times$&$\times$&$\times$&$\times$&&&&&&&&&&$\times$  \\
\hline\hline
\end{tabular}

\bigskip
\bigskip

\bigskip
\bigskip

\noindent By identifying $SU(4)$ with the diagonal subgroup of
$SU(4)_{2,3}$, breaking $SU(4)_1$ to $SU(2)_L^{'}\times
SU(2)_R^{'}$, then identifying
$SU(2)_L$ with the diagonal subgroup
of $SU(2)_{6,7,8}$ and $SU(2)_L^{'}$
and $SU(2)_R$ with the diagonal subgroup of
$SU(2)_{10,11,12}$ and $SU(2)_R^{'}$
then leads to a three-family model as explained already in \cite{FK2}.

\bigskip

It is convenient to represent the chiral fermions in a quiver diagram\cite{DM}
as shown in the Figure:

\bigskip
\bigskip
\bigskip
\bigskip

\begin{center}
***Insert figure and caption here***

\end{center}

\bigskip
\bigskip
\bigskip
\bigskip

\noindent This model is especially interesting because, uniquely among
the large number of models examined in this study, the prescribed scalars are 
sufficient to break the gauge symmetry to that of the standard model.

\bigskip
\bigskip

\newpage

\noindent \underline{Group 24/8; also designated $Q \times Z_3$}

\bigskip

\noindent The multiplication tables of $D_4$ and $Q$ and hence the
multiplication tables of 24/7 and 24/8 are identical. Model building
for 24/8 is therefore the same as 24/7 and merits
no additional discussion.

\bigskip
\bigskip 

\noindent \underline{Group 24/9; also designated $D_3 \times Z_4$}

\bigskip

This group generates one of the richest sets of chiral model in the class of models discussed
in this paper. The group has as irreps eight singlets $(1_1\alpha^j, 1_2\alpha^j)$ 
and four doublets $2\alpha^j$ (j = 0, 1, 2, 3),
where $\alpha = exp (i \pi / 4)$.

\bigskip

\noindent The embedding ${\bf 4} = (1_1 \alpha^{a_1}, 1_2\alpha^{a_2}, 2 \alpha^{a_3})$
must satisfy $a_1 \neq 0$ (for ${\cal N} = 0$) and
$a_1 + a_2 = -2a_3 $ (mod 4) (to ensure reality of ${\bf 6} = ({\bf 4} \times {\bf 4})_a$). 
There are several interesting possibilities including:
$(1_1\alpha, 1_2\alpha, 2\alpha))$,
$(1_1\alpha, 1_2\alpha^3, 2\alpha^2))$,
$(1_1\alpha^2, 1_2, 2\alpha^3))$,
$(1_1\alpha^2, 1_2, 2\alpha))$, and
$(1_1\alpha^2, 1_2\alpha^2, 2))$. The third and fourth cases are equivalent as can be seen by letting $\alpha$ go to ${\alpha}^{-1}$, 
and the last case has only real fermions since ${\alpha}^{2}=-1$.

\bigskip

\bigskip

The fermions for ${\bf 4=(1}_{{\bf 1}}{\bf \alpha }^{{\bf 2}}{\bf
,1}%
_{2}{\bf \alpha }^{{\bf 2}},{\bf 2)}$ are vectorlike.

\bigskip
\bigskip

Moving on to 24/9 with ${\bf 4=(1}_{{\bf 1}}{\bf \alpha ,{\bf
1}_{2}\alpha }%
^{{\bf 3}},{\bf 2{\bf \alpha }^{{\bf 2}})}$ we find the fermions are
chiral
and fall into the irrep:

\bigskip

\begin{tabular}{|c||c|c|c|c|c|c|c|c|c|c|c|c|}
\hline
 & $1_{1}$ & $1_{2}$ & 2 & $1_{1}\alpha $ & $1_{2}\alpha $ &
$2\alpha $ & $1_{1}\alpha^{2}$ & $1_{2}\alpha^{2}$ & $2\alpha ^{2}$ 
&$1_{1}\alpha^{3}$ & $1_{2}\alpha^{3}$ & $2\alpha^{3}$ \\ \hline\hline
$1_{1}$ &  &  &  & $\times $ &  &  &  &  & $\times $ &  & $\times $ &
\\
\hline
$1_{2}$ &  &  &  &  & $\times $ &  &  &  & $\times $ & $\times $ &  &
\\
\hline
2 &  &  &  &  &  & $\times $ & $\times $ & $\times $ & $\times $ &  &  &
$%
\times $ \\ \hline\hline
$1_{1}\alpha $ &  & $\times $ &  &  &  &  & $\times $ &  &  &  &  &
$\times$
\\ \hline
$1_{2}\alpha $ & $\times $ &  &  &  &  &  &  & $\times $ &  &  &  &
$\times $
\\ \hline
2$\alpha $ &  &  & $\times $ &  &  &  &  &  & $\times $ & $\times $ & $%

\times $ & $\times $ \\ \hline\hline
$1_{1}\alpha^{2}$ &  &  & $\times $ &  & $\times $ &  &  &  &  &
$\times $
&  &  \\ \hline
$1_{2}\alpha^{2}$ &  &  & $\times $ & $\times $ &  &  &  &  &  &  &
$\times
$ &  \\ \hline
2$\alpha^{2}$ & $\times $ & $\times $ & $\times $ &  &  & $\times $ &
&  &
&  &  & $\times $ \\ \hline\hline
$1_{1}\alpha^{3}$ & $\times $ &  &  &  &  & $\times $ &  & $\times $ &
&
&  &  \\ \hline
$1_{1}\alpha^{3}$ &  & $\times $ &  &  &  & $\times $ & $\times $ &  &
&
&  &  \\ \hline
$2\alpha^{3}$ &  &  & $\times $ & $\times $ & $\times $ & $\times $ &
&  &
$\times $ &  &  &  \\ \hline\hline
\end{tabular}

\bigskip
\bigskip

\noindent With the embedding ${\bf 4} = (1_1\alpha, 1_2\alpha, 2\alpha)$ the
chiral fermions are:

\bigskip
\bigskip

\begin{tabular}{|c||c|c|c|c|c|c|c|c|c|c|c|c|}
\hline
& $1_{1}$&$1_{2}$ & 2 & $1_{1}\alpha $ & $1_{2}\alpha $ &
$2\alpha $ & $1_{1}\alpha^{2}$ & $1_{2}\alpha^{2}$ & $2\alpha^{2}$ &
$1_{1}\alpha^{3}$ & $1_{2}\alpha^{3}$ & $2\alpha^{3}$ \\ 

\hline\hline

1$_{1}$ &  &  &  & $\times $ & $\times$  &$\times$  &  &  & &  & &\\

\hline

1$_{2}$ &  &  &  & $\times $ & $\times$  &$\times$  &  &  & &  & &\\

\hline

$2$ &  &  &  & $\times $ & $\times$  &$\times\times\times$  &  &  & &  & &\\

\hline\hline

$1_{1}\alpha$ &  & &  &  &  &  & $\times $ &$\times$  &$\times$  &  &  &\\ \hline
$1_{2}\alpha$ &  & &  &  &  &  & $\times $ &$\times$  &$\times$  &  &  &\\ \hline
$2\alpha$ &  & &  &  &  &  & $\times $ &$\times$  & $\times\times\times$  &  &  &\\ 
\hline\hline

$1_{1}\alpha ^{2}$ &  &  & &  & &  &  &  &  &$\times $& $\times$ & $\times$ \\ \hline
$1_{2}\alpha ^{2}$ &  &  & &  & &  &  &  &  &$\times $& $\times$ & $\times$ \\ \hline
$2\alpha ^{2}$ &  &  & &  & &  &  &  &  &$\times $& $\times$ & $\times\times\times$ \\ 
\hline\hline
$1_{1}\alpha^{3}$&$\times$&$\times$&$\times$&&&&&&&&&  \\ \hline
$1_{2}\alpha^{3}$&$\times$&$\times$&$\times$&&&&&&&&&  \\ \hline
$2\alpha^{3}$&$\times$&$\times$&$\times\times\times$&&&&&&&&&  \\

\hline\hline
\end{tabular}

\bigskip
\bigskip

\noindent Identifying $SU(2)_L$ with the diagonal subgroup of $SU(2)_{1,2,3,4}$,
$SU(2)_R$ with the diagonal subgroup of $SU(2)_{5,6,7,8}$
and the ${\bf 4}$ of $SU(4)$ with the ${\bf 4}$ of $SU(4)_{2,3}$
and the ${\bf \bar{4}}$ of $SU(4)_{1,4}$ leads to eight chiral families.

\bigskip
\bigskip

\noindent Taking the embedding ${\bf 4} = (1_1\alpha^2, 1_2, 2\alpha)$
gives as chiral fermions:

\bigskip
\bigskip

\bigskip

\begin{tabular}{||c||c|c|c||c|c|c||c|c|c||c|c|c||}
\hline
 & $1_1$ & $1_2$ & $2$ & $1_1\alpha$ & $1_2\alpha$& $2\alpha$ & $1_1\alpha^2$ & $1_2\alpha^2$ 
&$2\alpha^2$&$1_1\alpha^3$&$1_2\alpha^3$&$2\alpha^3$  \\
\hline\hline
$1_1$&&$\times$&&&&$\times$&$\times$&&&&& \\
\hline
$1_2$&$\times$&&&&&$\times$&&$\times$&&& & \\
\hline
$2$&&&$\times$&$\times$&$\times$&$\times$& &&$\times$&&&\\
\hline\hline
$1_1\alpha$&&&&&$\times$&&&&$\times$ &$\times$&&\\
\hline
$1_2\alpha$&&&&$\times$&&&&&$\times$&&$\times$& \\
\hline
$2\alpha$&&&&&&$\times$&$\times$&$\times$&$\times$&&&$\times$ \\
\hline\hline
$1_1\alpha^2$ &$\times$&&&&&&&$\times$&&&&$\times$ \\
\hline
$1_2\alpha^2$ &&$\times$& &&& &$\times$&& &&&$\times$ \\
\hline
$2\alpha^2$ &&&$\times$ &&& &&&$\times$ &$\times$ &$\times$& $\times$ \\
\hline\hline
$1_1\alpha^3$ &&&$\times$ &$\times$&& &&& &&$\times$& \\
\hline
$1_2\alpha^3$ &&&$\times$ &&$\times$& &&& &$\times$&& \\
\hline
$2\alpha^3$ &$\times$&$\times$&$\times$ &&&$\times$ &&& &&&$\times$ \\
\hline
\end{tabular}

\bigskip
\bigskip

\noindent We identify $SU(2)_L$, $SU(2)_R$ with the diagonal subgroups
of $SU(2)_{1,2}$ and $SU(2)_{3,4}$, respectively
and break completely $SU(2)_{5,6,7,8}$. The generalized
color embedding ${\bf 4} \equiv {\bf 4_1} \equiv {\bf 4_2} \equiv {\bf \bar{4}_3} \equiv {\bf \bar{4}_4}$
leads to four chiral families. This can be reduced to three families
by further symmetry breaking
using the same idea as in \cite{FK2}.

\bigskip
\bigskip
\bigskip

An even more interesting embedding for 24/9 is to set ${\bf 4} = (2\alpha,2\alpha)$
which gives a real ${\bf 6}$ as required (since $\alpha^2=-1$ is real).
The table for fermions is:

\begin{tabular}{||c||c|c|c||c|c|c||c|c|c||c|c|c||}
\hline
 & $1_1$ & $1_2$ & $2$ & $1_1\alpha$ & $1_2\alpha$& $2\alpha$ & $1_1\alpha^2$ & $1_2\alpha^2$ 
&$2\alpha^2$&$1_1\alpha^3$&$1_2\alpha^3$&$2\alpha^3$  \\
\hline\hline
$1_1$&&&&&&$\times\times$&&&&&& \\
\hline
$1_2$&&&&&&$\times\times$&&&&& & \\
\hline
$2$&&&&$\times\times$&$\times\times$&$\times\times$&&&&&&\\
\hline\hline
$1_1\alpha$&&&&&&&&&$\times\times$&&&\\
\hline
$1_2\alpha$&&&&&&&&&$\times\times$&&& \\
\hline
$2\alpha$&&&&&&&$\times\times$&$\times\times$&$\times\times$&&& \\
\hline\hline
$1_1\alpha^2$&&&&&&&&&&&&$\times\times$ \\
\hline
$1_2\alpha^2$&&&&&&&&&&&&$\times\times$ \\
\hline
$2\alpha^2$&&&&&&&&&&$\times\times$&$\times\times$&$\times\times$ \\
\hline\hline
$1_1\alpha^3$&&&$\times\times$&&&&&&&&& \\
\hline
$1_2\alpha^3$&&&$\times\times$&&&&&&&&& \\
\hline
$2\alpha^3$&$\times\times$&$\times\times$&$\times\times$&&&&&&&&& \\
\hline
\end{tabular}

\bigskip
\bigskip
Identifying $SU(2)_L$ with the diagonal subgroup of $SU(2)_{1,3,5,7}$,
$SU(2)_R$ with the diagonal subgroup of $SU(2)_{2,4,6,8}$, 
breaking $SU(4)_{1,3}$ and keeping the unbroken $SU(4)$
which is the diagonal subgroup of $SU(4)_{2,4}$ gives rise to 
eight chiral families:
\begin{equation}
8[(2, 1, \bar{4}) + (1, 2, 4)]
\end{equation}

The possibility of achieving the relevant symmetry breaking will be
examined below in Section V.

\bigskip

\newpage

\bigskip

\noindent \underline{Group 24/13; also designated $Q \tilde{\times} Z_3$}

\bigskip

This group has three singlets $1_1, 1_2, 1_3$, three doublets
$2_1, 2_2, 2_3$ and one triplet $3$. For $N = 2$ the gauge group is
therefore $SU(2)^3 \times SU(4)^3 \times SU(6)$.

\bigskip

\noindent With the embedding ${\bf 4} = (2_1, 2_2)$ the chiral fermions
are:

\bigskip
\bigskip

\bigskip

\begin{tabular}{||c||c|c|c||c|c|c||c||}
\hline
 & $1_1$ & $1_2$ & $1_3$ & $2_1$ & $2_2$& $2_3$ & $3$  \\
\hline\hline
$1_1$&&&&$\times$&$\times$&& \\
\hline
$1_2$&&&&&$\times$&$\times$& \\
\hline
$1_3$&&&&$\times$&&$\times$& \\
\hline\hline
$2_1$&$\times$&$\times$&&&&&$\times\times$ \\
\hline
$2_2$&&$\times$&$\times$&&&&$\times\times$ \\
\hline
$2_3$&$\times$&&$\times$&&&&$\times\times$ \\
\hline\hline
$3$ &&&&$\times\times$&$\times\times$&$\times\times$& \\
\hline
\end{tabular}

\bigskip
\bigskip

\noindent If we identify $SU(2)_L \equiv SU(2)_3$,
$SU(2)_R \equiv SU(2)_2$, and break $SU(2)_1$ there
are two chiral families for
${\bf 4} \equiv {\bf 4_1} \equiv {\bf \bar{4}_2} \equiv {\bf \bar{4}_3}$.

\bigskip

\newpage

\bigskip

\noindent If, instead, we embed ${\bf 4} = (2_2, 2_3)$ the
fermions fall according to the following table:

\bigskip
\bigskip
\bigskip

\begin{tabular}{||c||c|c|c||c|c|c||c||}
\hline
 & $1_1$ & $1_2$ & $1_3$ & $2_1$ & $2_2$& $2_3$ & $3$  \\
\hline\hline
$1_1$&&&&&$\times$&$\times$& \\
\hline
$1_2$&&&&$\times$&&$\times$& \\
\hline
$1_3$&&&&$\times$&$\times$&& \\
\hline\hline
$2_1$&&$\times$&$\times$&&&&$\times\times$ \\
\hline
$2_2$&$\times$&&$\times$&&&&$\times\times$ \\
\hline
$2_3$&$\times$&$\times$&&&&&$\times\times$ \\
\hline\hline
$3$ &&&&$\times\times$&$\times\times$&$\times\times$& \\
\hline
\end{tabular}

\bigskip
\bigskip

\noindent This model is manifestly non-chiral because of the
total symmetry of the table.

\bigskip
\bigskip

\newpage

\noindent \underline{Group 24/14; also designated $Z_8 \tilde{\times} Z_3$}

\bigskip
\bigskip

\noindent There are eight singlets and four doublets, with multiplication
table as in Appendix A. With the embedding ${\bf 4} = (2_2, 2_4)$ one
arrives at the fermions:

\bigskip
\bigskip

\begin{tabular}{||c||c|c|c|c|c|c|c|c||c|c|c|c||}
\hline
 & $1_1$ & $1_2$ & $1_3$ & $1_4$ & $1_5$& $1_6$ & $1_7$ & $1_8$ & $2_1$ & $2_2$ & $2_3$ & $2_4$ \\
\hline\hline
$1_1$&&&&&&&&&&$\times$&&$\times$ \\
\hline
$1_2$&&&&&&&&&$\times$&&$\times$& \\
\hline
$1_3$&&&&&&&&&&$\times$&&$\times$ \\
\hline
$1_4$&&&&&&&&&$\times$&&$\times$& \\
\hline
$1_5$&&&&&&&&&&$\times$&&$\times$ \\
\hline
$1_6$&&&&&&&&&$\times$&&$\times$& \\
\hline
$1_7$ &&&&&&&&&&$\times$&&$\times$ \\
\hline
$1_8$ &&&&&&&&&$\times$&&$\times$&  \\
\hline\hline
$2_1$ && $\times$ && $\times$ && $\times$ && $\times$ && $\times$ && $\times$ \\
\hline
$2_2$ & $\times$ && $\times$ && $\times$ && $\times$ && $\times$ && $\times$ & \\
\hline
$2_3$ && $\times$ && $\times$ && $\times$ && $\times$ && $\times$ && $\times$ \\
\hline
$2_4$ & $\times$ && $\times$ && $\times$ && $\times$ && $\times$ && $\times$ & \\
\hline
\end{tabular}

\bigskip
\bigskip

\noindent This arrangement has zero families.

\bigskip

\bigskip

\bigskip

A chiral embedding is ${\bf 4} = (2_1, 2_2)$ giving rise to the fermions:

\bigskip
\bigskip

\begin{tabular}{||c||c|c|c|c|c|c|c|c||c|c|c|c||}
\hline
 & $1_1$ & $1_2$ & $1_3$ & $1_4$ & $1_5$& $1_6$ & $1_7$ & $1_8$ & $2_1$ & $2_2$ & $2_3$ & $2_4$ \\
\hline\hline
$1_1$&&&&&&&&&$\times$&$\times$&& \\
\hline
$1_2$&&&&&&&&&&$\times$&$\times$& \\
\hline
$1_3$&&&&&&&&&&&$\times$&$\times$ \\
\hline
$1_4$&&&&&&&&&$\times$&&&$\times$ \\
\hline
$1_5$&&&&&&&&&$\times$&$\times$&& \\
\hline
$1_6$&&&&&&&&&&$\times$&$\times$& \\
\hline
$1_7$ &&&&&&&&&&&$\times$&$\times$ \\
\hline
$1_8$ &&&&&&&&&$\times$&&&$\times$  \\
\hline\hline
$2_1$&$\times$&$\times$&&&$\times$&$\times$&&&$\times$&$\times$&& \\
\hline
$2_2$&&$\times$&$\times$&&&$\times$&$\times$&&&$\times$&$\times$& \\
\hline
$2_3$ &&&$\times$&$\times$&&&$\times$&$\times$&&&$\times$&$\times$ \\
\hline
$2_4$&$\times$&&&$\times$&$\times$&&&$\times$&$\times$&&&$\times$ \\
\hline
\end{tabular}

\bigskip
\bigskip

If we identify $SU(2)_L$ as the diagonal subgroup
of $SU(2)_{1,2,5,6}$ and $SU(2)_R$ as the diagonal
subgroup of $SU(2)_{3,4,7,8}$, then identify the {\bf 4} of $SU(4)$
with the ${\bf 4}$ of $SU(4)_{2,3}$
and the ${\bf \bar{4}}$ of $SU(4)_{1,4}$, this
model has eight chiral families under
$SU(2)_L \times SU(2)_R \times SU(4)$.

\bigskip
\bigskip

\noindent \underline{Group 24/15; also designated $D_4 \tilde{\times} Z_3$}

\bigskip
\bigskip
\bigskip

\noindent The group 24/15 has nine inequivalent irreducible representations,
four singlets and five doublets.

\bigskip
\bigskip

\noindent With the embedding ${\bf 4} = (2_3, 2_5)$, the fermion table is:

\bigskip
\bigskip

\begin{tabular}{||c||c|c|c|c||c|c|c|c|c||}
\hline
 & $1_1$ & $1_2$ & $1_3$ & $1_4$ & $2_1$ & $2_2$ & $2_3$ & $2_4$ & $2_5$ \\
\hline\hline
$1_1$&&&&&&&$\times$&&$\times$ \\
\hline
$1_2$&&&&&&&&$\times$&$\times$ \\
\hline
$1_3$&&&&&&&$\times$&&$\times$ \\
\hline
$1_4$&&&&&&&&$\times$&$\times$ \\
\hline\hline
$2_1$&&&&&&&$\times$&$\times\times$&$\times$ \\
\hline
$2_2$&&&&&&&$\times\times$&$\times$&$\times$ \\
\hline
$2_3$ &&$\times$&&$\times$&$\times\times$&$\times$&&& \\
\hline
$2_4$ &$\times$&&$\times$&&$\times$&$\times\times$&&&  \\
\hline
$2_5$ &$\times$& $\times$ &$\times$& $\times$ &$\times$& $\times$ &&&  \\
\hline
\end{tabular}

\bigskip
\bigskip

\noindent Identifying $SU(2)_L \equiv SU(2)_{1,3}$ and
$SU(2)_R \equiv SU(2)_{2,4}$ gives rise to two chiral families.

\bigskip

\bigskip

\newpage

\bigskip

Another chiral embedding is ${\bf 4} = (1_2, 1_3, 2_3)$ which
gives the chiral fermions:

\bigskip
\bigskip

\begin{tabular}{||c||c|c|c|c||c|c|c|c|c||}
\hline
 & $1_1$ & $1_2$ & $1_3$ & $1_4$ & $2_1$ & $2_2$ & $2_3$ & $2_4$ & $2_5$ \\
\hline\hline
$1_1$&&$\times$&$\times$&&&&$\times$&& \\
\hline
$1_2$&$\times$&&&$\times$&&&&$\times$& \\
\hline
$1_3$&$\times$&&&$\times$&&&$\times$&& \\
\hline
$1_4$&&$\times$&$\times$&&&&&$\times$& \\
\hline\hline
$2_1$&&&&&$\times$&$\times$&&$\times$&$\times$ \\
\hline
$2_2$&&&&&$\times$&$\times$&$\times$&&$\times$ \\
\hline
$2_3$&&$\times$&&$\times$&$\times$&&$\times$&$\times$& \\
\hline
$2_4$&$\times$&&$\times$&&&$\times$&$\times$&$\times$&  \\
\hline
$2_5$&&&&&$\times$&$\times$&&&$\times\times$  \\
\hline
\end{tabular}

\bigskip

\noindent Identifying $SU(2)_L$ with the diagonal subgroup
of $1_1$ and $1_3$, $SU(2)_R$ with $1_2$ and $1_4$, and then identifying 
$2_3={\bf 4}$ and $2_4 = {\bf \bar{4}}$ and finally breaking the other
three $SU(4)$'s gives rise
to six chiral families.

\bigskip

\newpage

\bigskip

\noindent As an alternative 24/15 model we can embed
${\bf 4} = (2_3, 2_3)$ and obtain:

\bigskip
\bigskip

\begin{tabular}{||c||c|c|c|c||c|c|c|c|c||}
\hline
 & $1_1$ & $1_2$ & $1_3$ & $1_4$ & $2_1$ & $2_2$ & $2_3$ & $2_4$ & $2_5$ \\
\hline\hline
$1_1$&&&&&&&$\times\times$&& \\
\hline
$1_2$&&&&&&&&$\times\times$& \\
\hline
$1_3$&&&&&&&$\times\times$&& \\
\hline
$1_4$&&&&&&&&$\times\times$& \\
\hline\hline
$2_1$&&&&&&&&$\times\times$&$\times\times$ \\
\hline
$2_2$&&&&&&&$\times\times$&&$\times\times$ \\
\hline
$2_3$ &&$\times\times$&&$\times\times$&$\times\times$&&&& \\
\hline
$2_4$ &$\times\times$&&$\times\times$&&&$\times\times$&&&  \\
\hline
$2_5$ &&&&&$\times\times$& $\times\times$ &&&  \\
\hline
\end{tabular}

\bigskip
\bigskip

\noindent With $SU(2)_L$, $SU(2)_R$ as diagonal subgroups of
$SU(2)_1 \times SU(2)_3$ and
$SU(2)_2 \times SU(2)_4$ respectively,
and breaking completely $SU(4)_4$,
this leads to four chiral families.

\bigskip
\bigskip

\newpage

\underline{{\bf g = 27.}}

\bigskip

\noindent The non-pseudoreal groups number two and both are SDPGs. In the
notation of Thomas and Wood\cite{books}
they are: $27/4,5$.
So we now treat these in the order they are
enumerated by Thomas and Wood.

\bigskip

\noindent \underline{Group 27/4; also designated $Z_9 \tilde{\times} Z_3$}

\bigskip

\noindent 27/4 has nine singlet $1_1, ...., 1_9$ and two triplet $3_1, 3_2$
irreducible representations.

\noindent 
We may choose the embedding $4 = (1_2, 3_1)$. The chiral fermions are:

\bigskip

\begin{tabular}{||c||c|c|c|c|c|c|c|c|c||c|c||}
\hline
& $1_1$ & $1_2$ & $1_3$ & $1_4$ & $1_5$ & $1_6$ & $1_7$ & $1_8$ & $1_9$ & $3_1$ & $3_2$ \\
\hline\hline
$1_1$&&$\times$&&&&&&&&$\times$& \\
\hline
$1_2$&&&$\times$&&&&&&&$\times$& \\
\hline
$1_3$&$\times$&&&&&&&&&$\times$& \\
\hline
$1_4$&&&&&$\times$&&&&&$\times$& \\
\hline
$1_5$&&&&&&$\times$&&&&$\times$& \\
\hline
$1_6$&&&&$\times$&&&&&&$\times$& \\
\hline
$1_7$ &&&&&&&&$\times$&&$\times$& \\
\hline
$1_8$ &&&&&&&&&$\times$&$\times$&  \\
\hline
$1_9$ &&&&&&&$\times$&&&$\times$&\\
\hline\hline
$3_1$ &&&&&&&&&&$\times$&$\times\times\times$ \\
\hline
$3_2$ &$\times$&$\times$&$\times$&$\times$&$\times$&$\times$&$\times$&
$\times$&$\times$&&$\times$ \\
\hline
\hline
\end{tabular}

\bigskip
\bigskip

\noindent Putting $N=2$, the gauge group is $SU(2)^9 \times SU(6)_1 \times SU(6)_2$
and the chiral fermions are, from the above table:

\begin{equation}
( \sum_{i=1}^{i=9} 2_i, \bar{6}_1 ) + (6_1, \bar{6}_1 + 3 (\bar{6_2}) )
+ (6_2, \sum_{i=1}^{i=9} 2_i ) + (6_2, \bar{6_2} )
\end{equation}

\noindent Though asymmetric in representations, this result is anomaly-free with respect both to $SU(6)_1$ and
$SU(6)_2$.

\bigskip
\bigskip

\bigskip
\bigskip
\noindent \underline{Group 27/5; also designated $(Z_3 \times Z_3) \tilde{\times} Z_3$}

\bigskip

\noindent The multiplication tables, and hence the model-building, are identical
for 27/4 and 27/5. The group 27/5 merits no further separate discussion.

\bigskip
\bigskip
\bigskip

\newpage

\underline{{\bf g = 30.}}

\bigskip

\noindent The non-pseudoreal groups number two, and neither is an SDPG. In the
notation of Thomas and Wood\cite{books}, they are: $30/2,3$.
So we now treat these in the order they are
enumerated by Thomas and Wood.

\bigskip

\noindent \underline{Group 30/2; also designated $D_5 \times Z_3$}

\bigskip

\noindent 30/2 has six singlets $1\alpha^i, 1^{'}\alpha^i$ and six
doublets $2\alpha^i, 2^{'}\alpha^i$ with $\alpha = exp(i \pi/3)$ and 
i = 0, 1, 2.

\bigskip

Choosing ${\bf 4} = (1\alpha, 1^{'}, 2\alpha)$ yields as fermions:

\bigskip
\bigskip

\begin{tabular}{||c||c|c|c|c||c|c|c|c||c|c|c|c||}
\hline
 & $1$ & $1^{'}$ & $2$ & $2^{'}\alpha$ & $1\alpha$& $1^{'}\alpha$ & $2\alpha$ 
& $2^{'}\alpha$ & $1\alpha^2$ & $1^{'}\alpha^2$ & $2\alpha^2$ & $2^{'}\alpha^2$ \\
\hline\hline
$1$&&$\times$&&&$\times$&&$\times$&&&&& \\
\hline
$1^{'}$&$\times$&&&&&$\times$&$\times$&&&&& \\
\hline
$2$&&&$\times$&&$\times$&$\times$&$\times$&$\times$&&&& \\
\hline
$2^{'}$&&&&$\times$&&&$\times$&$\times\times$&&&& \\
\hline\hline
$1\alpha$&&&&&&$\times$&&&$\times$&&$\times$& \\
\hline
$1^{'}\alpha$&&&&&$\times$&&&&&$\times$&$\times$& \\
\hline
$2\alpha$&&&&&&&$\times$&&$\times$&$\times$&$\times$&$\times$ \\
\hline
$2^{'}\alpha$&&&&&&&&$\times$&&&$\times$&$\times\times$  \\
\hline\hline
$1\alpha^2$&$\times$&&$\times$&&&&&&&$\times$&& \\
\hline
$1^{'}\alpha^2$ & &$\times$ & $\times$ &&&& && $\times$ &&& \\
\hline
$2\alpha^2$ &$\times$& $\times$ &$\times$& $\times$ && && & && $\times$ & \\
\hline
$2^{'}\alpha^2$ & && $\times$ &$\times\times$&  &&  && & && $\times$  \\
\hline
\end{tabular}

\bigskip
\bigskip

\noindent Identify $SU(2)_L$ with the diagonal subgroup of
$SU(2)_1 \times SU(2)_2$ (associated with $1, 1^{'}$) and
$SU(2)_R$ with the diagonal subgroup of
$SU(2)_5 \times SU(2)_6$ (associated with $1\alpha^2, 1^{'}\alpha^2$);
break the $SU(4)$s associated with $2, 2\alpha^2$ to arrive at
two chiral families.

\bigskip
\bigskip

\noindent \underline{Group 30/3; also designated $D_3 \times Z_5$}

\bigskip
\bigskip
\bigskip

\noindent This group has irreps which comprise ten singlets
and five doublets and yields, for $N = 2$, the gauge group
$SU(2)^{10} \times SU(4)^5$.

\bigskip
\bigskip

\noindent As we have encountered for groups $D_3 \times Z_p$
(with $g = 6p$) the embedding
${\bf 4} = (1\alpha^{a_1}, 1^{'}\alpha^{a_2}, 2\alpha^{a_3})$
must satisfy $a_1 + a_2 = - 2 a_3$ (mod p) for consistency,
as well as $a_1 \neq 0$ to ensure ${\cal N} = 0$.

\bigskip

\noindent There are several interesting such examples, one of which
is ${\bf 4} = (1\alpha, 1^{'}, 2\alpha^2)$ which gives as fermions:

\bigskip
\bigskip

\begin{tabular}{||c||c|c|c||c|c|c||c|c|c||c|c|c||c|c|c|}
\hline
 & $1$ & $1^{'}$ & $2$ & $1\alpha$ & $1^{'}\alpha$& $2\alpha$ & $1\alpha^2$ & $1^{'}\alpha^2$ & $2\alpha^2$ 
& $1\alpha^3$ & $1^{'}\alpha^3$ & $2\alpha^3$ & $1\alpha^4$& $1^{'}\alpha^4$ & $2\alpha^4$ \\
\hline\hline
$1$&&$\times$&&$\times$&&&&&$\times$&&&&&& \\
\hline
$1^{'}$&$\times$&&&&$\times$&&&&$\times$&&&&&& \\
\hline
$2$&&&$\times$&&&$\times$&$\times$&$\times$&$\times$&&&&&& \\
\hline\hline
$1\alpha$&&&&&$\times$&&$\times$&&&&&$\times$&&& \\
\hline
$1^{'}\alpha$&&&&$\times$&&&&$\times$&&&&$\times$&&& \\
\hline
$2\alpha$&&&&&&$\times$&&&$\times$&$\times$&$\times$&$\times$&&& \\
\hline\hline
$1\alpha^2$&&&&&&&&$\times$&&$\times$&&&&&$\times$  \\
\hline
$1^{'}\alpha^2$&&&&&&&$\times$&&&&$\times$& &&&$\times$ \\
\hline
$2\alpha^2$ & && &&&& && $\times$ &&&$\times$&$\times$&$\times$&$\times$ \\
\hline\hline
$1\alpha^3$ && &$\times$&&& && &&& $\times$&& $\times$ && \\
\hline
$1^{'}\alpha^3$ & && $\times$ &&  &&  && & $\times$&&&& $\times$&  \\
\hline
$2\alpha^3$&$\times$&$\times$&$\times$ &&& &&& &&&$\times$ &&&$\times$ \\
\hline\hline
$1\alpha^4$ &$\times$&& &&&$\times$ &&& &&& &&$\times$& \\
\hline
$1^{'}\alpha^4$ &&$\times$& &&&$\times$ &&& &&& &$\times$&& \\
\hline
$2\alpha^4$ &&&$\times$ &$\times$&$\times$&$\times$ &&& &&& &&&$\times$ \\
\hline
\end{tabular}

\bigskip
\bigskip
\bigskip

\noindent In an obvious notation, the chiral fermions are:

\bigskip

\begin{equation}
(2_1 + 2_2, \bar{4}_3 + 4_4) +
(2_3 + 2_4, \bar{4}_4 + 4_5) +
(2_5 + 2_6, \bar{4}_5 + 4_1) +
(2_7 + 2_8, \bar{4}_1 + 4_2) +
(2_9 + 2_{10}, \bar{4}_2 + 4_3)
\label{30/3}
\end{equation}

\bigskip

\noindent By identifying, for example (there are equivalent cyclic permutations)
$SU(2)_L$ as the diagonal subgroup of
$SU(2)_1 \times SU(2)_2 \times SU(2)_7 \times SU(2)_8$,
$SU(2)_R$ as the diagonal subgroup
of
$SU(2)_5 \times SU(2)_6 \times SU(2)_9 \times SU(2)_{10}$,
generalized color $SU(4)$ as the diagonal subgroup of
$SU(4)_1 \times SU(4)_3$,
and breaking completely $SU(4)_{2,4,5}$
give rise to four chiral families.

\bigskip
\bigskip

\noindent We can examine the infinite series $D_3 \times Z_p$ for
$p \geq 3$ (as necessary for non-pseudoreality). The order is $g = 6p$.
By generalizing the above discussions of 18/3 ($D_3 \times Z_3$),
24/9 ($D_3 \times Z_4$) and 30/3 ($D_3 \times Z_5$) we find
that with the same type of embedding one arrives at a maximal
number of 2[p] chiral families where [x] is the largest integer
not greater than x.
For example, with p = 3, 4, 5, 6, 7, 8, 9, 10,....
one obtains 2, 4, 4, 6, 6, 8, 8, 10.... chiral families resspectively.
This is an example of accessing the more difficult nonabelian $\Gamma$
with $g \geq 32$
at least for orders $g = 6p \geq 36$.

\bigskip
\bigskip

That completes the analysis of the occurrence of chiral fermions for $\Gamma$
with $g \leq 31$. For the cases where there are $ \geq 3$ chiral families, it remains
to check whether the spectrum of complex scalars is sufficient to allow
spontaneous symmetry breaking to the Standard Model gauge group.

This is the subject of the next two sections.

\bigskip
\bigskip
\bigskip

\newpage

\bigskip
\bigskip
\bigskip

\section{The Scalar Sector}

In order to carry out the spontaneous symmetry breaking (SSB) in the
chiral
models we found in the last section, we must first extract the scalar
sector
from eq. (5), where the $6$ is gotten from the embedding of $({\bf
4}\times
{\bf 4})_{A}$ which in turn follows from the embedding of the ${\bf
4}$. We
only consider models of phenomenological interest, i.e., those which
potentially have three or more families, but preferably three. With
this
perspective in mind we first collect the models, they are:

\bigskip

\underline{16/8 with ${\bf 4}=({\bf 2}_{1},{\bf 2}_{1})$} and $\chi = 2^8$ with
$N=2$.

\underline{16/8 with ${\bf 4}=({\bf 1}_{2},{\bf 1}_{5},{\bf 2}_{1})$} and
$\chi = 2^7$
with $N=2$.

\underline{16/11 with ${\bf 4}=({\bf 1}_{2},{\bf 1}_{2},{\bf 1}_{2},{\bf 1}_{2})$} and
$\chi = 432$
with $N=3$.

\underline{16/11 with ${\bf 4}=({\bf 1}_{2},{\bf 1}_{2},{\bf 1}_{2},{\bf 1}_{4})$} and
$\chi = 216$
with $N=3$.

\underline{16/13 with ${\bf 4}=({\bf 1}_{3},{\bf 1}_{4},{\bf 2}_{1})$} and
$\chi = 2^6$
with $N=2$.

\underline{16/13 with ${\bf 4}=({\bf 2}_{1},{\bf 2}_{2})$} and $\chi = 2^6$ with
$N=2$

\underline{16/13 with ${\bf 4}=({\bf 2}_{1},{\bf 2}_{1})$} and $\chi = 2^7$ with
$N=2$.

\underline{18/3 with ${\bf 4}=({\bf 1}\alpha ,{\bf 1}^{\prime },{\bf 2}\alpha )$}
and $\chi = 192$ with $N=2$.

\underline{20/5 with ${\bf 4}=({\bf 1}_{2},{\bf 1}_{2},{\bf 1}_{2},{\bf 1}_{2})$}
and $\chi = 144$ with $N=3$.

\underline{20/5 with ${\bf 4}=({\bf 1}_{2},{\bf 1}_{2},{\bf 1}_{2},{\bf 1}_{4})$}
and $\chi=72$ with $N=3$.

\underline{21/2 with ${\bf 4}=({\bf 1}_{2},{\bf 3}_{1})$} and $\chi = 108$ with
$N=2$.

\underline{24/7 with ${\bf 4}=({\bf 1}\alpha ,{\bf 1}^{\prime },{\bf 2}\alpha )$} 
and $\chi = 240$ with $N=2$.

\underline{24/9 with ${\bf 4}=({\bf 1}_{1}\alpha ,{\bf 1}_{2}\alpha^3 ,{\bf 2}\alpha^2)$} 
and $\chi = 320$ with $N=2$.

\underline{24/9 with ${\bf 4}=({\bf 1}_{1}\alpha ,{\bf 1}_{2}\alpha ,{\bf 2}\alpha)$}
and $\chi = 320$ with $N=2$.

\underline{24/9 with ${\bf 4}=({\bf 1}_{1}\alpha^2 ,{\bf 1}_{2}, {\bf 2}\alpha)$}
and $\chi = 192$ with $N=2$.

\underline{24/9 with ${\bf 4}=({\bf 2}\alpha ,{\bf 2}\alpha )$} and $\chi = 384$
with $ N=2$.

\underline{24/13 with ${\bf 4}=({\bf 2}_{1},{\bf 2}_{2})$} and $\chi = 48$ with
$N=2$.

\underline{24/14 with ${\bf 4}=({\bf 2}_{1},{\bf 2}_{2})$} and $\chi = 192$ with
$N=2$.

\underline{24/15 with ${\bf 4}=({\bf 1}_{2},{\bf 1}_{3},{\bf 2}_{3})$} and $\chi = 2^7$ with
$N=2$.

\underline{24/15 with ${\bf 4}=({\bf 2}_{3},{\bf 2}_{5})$} and $\chi = 2^7$ with
$N=2$.

\underline{24/15 with ${\bf 4}=({\bf 2}_{3},{\bf 2}_{3})$} and $\chi = 2^8$ with
$N=2$.

\underline{27/4 with ${\bf 4}=({\bf 1}_{2},{\bf 3}_{1})$} and $\chi = 324$ with
$N=2$.

\underline{30/2 with ${\bf 4}=({\bf 1}\alpha ,{\bf 1}^{\prime },{\bf 2}\alpha )$}
and $\chi = 336$ with $N=2$.

\underline{30/3 with ${\bf 4}=({\bf 1}\alpha ,{\bf 1}^{\prime },{\bf 2}\alpha^{2})$}
and $\chi = 320$ with $N=2$.

\bigskip

First we consider \underline{16/8 with ${\bf 4}=({\bf 1}_{2},{\bf 1}_{2},{\bf
2}_{1})$},
where we have included this example to demonstrate improper embedding.
This
representation is complex and would be expected to lead to chiral
fermions,
but ${\bf 6}=({\bf 4}\times {\bf 4})_{A}={\bf 1}_{1}+2({\bf 2}_{1}+{\bf
2}
_{1})+({\bf 1}_{5}+{\bf 1}_{6}+{\bf 1}_{7}+{\bf 1}_{8})_{A}$ is complex
(for
any choice of singlet in the last parenthetical expression), and
therefore
the embedding ${\bf 4}=({\bf 1}_{2},{\bf 1}_{2},{\bf 2}_{1})$ is
improper
and we need not consider this or other such models further.

\bigskip

Let us define the chirality measure $\chi $ of a model as the number of
chiral fermion states. 
This variable applies to any irreps and provides
a somewhat finer measure of chirality than the number of families. 
As spontaneous symmetry breaking (SSB) proceeds,
$%
\chi $ decreased (except under unusual circumstances). For instance, the
standard model and minimal $SU(5)$ both have $\chi =45$ initially. By
the
time the symmetry is broken to $SU(3)\times U_{EM}(1)$, $\chi =3$ since
the
neutrino's cannot acquire mass due to global $B-L$ symmetry. On the
other
hand, three family $SO(10)$ and $E_{6}$ models start with $\chi =48$ and
$%
\chi =81$ respectively but both break to $\chi =0.$

In model building with AdSCFTs we are faced with a number of choices. if
we
require the initial model be chiral before SSB, then we need $\chi \geq
45$%
initially. However, since the scale of SSB  $M_{AdS}$ in these models
can
be relatively low (few 10s of $TeV$), vector like models are more
appealing
than usual, and we could allow an initial $\chi =0$ without resorting to
incredibly detailed fine tunings. Our prejudice is to still require a
chiral
model with $\chi \geq 45$ initially in order to gain some control in
model
building, but we want to make it clear that, even though
we have not displayed them explicitly,   
 the entire class of
vectorlike
model based on the nonabelian orbifold classification given here would
be
worthy of detailed study. There are also models (chiral or vectorlike)
that
break from $G_{AdS}$ to $SU(3)\times U_{EM}(1)$ but without going
through $SU(3)\times SU(2)\times U(1)$ directly. As $M_{AdS}$ may be not
far
above $M_{Z}$, there may be models in this class that could be in
agreement
with current data, but again we restrict most of our discussion to
chiral
models that break through the standard model. What is encouraging is the
fact that orbifold AdS/CFTs provide such a wealth of potentially
interesting
models.

\bigskip
\bigskip

\newpage

\bigskip
\bigskip

\underline{16/8 with ${\bf 4}=({\bf 2}_{1},{\bf 2}_{1})$}. Here ${\bf 6}=\ 3({\bf
1}%
_{5})+{\bf 1}_{6}+{\bf 1}_{7}+{\bf 1}_{8}\ \ $ which is real so the
embedding is proper and the scalar sector is:

\bigskip

\begin{tabular}{|c||c|c|c|c|c|c|c|c|c|c|}
\hline
$\otimes $ & 1$_{1}$ & 1$_{2}$ & 1$_{3}$ & 1$_{4}$ & 1$_{5}$ & 1$_{6}$ &
1$%
_{7}$ & 1$_{8}$ & 2 & 2$^{\prime }$ \\ \hline\hline
1$_{1}$ &  &  &  &  & $\times \times \times $ & $\times $ & $\times $ &
$%
\times $ &  &  \\ \hline
1$_{2}$ &  &  &  &  & $\times $ & $\times $ & $\times $ & $\times $ &
&  \\
\hline
1$_{3}$ &  &  &  &  & $\times $ & $\times $ & $\times $ & $\times $ &
&  \\
\hline
1$_{4}$ &  &  &  &  & $\times $ & $\times $ & $\times $ & $\times \times

\times $ &  &  \\ \hline
1$_{5}$ & $\times \times \times $ & $\times $ & $\times $ & $\times $ &
&
&  &  &  &  \\ \hline
1$_{6}$ & $\times $ & $\times $ & $\times $ & $\times $ &  &  &  &  &
&  \\
\hline
1$_{7}$ & $\times $ & $\times $ & $\times $ & $\times $ &  &  &  &  &
&  \\
\hline
1$_{8}$ & $\times $ & $\times $ & $\times $ & $\times \times \times $ &
&
&  &  &  &  \\ \hline
2 &  &  &  &  &  &  &  &  &  &
\begin{tabular}{c}
$\times \times \times $ \\
$\times \times \times $%
\end{tabular}
\\ \hline
2$^{\prime }$ &  &  &  &  &  &  &  &  &
\begin{tabular}{c}
$\times \times \times $ \\
$\times \times \times $%
\end{tabular}
&  \\ \hline
\end{tabular}

\bigskip
\bigskip

\bigskip

\underline{16/8 with ${\bf 4}=({\bf 1}_{2},{\bf 1}_{4+i},{\bf 2}_{1})$} and ${\bf
6}=(%
{\bf 1}_{x(i)},{\bf 2},{\bf 2}^{{\bf \prime }}{\bf ,({\bf 1}_{5}+{\bf
1}_{6}+%
{\bf 1}_{7}+{\bf 1}_{8})}_{{\bf A}}{\bf )}$ where $x=6,5,8,$ or 7 for $
i=1,2,3,4$. The fermionic sectors of these models are identical up to
permutation, but there are two potential types of scalar sectors, depending on
whether
${\bf 1}_{x(i)}$ is the same as or different from the antisymmetric
product $
({\bf 2}_{1}\times {\bf 2}_{1})_{A}$ . Let us relabel the singlets so
$({\bf
2}_{1}\times {\bf 2}_{1})_{A}={\bf {\bf 1}_{6},}$ and then choose ${\bf
1}%
_{x(i)}$ to be either ${\bf {\bf 1}_{5}}$ or ${\bf {\bf 1}_{6}}$. Now the
two inequivalent scalar sectors (In this instance, it is easier to analyse both models and show that neither 
phenomenology is interesting, rather than untangle the correct antisymmetric 
singlet in $({\bf 2}_{1}\times {\bf 2}_{1})_{A}$. See the next section.) are:

\begin{tabular}{|c||c|c|c|c|c|c|c|c|c|c|}
\hline
$\otimes $ & 1$_{1}$ & 1$_{2}$ & 1$_{3}$ & 1$_{4}$ & 1$_{5}$ & 1$_{6}$ &
1$
_{7}$ & 1$_{8}$ & 2 & 2$^{\prime }$ \\ \hline\hline
1$_{1}$ &  &  &  &  & $\times (5)$ & (6) &  &  & $\times $ & $\times $
\\
\hline
1$_{2}$ &  &  &  &  & (6) & $\times (5)$ &  &  & $\times $ & $\times $
\\
\hline
1$_{3}$ &  &  &  &  &  &  & $\times (5)$ & (6) & $\times $ & $\times $
\\
\hline
1$_{4}$ &  &  &  &  &  &  & (6) & $\times (5)$ & $\times $ & $\times $
\\
\hline
1$_{5}$ & $\times (5)$ & (6) &  &  &  &  &  &  & $\times $ & $\times $
\\
\hline
1$_{6}$ & (6) & $\times (5)$ &  &  &  &  &  &  & $\times $ & $\times $
\\
\hline
1$_{7}$ &  &  & $\times (5)$ & (6) &  &  &  &  & $\times $ & $\times $
\\
\hline
1$_{8}$ &  &  & (6) & $\times (5)$ &  &  &  &  & $\times $ & $\times $
\\
\hline
2 & $\times $ & $\times $ & $\times $ & $\times $ & $\times $ & $\times
$ & $%
\times $ & $\times $ &  & $\times \times $ \\ \hline
2$^{\prime }$ & $\times $ & $\times $ & $\times $ & $\times $ & $\times
$ & $%
\times $ & $\times $ & $\times $ & $\times \times $ &  \\ \hline
\end{tabular}

\bigskip

\bigskip where(5) is replaced by an $"\times "$ and (6) by a blank if
${\bf 1%
}_{x(i)}={\bf {\bf 1}_{5}}$ and vis versa if ${\bf 1}_{x(i)}={\bf {\bf
1}%
_{6}.}$

\bigskip

\newpage

\bigskip

\underline{16/11 with ${\bf 4}=({\bf 1}_{2},{\bf 1}_{2},{\bf 1}_{2},{\bf 1}_{2})$}
and ${\bf 6}=({\bf 1}_{3},{\bf 1}_{3},{\bf 1}_{3},{\bf 1}_{3},{\bf
1}_{3},{\bf 1}%
_{3}{\bf \ )}$

\bigskip
\begin{tabular}{|c||c|c|c|c|c|c|c|c|c|c|}
\hline
$\otimes $ & 1$_{1}$ & 1$_{2}$ & 1$_{3}$ & 1$_{4}$ & 1$_{5}$ & 1$_{6}$ &
1$%
_{7}$ & 1$_{8}$ & 2 & 2$^{\prime }$ \\ \hline\hline
1$_{1}$ &  &  & ($\times )^{6}$ &  &  &  &  &  &  &  \\ \hline
1$_{2}$ &  & ($\times )^{6}$ &  &  &  &  &  &  &  &  \\ \hline
1$_{3}$ & ($\times )^{6}$ &  &  &  &  &  &  &  &  &  \\ \hline
1$_{4}$ &  &  &  & ($\times )^{6}$ &  &  &  &  &  &  \\ \hline
1$_{5}$ &  &  &  &  &  &  & ($\times )^{6}$ &  &  &  \\ \hline
1$_{6}$ &  &  &  &  &  & ($\times )^{6}$ &  &  &  &  \\ \hline
1$_{7}$ &  &  &  &  & ($\times )^{6}$ &  &  &  &  &  \\ \hline
1$_{8}$ &  &  &  &  &  &  &  & ($\times )^{6}$ &  &  \\ \hline
2 &  &  &  &  &  &  &  &  & ($\times )^{6}$ &  \\ \hline
2$^{\prime }$ &  &  &  &  &  &  &  &  &  & ($\times )^{6}$ \\ \hline
\end{tabular}

\bigskip
\bigskip

\newpage

\bigskip
\bigskip

\underline{16/11 with ${\bf 4}=({\bf 1}_{2},{\bf 1}_{2},{\bf 1}_{2},{\bf 1}_{4})$}
and ${\bf 6}=({\bf 1}_{1},{\bf 1}_{1},{\bf 1}_{1},{\bf 1}_{3},{\bf
1}_{3},{\bf 1}%
_{3}{\bf \ )}$

\bigskip

\bigskip
\begin{tabular}{|c||c|c|c|c|c|c|c|c|c|c|}
\hline
$\otimes $ & 1$_{1}$ & 1$_{2}$ & 1$_{3}$ & 1$_{4}$ & 1$_{5}$ & 1$_{6}$ &
1$%
_{7}$ & 1$_{8}$ & 2 & 2$^{\prime }$ \\ \hline\hline
1$_{1}$ & ($\times )^{3}$ &  & ($\times )^{3}$ &  &  &  &  &  &  &  \\
\hline
1$_{2}$ &  & ($\times )^{6}$ &  &  &  &  &  &  &  &  \\ \hline
1$_{3}$ & ($\times )^{3}$ &  & ($\times )^{3}$ &  &  &  &  &  &  &  \\
\hline
1$_{4}$ &  &  &  & ($\times )^{6}$ &  &  &  &  &  &  \\ \hline
1$_{5}$ &  &  &  &  & ($\times )^{3}$ &  & ($\times )^{3}$ &  &  &  \\
\hline
1$_{6}$ &  &  &  &  &  & ($\times )^{6}$ &  &  &  &  \\ \hline
1$_{7}$ &  &  &  &  & ($\times )^{3}$ &  & ($\times )^{3}$ &  &  &  \\
\hline
1$_{8}$ &  &  &  &  &  &  &  & ($\times )^{6}$ &  &  \\ \hline
2 &  &  &  &  &  &  &  &  & ($\times )^{6}$ &  \\ \hline
2$^{\prime }$ &  &  &  &  &  &  &  &  &  & ($\times )^{6}$ \\ \hline
\end{tabular}

\bigskip
\underline{$16/13$ with ${\bf 4=({\bf 1}_{3},{\bf 1}_{4},2}_{1}{\bf )}$}
and ${\bf6}=({\bf 1}_{2},{\bf 1}_{c},{\bf 2}_{1},{\bf 2}_{3})$, 
where ${\bf 1}_{c}= ({\bf 2}_{1} \times {\bf 2}_{1})_{{\bf A}}$ 
so we have ${\bf 1}_{c}$ is
either
${\bf 1}_{2}$ or ${\bf 1}_{3}.$

\bigskip

\bigskip

\bigskip
\begin{tabular}{|c||c|c|c|c|c|c|c|}
\hline
$\otimes $ & 1$_{1}$ & 1$_{2}$ & 1$_{3}$ & 1$_{4}$ & 2$_{1}$ & 2$_{2}$ &
2$%
_{3}$ \\ \hline\hline
1$_{1}$ &  & $\times (2)$ & (3) &  & $\times $ &  & $\times $ \\ \hline
1$_{2}$ & $\times (2)$ &  &  & (3) & $\times $ &  & $\times $ \\ \hline
1$_{3}$ & (3) &  &  & $\times (2)$ & $\times $ &  & $\times $ \\ \hline
1$_{4}$ &  & (3) & $\times (2)$ &  & $\times $ &  & $\times $ \\ \hline
2$_{1}$ & $\times $ & $\times $ & $\times $ & $\times $ &  & $\times
\times $
& $\times \times $ \\ \hline
2$_{2}$ &  &  &  &  & $\times \times $ & $\times \times $ & $\times
\times $
\\ \hline
2$_{3}$ & $\times $ & $\times $ & $\times $ & $\times $ & $\times \times
$ &
$\times \times $ &  \\ \hline
\end{tabular}

\bigskip

\bigskip

\bigskip

\underline{$16/13$ with ${\bf 4=(2}_{1}{\bf ,2}_{2}{\bf )}$} and ${\bf 6}=({\bf
1}_{a},%
{\bf 1}_{b},{\bf 2}_{1},{\bf 2}_{3})$, where ${\bf 1}_{a}={\bf
(2}_{1}{\bf
\times 2}_{1}{\bf )}_{{\bf A}}=(1_2+1_3+2_2)_{{\bf A}}$ and ${\bf 1}_{b}={\bf (2}_{2}{\bf \times
2}%
_{2}{\bf )}_{{\bf A}}+(1_1+1_2+1_3+1_4)_{{\bf A}}$.

\bigskip

\bigskip

\bigskip

\begin{tabular}{|c||c|c|c|c|c|c|c|}
\hline
$\otimes $ & 1$_{1}$ & 1$_{2}$ & 1$_{3}$ & 1$_{4}$ & 2$_{1}$ & 2$_{2}$ &
2$%
_{3}$ \\ \hline\hline
1$_{1}$ & (1) & (2) & (3) & (4) & $\times $ &  & $\times $ \\ \hline
1$_{2}$ & (2) & (1) & (4) & (3) & $\times $ &  & $\times $ \\ \hline
1$_{3}$ & (3) & (4) & (1) & (2) & $\times $ &  & $\times $ \\ \hline
1$_{4}$ & (4) & (3) & (2) & (1) & $\times $ &  & $\times $ \\ \hline
2$_{1}$ & $\times $ & $\times $ & $\times $ & $\times $ & (1)(4) &
$\times
\times $ & (2)(3) \\ \hline
2$_{2}$ &  &  &  &  & $\times \times $ & $
\begin{array}{l}
(1)(2) \\
(3)(4)
\end{array}
$ & $\times \times $ \\ \hline
2$_{3}$ & $\times $ & $\times $ & $\times $ & $\times $ & (2)(3) &
$\times
\times $ & (1)(4) \\ \hline
\end{tabular}

\bigskip

\noindent where we insert $\times $s at the locations in parenthesis when the
singlets
are chosen properly from the antisymmetric products of the doublets.
There are three inequivalent choices, either (i) put $\times \times$ at
location (2), or (ii) put an $\times $ at (2) and one at (3), or (iii)
put $%
\times $ at (2) and $\times $ at (1). All other choices lead to
equivalent
models. Thus, without detailed knowledge of the antisymmetric products,
we
can still reduce the analysis to the consideration of these three cases.

\bigskip

\newpage

\bigskip

\bigskip

\underline{$16/13$ with ${\bf 4=(2}_{1}{\bf ,2}_{1}{\bf )}$} and ${\bf 6}=({\bf
1}_{2},%
{\bf 1}_{2},{\bf 1}_{2},{\bf 1}_{3},{\bf 2}_{2})$ (which is equivalent
to $%
{\bf 6}=({\bf 1}_{2},{\bf 1}_{3},{\bf 1}_{3},{\bf 1}_{3},{\bf 2}_{2})$
for
SSB up to a relabeling of irreps).

\bigskip

\bigskip

\bigskip
\begin{tabular}{|c||c|c|c|c|c|c|c|}
\hline
$\otimes $ & 1$_{1}$ & 1$_{2}$ & 1$_{3}$ & 1$_{4}$ & 2$_{1}$ & 2$_{2}$ &
2$%
_{3}$ \\ \hline\hline
1$_{1}$ &  & $\times \times \times $ & $\times $ &  &  & $\times $ &  \\

\hline
1$_{2}$ & $\times \times \times $ &  &  &  &  & $\times $ &  \\ \hline
1$_{3}$ & $\times $ &  &  & $\times \times \times $ &  & $\times $ &  \\

\hline
1$_{4}$ &  &  & $\times \times \times $ &  &  & $\times $ &  \\ \hline
2$_{1}$ &  &  &  &  & $\times $ &  & $
\begin{array}{l}
\times \times  \\
\times \times \times
\end{array}
$ \\ \hline
2$_{2}$ & $\times $ & $\times $ & $\times $ & $\times $ &  & $
\begin{array}{l}
\times \times  \\
\times \times
\end{array}
$ &  \\ \hline
2$_{3}$ &  &  &  &  & $
\begin{array}{l}
\times \times  \\
\times \times \times
\end{array}
$ &  & $\times $ \\ \hline
\end{tabular}

\bigskip

\newpage

\bigskip
\bigskip

\underline{18/3 with $4=(1^{\prime }\alpha ,1^{\prime },2\alpha )$} and
$6=(1^{\prime
}\alpha ,2\alpha ,2\alpha ^{2},1^{\prime }\alpha ^{2})$

\bigskip

\begin{tabular}{|c||c|c|c|c|c|c|c|c|c|}
\hline
$\otimes $ & 1 & 1$^{\prime }$ & 2 & 1$\alpha $ & 1$^{\prime }\alpha $ &
2$%
\alpha $ & 1$\alpha ^{2}$ & 1$^{\prime }\alpha ^{2}$ & 2$\alpha ^{2}$ \\

\hline\hline
1 &  &  &  &  & $\times $ & $\times $ &  & $\times $ & $\times $ \\
\hline
1$^{\prime }$ &  &  &  & $\times $ &  & $\times $ & $\times $ &  &
$\times $
\\ \hline
2 &  &  &  & $\times $ & $\times $ & $\times \times $ & $\times $ &
$\times $
& $\times \times $ \\ \hline
1$\alpha $ &  & $\times $ & $\times $ &  &  &  &  & $\times $ & $\times
$ \\
\hline
1$^{\prime }\alpha $ & $\times $ &  & $\times $ &  &  &  & $\times $ &
& $%
\times $ \\ \hline
2$\alpha $ & $\times $ & $\times $ & $\times \times $ &  &  &  & $\times
$ &
$\times $ & $\times \times $ \\ \hline
1$\alpha ^{2}$ &  & $\times $ & $\times $ &  & $\times $ & $\times $ &
&  &
\\ \hline
1$^{\prime }\alpha ^{2}$ & $\times $ &  & $\times $ & $\times $ &  &
$\times
$ &  &  &  \\ \hline
2$\alpha ^{2}$ & $\times $ & $\times $ & $\times \times $ & $\times $ &
$%
\times $ & $\times \times $ &  &  &  \\ \hline
\end{tabular}

\bigskip

\newpage

\bigskip

\bigskip

\underline{20/5 with ${\bf 4}=({\bf 1}_{2},{\bf 1}_{2},{\bf 1}_{2},{\bf 1}_{2})$}
and ${\bf 6}=({\bf 1}_{3},{\bf 1}_{3},{\bf 1}_{3},{\bf 1}_{3},{\bf
1}_{3},{\bf 1}%
_{3}{\bf \ )}$ is very much like the 16/11 model with the similar
embedding.
Here

\bigskip

\begin{tabular}{|c||c|c|c|c|c|}
\hline
$\otimes $ & 1$_{1}$ & 1$_{2}$ & 1$_{3}$ & 1$_{4}$ & 4 \\ \hline\hline
1$_{1}$ &  &  & ($\times )^{6}$ &  &  \\ \hline
1$_{2}\ $ &  &  &  & ($\times )^{6}$ &  \\ \hline
1$_{3}$ & ($\times )^{6}$ &  &  &  &  \\ \hline
1$_{4}$ &  & ($\times )^{6}$ &  &  &  \\ \hline
4 &  &  &  &  &  \\ \hline
\end{tabular}

\bigskip

and a VEV for any of these renders the entire fermion sector vectorlike.

\bigskip

\newpage

\bigskip
\bigskip

For \underline{20/5 with ${\bf 4}=({\bf 1}_{2},{\bf 1}_{2},{\bf 1}_{2},{\bf
1}_{4})$}
and ${\bf 6}=({\bf 1}_{1},{\bf 1}_{1},{\bf 1}_{1},{\bf 1}_{3},{\bf
1}_{3},%
{\bf 1}_{3}{\bf \ )}$ we have the scalars:

\bigskip

\bigskip

\begin{tabular}{|c||c|c|c|c|c|}
\hline
$\otimes $ & 1$_{1}$ & 1$_{2}$ & 1$_{3}$ & 1$_{4}$ & 4 \\ \hline\hline
1$_{1}$ & $\times \times \times $ &  & $\times \times \times $ &  &  \\
\hline
1$_{2}\ $ &  & $\times \times \times $ &  & $\times \times \times $ &
\\
\hline
1$_{3}$ & $\times \times \times $ &  & $\times \times \times $ &  &  \\
\hline
1$_{4}$ &  & $\times \times \times $ &  & $\times \times \times $ &  \\
\hline
4 &  &  &  &  & ($\times $)$^{6}$ \\ \hline
\end{tabular}

\bigskip

\newpage

\bigskip
\bigskip

\underline{21/2 with ${\bf 4}=({\bf 1}_{2},{\bf 3}_{1})$} and ${\bf 6}%
=\ {\bf 3}_{1}+{\bf 3}_{2}\ $ with
$N=2$.
(All other embeddings of the ${\bf 4}$ with chiral fermions and $\aleph
=0$
SUSY are permutations and therefore equivalent to this model.). Here
${\bf 6}$ is real so the embedding is proper
and
the scalar sector is:

\bigskip

\begin{tabular}{|c||c|c|c|c|c|}
\hline
$\otimes $ & 1$_{1}$ & 1$_{2}$ & 1$_{3}$ & 3$_{1}$ & 3$_{2}$ \\
\hline\hline
1$_{1}$ &  &  &  & $\times $ & $\times $ \\ \hline
1$_{2}\ $ &  &  &  & $\times $ & $\times $ \\ \hline
1$_{3}$ &  &  &  & $\times $ & $\times $ \\ \hline
3$_{1}$ & $\times $ & $\times $ & $\times $ & $\times \times $ & $\times

\times \times $ \\ \hline
3$_{2}$ & $\times $ & $\times $ & $\times $ & $\times \times \times $ &
$%
\times \times $ \\ \hline
\end{tabular}

\bigskip

\bigskip

\newpage

\underline{24/7 or equivalently 24/8 (since they have isomorphic irrep product
tables)
with ${\bf 4=(1}_{{\bf 1}}{\bf \alpha ,1}_{2},{\bf 2}\alpha {\bf )}$} and
$%
{\bf 6}=({\bf 1}_{2}{\bf \alpha ,1}_{2}\alpha ^{2},{\bf 2}\alpha ,{\bf
2}%
\alpha ^{2})$

\bigskip

\bigskip
\begin{tabular}{|c||c|c|c|c|c|c|c|c|c|c|c|c|c|c|c|}
\hline
$\otimes $ & 1$_{1}$ & 1$_{2}$ & 1$_{3}$ & 1$_{4}$ & 2 & 1$_{1}\alpha $
& 1$%
_{2}\alpha $ & 1$_{3}\alpha $ & 1$_{4}\alpha $ & 2$\alpha $ &
1$_{1}\alpha
^{2}$ & 1$_{2}\alpha ^{2}$ & 1$_{3}\alpha ^{2}$ & 1$_{4}\alpha ^{2}$ &
2$%
\alpha ^{2}$ \\ \hline\hline
1$_{1}$ &  &  &  &  &  &  & $\times $ &  &  & $\times $ &  & $\times $
&  &
& $\times $ \\ \hline
1$_{2}$ &  &  &  &  &  & $\times $ &  &  &  & $\times $ & $\times $ &
&  &
& $\times $ \\ \hline
1$_{3}$ &  &  &  &  &  &  &  &  & $\times $ & $\times $ &  &  &  &
$\times $
& $\times $ \\ \hline
1$_{4}$ &  &  &  &  &  &  &  & $\times $ &  & $\times $ &  &  & $\times
$ &
& $\times $ \\ \hline
2 &  &  &  &  &  & $\times $ & $\times $ & $\times $ & $\times $ &
$\times $
& $\times $ & $\times $ & $\times $ & $\times $ & $\times $ \\ \hline
1$_{1}\alpha $ &  & $\times $ &  &  & $\times $ &  &  &  &  &  &  &
$\times $
&  &  & $\times $ \\ \hline
1$_{2}\alpha $ & $\times $ &  &  &  & $\times $ &  &  &  &  &  & $\times
$ &
&  &  & $\times $ \\
\hline
\end{tabular}

\bigskip

\newpage

\bigskip

The scalars for \underline{24/9 with ${\bf 4=(1}_{{\bf 1}}{\bf \alpha ,{\bf
1}%
_{2}\alpha }^{{\bf 3}},{\bf 2{\bf \alpha }^{{\bf 2}})}$} and ${\bf
6}=({\bf 1}%
_{2}{\bf ,1}_{2},{\bf 2}\alpha ,{\bf 2}\alpha ^{3})$ are:

\bigskip

\begin{tabular}{|c||c|c|c|c|c|c|c|c|c|c|c|c|}
\hline
$\otimes $ & 1$_{1}$ & 1$_{2}$ & 2 & 1$_{1}\alpha $ & 1$_{2}\alpha $ &
2$%
\alpha $ & 1$_{1}\alpha ^{2}$ & 1$_{2}\alpha ^{2}$ & 2$\alpha ^{2}$ &
1$%
_{1}\alpha ^{3}$ & 1$_{2}\alpha ^{3}$ & 2$\alpha ^{3}$ \\ \hline\hline
1$_{1}$ &  & $\times \times $ &  &  &  & $\times $ &  &  &  &  &  &
$\times $
\\ \hline
1$_{2}$ & $\times \times $ &  &  &  &  & $\times $ &  &  &  &  &  &
$\times $
\\ \hline
2 &  &  & $\times \times $ & $\times $ & $\times $ & $\times $ &  &  &
& $%
\times $ & $\times $ & $\times $ \\ \hline
1$_{1}\alpha $ &  &  & $\times $ &  & $\times \times $ &  &  &  &
$\times $
&  &  &  \\ \hline
1$_{2}\alpha $ &  &  & $\times $ & $\times \times $ &  &  &  &  &
$\times $
&  &  &  \\ \hline
2$\alpha $ & $\times $ & $\times $ & $\times $ &  &  & $\times \times $
& $%
\times $ & $\times $ & $\times $ &  &  &  \\ \hline
1$_{1}\alpha ^{2}$ &  &  &  &  &  & $\times $ &  & $\times \times $ &
&  &
& $\times $ \\ \hline
1$_{2}\alpha ^{2}$ &  &  &  &  &  & $\times $ & $\times \times $ &  &
&  &
& $\times $ \\ \hline
2$\alpha ^{2}$ &  &  &  & $\times $ & $\times $ & $\times $ &  &  &
$\times
\times $ & $\times $ & $\times $ & $\times $ \\ \hline
1$_{1}\alpha ^{3}$ &  &  & $\times $ &  &  &  &  &  & $\times $ &  &
$\times
\times $ &  \\ \hline
1$_{1}\alpha ^{3}$ &  &  & $\times $ &  &  &  &  &  & $\times $ &
$\times
\times $ &  &  \\ \hline
2$\alpha ^{3}$ & $\times $ & $\times $ & $\times $ &  &  &  & $\times $
& $%
\times $ & $\times $ &  &  & $\times \times $ \\ \hline
\end{tabular}

\bigskip

\newpage

\bigskip

The scalars for \underline{24/9 with 
${\bf 4= ({1}_{1}\alpha , 1_{2}\alpha, 2\alpha)}$} 
and 
${\bf 6}=({\bf 1_2\alpha, 2,1_2\alpha , 2 \alpha^2})$
are:
\bigskip

\bigskip

\bigskip

\begin{tabular}{|c||c|c|c|c|c|c|c|c|c|c|c|c|}
\hline
$\otimes $ & 1$_{1}$ & 1$_{2}$ & 2 & 1$_{1}\alpha $ & 1$_{2}\alpha $ &
2$%
\alpha $ & 1$_{1}\alpha ^{2}$ & 1$_{2}\alpha ^{2}$ & 2$\alpha ^{2}$ &
1$%
_{1}\alpha ^{3}$ & 1$_{2}\alpha ^{3}$ & 2$\alpha ^{3}$ \\ \hline\hline
1$_{1}$ &  &  &  &  &  &  &  & $\times \times $ & $\times \times $ &  &
&
\\ \hline
1$_{2}$ &  &  &  &  &  &  & $\times \times $ &  & $\times \times $ &  &
&
\\ \hline
2 &  &  &  &  &  &  & $\times \times $ & $\times \times $ & $
\begin{array}{l}
\times \times  \\
\times \times
\end{array}
$ &  &  &  \\ \hline
1$_{1}\alpha $ &  &  &  &  &  &  &  &  &  &  & $\times \times $ &
$\times
\times $ \\ \hline
1$_{2}\alpha $ &  &  &  &  &  &  &  &  &  & $\times \times $ &  &
$\times
\times $ \\ \hline
2$\alpha $ &  &  &  &  &  &  &  &  &  & $\times \times $ & $\times
\times $
& $
\begin{array}{l}
\times \times  \\
\times \times
\end{array}
$ \\ \hline
1$_{1}\alpha ^{2}$ &  & $\times \times $ & $\times \times $ &  &  &  &
&  &
&  &  &  \\ \hline
1$_{2}\alpha ^{2}$ & $\times \times $ &  & $\times \times $ &  &  &  &
&  &
&  &  &  \\ \hline
2$\alpha ^{2}$ & $\times \times $ & $\times \times $ & $
\begin{array}{l}
\times \times  \\
\times \times
\end{array}
$ &  &  &  &  &  &  &  &  &  \\ \hline
1$_{1}\alpha ^{3}$ &  &  &  &  & $\times \times $ & $\times \times $ &
&  &
&  &  &  \\ \hline
1$_{1}\alpha ^{3}$ &  &  &  & $\times \times $ &  & $\times \times $ &
&  &
&  &  &  \\ \hline
2$\alpha ^{3}$ &  &  &  & $\times \times $ & $\times \times $ & $
\begin{array}{l}
\times \times  \\
\times \times
\end{array}
$ &  &  &  &  &  &  \\ \hline
\end{tabular}

\bigskip

\newpage

\bigskip

For \underline{24/9 with ${\bf 4}=({\bf 1}_{1}\alpha ,{\bf 1}_{2},{\bf 2}\alpha )$} and
   ${\bf 6}=({\bf 1}_{2}\alpha ^{2},{\bf
2}\alpha
,{\bf 2}\alpha ^{-1},{\bf 1}_{2}\alpha ^{-2})$ where $\alpha ^{4}=1,$
the
scalar sector is:

\bigskip

\begin{tabular}{|c||c|c|c|c|c|c|c|c|c|c|c|c|}
\hline
$\otimes $ & 1$_{1}$ & 1$_{2}$ & 2 & 1$_{1}\alpha $ & 1$_{2}\alpha $ &
2$%
\alpha $ & 1$_{1}\alpha ^{2}$ & 1$_{2}\alpha ^{2}$ & 2$\alpha ^{2}$ &
1$%
_{1}\alpha ^{3}$ & 1$_{2}\alpha ^{3}$ & 2$\alpha ^{3}$ \\ \hline\hline
1$_{1}$ &  &  &  &  &  & $\times $ &  & $\times \times $ &  &  &  &
$\times $
\\ \hline
1$_{2}$ &  &  &  &  &  & $\times $ & $\times \times $ &  &  &  &  &
$\times $
\\ \hline
2 &  &  &  & $\times $ & $\times $ & $\times $ &  &  & $\times \times $
& $%
\times $ & $\times $ & $\times $ \\ \hline
1$_{1}\alpha $ &  &  & $\times $ &  &  &  &  &  & $\times $ &  & $\times

\times $ &  \\ \hline
1$_{2}\alpha $ &  &  & $\times $ &  &  &  &  &  & $\times $ & $\times
\times
$ &  &  \\ \hline
2$\alpha $ & $\times $ & $\times $ & $\times $ &  &  &  & $\times $ & $%

\times $ & $\times $ &  &  & $\times \times $ \\ \hline
1$_{1}\alpha ^{2}$ &  & $\times \times $ &  &  &  & $\times $ &  &  &
&  &
& $\times $ \\ \hline
1$_{2}\alpha ^{2}$ & $\times \times $ &  &  &  &  & $\times $ &  &  &
&  &
& $\times $ \\ \hline
2$\alpha ^{2}$ &  &  & $\times \times $ & $\times $ & $\times $ &
$\times $
&  &  &  & $\times $ & $\times $ & $\times $ \\ \hline
1$_{1}\alpha ^{3}$ &  &  & $\times $ &  & $\times \times $ &  &  &  & $%

\times $ &  &  &  \\ \hline
1$_{2}\alpha ^{3}$ &  &  & $\times $ & $\times \times $ &  &  &  &  & $%

\times $ &  &  &  \\ \hline
2$\alpha ^{3}$ & $\times $ & $\times $ & $\times $ &  &  & $\times
\times $
& $\times $ & $\times $ & $\times $ &  &  &  \\ \hline
\end{tabular}

\bigskip

\newpage

\bigskip

For \underline{24/9 with ${\bf 4}=({\bf 2}\alpha ,{\bf 2}\alpha )$} 
where ${\bf 6}=3({\bf 1}_{2}\alpha^{2})+{\bf 1}_{1}
\alpha^{2}+{\bf 2}\alpha^{2},$the scalars are:
\bigskip
%

\bigskip

\newpage

\bigskip

The next example of interest is \underline{24/13 with ${\bf 4}=({\bf 2}_{1},{\bf
2}%
_{2})\ $} and ${\bf 6}=\ \ {\bf 1}_{1}+{\bf 1}_{2}+{\bf 1}_{3}+{\bf
3}$
with scalars:

\bigskip

\begin{tabular}{|c||c|c|c|c|c|c|c|}
\hline
$\otimes $ & 1$_{1}$ & 1$_{2}$ & 1$_{3}$ & 2$_{1}$ & 2$_{2}$ & 2$_{3}$ &
3
\\ \hline\hline
1$_{1}$ & $\times $ & $\times $ & $\times $ &  &  &  & $\times $ \\
\hline
1$_{2}$ & $\times $ & $\times $ & $\times $ &  &  &  & $\times $ \\
\hline
1$_{3}$ & $\times $ & $\times $ & $\times $ &  &  &  & $\times $ \\
\hline
2$_{1}$ &  &  &  & $\times \times $ & $\times \times $ & $\times \times
$ &
\\ \hline
2$_{2}$ &  &  &  & $\times \times $ & $\times \times $ & $\times \times
$ &
\\ \hline
2$_{3}$ &  &  &  & $\times \times $ & $\times \times $ & $\times \times
$ &
\\ \hline
3 & $\times $ & $\times $ & $\times $ &  &  &  & $\times \times $ \\
\hline
\end{tabular}

\bigskip

\bigskip

There are two ineqivalent models for the group \underline{24/15}, they are  
\underline{${\bf 4}=({\bf 1}_{2},{\bf 1}_{3},{\bf 2}_{3})$} where ${\bf 6}={\bf
1}%
_{4}+{\bf 1}_{2[4]}+{\bf 2}_{3}+{\bf 2}_{4}$ and the scalars are:

\bigskip

\begin{tabular}{|c||c|c|c|c|c|c|c|c|c|}
\hline
$\otimes $ & 1$_{1}$ & 1$_{2}$ & 1$_{3}$ & 1$_{4}$ & 2$_{1}$ & 2$_{2}$ &
2$%
_{3}$ & 2$_{4}$ & 2$_{5}$ \\ \hline\hline
1$_{1}$ &  &  &  & $\times \times $ &  &  & $\times $ & $\times $ &  \\
\hline
1$_{2}$ &  &  & $\times \times $ &  &  &  & $\times $ & $\times $ &  \\
\hline
1$_{3}$ &  & $\times \times $ &  &  &  &  & $\times $ & $\times $ &  \\
\hline
1$_{4}$ & $\times \times $ &  &  &  &  &  & $\times $ & $\times $ &  \\
\hline
2$_{1}$ &  &  &  &  &  & $\times \times $ & $\times $ & $\times $ &
$\times
\times $ \\ \hline
2$_{2}$ &  &  &  &  & $\times \times $ &  & $\times $ & $\times $ &
$\times
\times $ \\ \hline
2$_{3}$ & $\times $ & $\times $ & $\times $ & $\times $ & $\times $ & $%

\times $ &  & $\times \times $ &  \\ \hline
2$_{4}$ & $\times $ & $\times $ & $\times $ & $\times $ & $\times $ & $%

\times $ & $\times \times $ &  &  \\ \hline
2$_{5}$ &  &  &  &  & $\times \times $ & $\times \times $ &  &  &
$\times
\times $ \\ \hline
\end{tabular}

\bigskip

if (${\bf 2}_{3}\times {\bf 2}_{3})_{A}={\bf 1}_{4}$ but if it is ${\bf
1}%
_{2}$ then the top $4\times 4$ changes to:

\bigskip

\begin{tabular}{||c|c|c|c|}
\hline\hline
& $\times $ &  & $\times $ \\ \hline
$\times $ &  & $\times $ &  \\ \hline
& $\times $ &  & $\times $ \\ \hline
$\times $ &  & $\times $ &  \\ \hline
\end{tabular}

\bigskip

\newpage

\bigskip
\bigskip

The other \underline{24/15} case has \underline{${\bf 4}=({\bf 2}_{3},{\bf 2}_{3})$} 
where ${\bf 6}=3({\bf 1}_{2})+{\bf 1}_{4}+{\bf 2}_{1}$ and the scalars
are (this time swapping ${\bf 1}_{2}\ $and ${\bf 1}_{4}$ gives equivalent
models):

\bigskip

%

\bigskip

\newpage

\bigskip
\bigskip

The next model to evaluate is \underline{27/4 with ${\bf 4}=({\bf 1}_{2},
{\bf3}_{1})$},
 where ${\bf 6}=\ {\bf 3}_{1}+{\bf 3}_{2}$ is real. The scalar
sector is:

\bigskip

\begin{tabular}{|c||c|c|c|c|c|c|c|c|c|c|c|}
\hline
$\otimes $ & 1$_{1}$ & 1$_{2}$ & 1$_{3}$ & 1$_{4}$ & 1$_{5}$ & 1$_{6}$ &
1$%
_{7}$ & 1$_{8}$ & 1$_{9}$ & 3$_{1}$ & 3$_{2}$ \\ \hline\hline
1$_{1}$ &  &  &  &  &  &  &  &  &  & $\times $ & $\times $ \\ \hline
1$_{2}$ &  &  &  &  &  &  &  &  &  & $\times $ & $\times $ \\ \hline
1$_{3}$ &  &  &  &  &  &  &  &  &  & $\times $ & $\times $ \\ \hline
1$_{4}$ &  &  &  &  &  &  &  &  &  & $\times $ & $\times $ \\ \hline
1$_{5}$ &  &  &  &  &  &  &  &  &  & $\times $ & $\times $ \\ \hline
1$_{6}$ &  &  &  &  &  &  &  &  &  & $\times $ & $\times $ \\ \hline
1$_{7}$ &  &  &  &  &  &  &  &  &  & $\times $ & $\times $ \\ \hline
1$_{8}$ &  &  &  &  &  &  &  &  &  & $\times $ & $\times $ \\ \hline
1$_{9}$ &  &  &  &  &  &  &  &  &  & $\times $ & $\times $ \\ \hline
3$_{1}$ & $\times $ & $\times $ & $\times $ & $\times $ & $\times $ & $%

\times $ & $\times $ & $\times $ & $\times $ &  & $\times \times \times
$ \\
\hline
3$_{2}$ & $\times $ & $\times $ & $\times $ & $\times $ & $\times $ & $%

\times $ & $\times $ & $\times $ & $\times $ & $\times \times \times $
&  \\
\hline
\end{tabular}

\bigskip

\newpage

\bigskip

And finally at order 30 we have for \underline{30/2 with  
${\bf 4}=({\bf 1}\alpha ,{\bf 1}^{\prime },{\bf 2}\alpha )$}
 and  ${\bf 6}=({\bf 1}^{\prime }\alpha +{\bf 2}\alpha +{\bf 2}\alpha
^{-1}+%
{\bf 1}^{\prime }\alpha ^{-1})$ where $\alpha ^{3}=1,$ a model with scalar sector:

\bigskip

\bigskip
\begin{tabular}{|c||c|c|c|c|c|c|c|c|c|c|c|c|}
\hline
$\otimes $ & 1 & 1$^{\prime }$ & 2 & $2^{\prime }$ & 1$\alpha $ &
1$^{\prime
}\alpha $ & 2$\alpha $ & $2^{\prime }\alpha $ & 1$\alpha ^{2}$ &
1$^{\prime
}\alpha ^{2}$ & 2$\alpha ^{2}$ & $2^{\prime }\alpha ^{2}$ \\
\hline\hline
1 &  &  &  &  &  & $\times $ & $\times $ &  &  & $\times $ & $\times $
&  \\
\hline
1$^{\prime }$ &  &  &  &  & $\times $ &  & $\times $ &  & $\times $ &  &
$%
\times $ &  \\ \hline
$2$ &  &  &  &  & $\times $ & $\times $ & $\times $ & $\times $ &
$\times $
& $\times $ & $\times $ & $\times $ \\ \hline
$2^{\prime }$ &  &  &  &  &  &  & $\times $ & $\times \times $ &  &  &
$%
\times $ & $\times \times $ \\ \hline
1$\alpha $ &  & $\times $ & $\times $ &  &  &  &  &  &  & $\times $ & $%

\times $ &  \\ \hline
1$^{\prime }\alpha $ & $\times $ &  & $\times $ &  &  &  &  &  & $\times
$ &
& $\times $ &  \\ \hline
2$\alpha $ & $\times $ & $\times $ & $\times $ & $\times $ &  &  &  &  &
$%
\times $ & $\times $ & $\times $ & $\times $ \\ \hline
$2^{\prime }\alpha $ &  &  & $\times $ & $\times \times $ &  &  &  &  &
&
& $\times $ & $\times \times $ \\ \hline
1$\alpha ^{2}$ &  & $\times $ & $\times $ &  &  & $\times $ & $\times $
&  &
&  &  &  \\ \hline
1$^{\prime }\alpha ^{2}$ & $\times $ &  & $\times $ &  & $\times $ &  &
$%
\times $ &  &  &  &  &  \\ \hline
2$\alpha ^{2}$ & $\times $ & $\times $ & $\times $ & $\times $ & $\times
$ &
$\times $ & $\times $ & $\times $ &  &  &  &  \\ \hline
$2^{\prime }\alpha ^{2}$ &  &  & $\times $ & $\times \times $ &  &  & $%

\times $ & $\times \times $ &  &  &  &  \\ \hline
\end{tabular}

\bigskip

\newpage

\bigskip
\bigskip

The other possibility at order 30 is 
\underline{30/3 with  ${\bf 4}=({\bf 1}\alpha ,
{\bf 1}^{\prime },{\bf 2}\alpha^{2})$}
 where ${\bf 6}={\bf 1}^{\prime }\alpha +{\bf
2}\alpha
^{2}+{\bf 2}\alpha ^{3}+{\bf 1}^{\prime }\alpha ^{4}$ and $\alpha
^{5}=1,$
where the scalars are:

\bigskip

\bigskip

\begin{tabular}{|c||c|c|c|c|c|c|c|c|c|c|c|c|c|c|c|}
\hline
$\otimes $ & 1 & 1$^{\prime }$ & 2 & 1$\alpha $ & 1$^{\prime }\alpha $ &
2$%
\alpha $ & 1$\alpha ^{2}$ & 1$^{\prime }\alpha ^{2}$ & 2$\alpha ^{2}$ &
1$%
\alpha ^{3}$ & 1$^{\prime }\alpha ^{3}$ & 2$\alpha ^{3}$ & 1$\alpha
^{4}$ & 1%
$^{\prime }\alpha ^{4}$ & 2$\alpha ^{4}$ \\ \hline\hline
1 &  &  &  &  & $\times $ &  &  &  & $\times $ &  &  & $\times $ &  & $%

\times $ &  \\ \hline
1$^{\prime }$ &  &  &  & $\times $ &  &  &  &  & $\times $ &  &  &
$\times $
& $\times $ &  &  \\ \hline
2 &  &  &  &  &  & $\times $ & $\times $ & $\times $ & $\times $ &
$\times $
& $\times $ & $\times $ &  &  & $\times $ \\ \hline
1$\alpha $ &  & $\times $ &  &  &  &  &  & $\times $ &  &  &  & $\times
$ &
&  & $\times $ \\ \hline
1$^{\prime }\alpha $ & $\times $ &  &  &  &  &  & $\times $ &  &  &  &
& $%
\times $ &  &  & $\times $ \\ \hline
2$\alpha $ &  &  & $\times $ &  &  &  &  &  & $\times $ & $\times $ & $%

\times $ & $\times $ & $\times $ & $\times $ & $\times $ \\ \hline
1$\alpha ^{2}$ &  &  & $\times $ &  & $\times $ &  &  &  &  &  & $\times
$ &
&  &  & $\times $ \\
\hline
\end{tabular}

\newpage

\section{Spontaneous Symmetry Breaking}

We are now in a position to carry out the spontaneous symmetry breaking for the models with 
fermions and scalars
given in the previous two sections. 
We restrict ourselves to chiral models with the potential of
at least three families
($\chi \geq 45$) and for the most part consider only models with $N = 2$, although we have 
included two $N = 3$
models. Again, we move progressively through the models of increasing order of $\Gamma$.
The model is completely fixed by $\Gamma$, the 
embedding of $\bf{4}$ in $\Gamma$, and 
the choice of $N$. the first
relevant model is:

\underline{16/8 with ${\bf 4=(2_{1}{\bf ,}2_{1})}$ and $N=2$}

The chiral fermions are 

$2[
(2,1,1,1,1,1,1,1;4,1)  +(1,1,1,1,2,1,1,1;1,4) 
+(1,2,1,1,1,1,1,1;4,1)  +(1,1,1,1,1,2,1,1;1,4) \\
+(1,1,2,1,1,1,1,1;4,1)  +(1,1,1,1,1,1,2,1;1,4) 
+(1,1,1,2,1,1,1,1;4,1)  +(1,1,1,1,1,1,1,2;1,4) \\
+(2,1,1,1,1,1,1,1;1,\bar{4})  +(1,1,1,1,2,1,1,1;\bar{4},1) 
+(1,2,1,1,1,1,1,1;1,\bar{4})  +(1,1,1,1,1,2,1,1;\bar{4},1) \\
+(1,1,2,1,1,1,1,1;1,\bar{4})  +(1,1,1,1,1,1,2,1;\bar{4},1) 
+(1,1,1,2,1,1,1,1;1,\bar{4}) + (1,1,1,1,1,1,1,2; \bar{4}, 1)]$ 
and $\chi = 2^8$. From the table of scalars
for this model, we find that if we break $SU(4) \times SU(4)$ 
to the diagonal $SU_{D}(4)$,
then the model becomes vectorlike.

All scalars that are nontrivial in the $SU(4)$s are of the form 
$(1,1,1,1,1,1,1,1;4,\bar{4})+h.c.$, 
and a VEV for any one can be rotated such that the
unbroken
symmetry is $SU_{D}(4)$. All other scalars are $SU_{i}(2)\times
SU_{j}(2)$
bilinears, hence we cannot break to a Pati-Salam (PS) model or any
standard
type chiral model.

\bigskip
\bigskip

\underline{16/8 with  ${\bf 4=(1_{2}{\bf ,}1_{4+i}{\bf ,}2_{1})}$ and $N=2,$}, where ${\bf
6=(1_{x(i)}%
{\bf ,}2_{1},2_{2,}({\bf 1}_{5}{\bf ,1}_{6},{\bf 1}}_{{\bf 7}}
{\bf ,1}_{8})_{{\bf A}}{\bf )}$ with $x=6,5,8,7$ for
$i=1,2,3,4.$

These models have only half the initial chirality of the previous
model ($%
\chi =2^{7})$, and the fermions are given  above if the overall factor
of 2
is removed. As above, we need to break one $SU(4)$, either will do. We
choose $SU_{2}(4).$ For the scalars shown, we can do this with, say, $%
(1,1,1,2,1,1,1,1;1,\bar{4})$ and $(1,1,1,1,1,1,1,2;1,4)$ VEVs. The
remaining
chiral fermion sector is 

$
(2,1,1,1,1,1;4)  +(1,1,1,2,1,1;\bar{4})  +(1,2,1,1,1,1;4) \\
 +(1,1,1,1,2,1;\bar{4}) +(1,1,2,1,1,1;4)  +(1,1,1,1,1,2;\bar{4})$

for
$G=\prod\limits_{k}SU_{k}(2)\times SU(4),$ with k=1,2,3,5,6,7.

There are only $SU_{i}(2)\times SU_{j}(2)$ bilinear scalars of the form$
(2_{i},2_{j})$ where $i=1,2,$ or $3$ and $j=4,5, $or $6$, who's VEVs reduce
chirality
further, so we cannot reach a three-family P-S model.

Note: what one would need is bilinears that allow us to break
$SU_{1}(2)%
\times SU_{2}(2)\times SU_{3}(2)$ to a diagonal subgroup $SU_{L}(2)$,
and
similarly for $SU_{4}(2)\times SU_{5}(2)\times SU_{6}(2)$ to
$SU_{R}(2)$.
This would then have been a three-family P-S model.

\bigskip

\bigskip

\bigskip

\underline{16/11 with ${\bf 4=({\bf 1}_{2}{\bf ,1}_{2},{\bf 1}_{2}{\bf ,1}_{2})}$
and $%
N=3$}

This model is highly chiral, with $\chi =432$,
and the chiral fermions are 
$6[(3,\bar{3},1,1,1,1,1,1;1,1)  +(1,1,1,1,3,\bar{3},1,1;1,1)  
+(1,3,\bar{3},1,1,1,1,1;1,1)  +(1,1,1,1,1,3,\bar{3},1;1,1) \\
+(1,1,3,\bar{3},1,1,1,1;1,1)  +(1,1,1,1,1,1,3,\bar{3};1,1)  
+(\bar{3},1,1,3,1,1,1,1;1,1)  +(1,1,1,1,\bar{3},1,1,3;1,1)
].$
We can ignore the $SU(6)\times SU(6)$  sector, since it can be broken
completely without affecting the chirality. If we then give VEVs to 
$(1,1,1,8,1,1,1,1)$ and $(1,1,1,1,1,1,1,8)$ representations of
$SU(3)^8$,
we arrive at $6[(3,\bar{3},1)+(1,3,\bar{3})+(1,1,3)+(\bar{3},1,1)]$ in
the
$SU_{i+1}(3)\times $ $SU_{i+2}(3)\times SU_{i+3}(3)$ sector for both
$i=0$
and $i=1$. The $i=0$ sector can be broken completely with
$(1,1,1,1,8,1)$%
-type VEVs plus $(1,1,1,3,1,\bar{3})$-type VEVs. The
remaining fermions falling nearly into six $E_{6}\longrightarrow
SU(3)\times %
SU(3)\times SU(3)$-type families. While close, this model is still
unsuccessful.

\bigskip

\bigskip

\bigskip 
\underline{16/11 with ${\bf 4=({\bf 1}_{2}{\bf ,1}_{2},{\bf 1}_{2}{\bf
,1}_{4})%
}$ and $N=3$}

The chiral fermion sector is exactly half the previous case. Again we
break $%
SU(6)\times SU(6)$ completely. Then breaking
$\prod\limits_{j=4}^{8}SU_{j}(3)
$ completely with $SU_{j}(3)$ octet VEVs gives us finally a chiral
fermion
sector $3[(3,\bar{3},1)+(1,3,\bar{3})+(1,1,3)+(\bar{3},1,1)]$. This is
tantalizingly close to the three-family model we seek, but still no
cigar!

\bigskip

\bigskip

\bigskip 

\bigskip

\bigskip

{\bf 16/13}: There are three potential models for this group.

Consider first the case with 

\underline{${\bf 4} =(2_{1},2_{1})$ and $N=2$. }

Here ${\bf 6}=(1_{2},1_{2},1_{2},1_{3},2_{2})$ and
the chiral fermions are 

\bigskip

2[$(2,1,1,1;4,1,1)+(1,2,1,1;1,1,4)+(1,1,2,1;1,1,4)+
(1,1,1,2;4,1,1)+(2,1,1,1;1,1,\bar{4})
+(1,2,1,1;\bar{4},1,1)+(1,1,2,1;\bar{4},1,1)+(1,1,1,2;1,1,\bar{4})$]

\bigskip

VEVs of the form
\mbox{$<$}%
$4_{2},\bar{4}_{2}>$ etc., can break  $SU_{2}(4)$ completely (this group
is
irrelevant,
since there are no chiral fermions with $SU_{2}(4)$ quantum numbers).
VEVs for ($4_{1},\bar{4}_{3})$ scalars then breaks $SU_{1}(4)\times
SU_{3}(4)
$ to $SU_{D}(4)$, such that the fermions become vectorlike. On the other
hand, VEVs for (2$_{4}$,$4_{2})+h.c.$ reduces the chiral sector to 

\bigskip

2[$%
(2,1,1;1,4)+(1,2,1;4,1)+(1,1,2;4,1)+2(1,1,1;1,4) \\$
$ +(2,1,1;\bar{4},1) 
+2(1,1,1;\bar{4},1)+(1,2,1;1,\bar{4})+(1,1,2;1,\bar{4})]$

and then a VEV for (2$_{3}$,$4_{2})+h.c$ reduces this farther to $%
2[(2,1;1,4)+(1,2;4,1)+(1,2;1,\bar{4})+(2,1;\bar{4},1)].$

As above a VEV for ($4_{1},\bar{4}_{3})$ scalars would render the model
vectorlike, while just breaking $SU_{3}(4)$ would give a one-family
model.
However, this needs VEVs for (2$_{1}$,$2_{4})$ and (2$_{2}$,$2_{3})$,
but no
scalars of this type exist in the model. We conclude the model has no
Pati-Salam type phenomology.

\bigskip

\bigskip

Consider next 

\underline{{\bf 16/13} with ${\bf 4}=({\bf 2}_{1}{\bf ,2}_{2})$ and $N=2$.}

This time ${\bf
6}$ is as given in Section 5, but undetermined up to the identification of
antisymmetric
singlets in ({\bf 2}$_{i}\times {\bf 2}_{i})_{A}$ with $i=1,2$. The
chiral
fermions are as in the ${\bf 4=(2_{1},2_{1})}$ case, but with the
overall
factor of 2 deleted. A useful strategy is to do a generic spontaneous
symmetry breaking analysis to try to obtain a realistic Pati-Salam type
phenomenology and then, if successful, one asks if the scalars to carry
out
the breaking are in the model. As above, $SU_{2}(4)$ is irrelevant and
so
can be ignored. If we identify $SU_{1}(2)\times SU_{4}(2)$ with
$SU_{L}(2)$
and $SU_{2}(2)\times SU_{3}(2)$ with $SU_{R}(2)$, we find 2$%
[(2,1;1,4)+(1,2;4,1)+(1,2;1,\bar{4})+(2,1;\bar{4},1)]$. Now breaking one
of
the remaining $SU(4)s$ completely gives two families, and this is the
best
one can do. Hence independent of what scalars are available, there is no
chance to get a model with three or more families.

The remaining 16/13 case is:

\underline{ ${\bf 4=(1_{3},1_{4},2_{1})}$ with $N=2$.}

Now ${\bf 6=(1_{2},2_{1},2_{3,}1}_{{\bf c}}{\bf )}$. 
but the chiral fermions are in the
same representations as the previous model, and so we can immediately
conclude on general grounds that there is no viable phenomenology for
this case.

\bigskip

{\bf \bigskip 18/3}

\bigskip

Now consider

\underline{18/3 with  ${\bf 4=(1\alpha ,1}^{{\bf \prime }}{\bf
;,2\alpha )}
$ and $N=3$.} 
This model has $\chi =192$ and chiral fermions 
$(2,1,1,1,1,1;1,4,1)+(1,2,1,1,1,1;1,4,1)+(1,1,2,1,1,1;\bar{4},1,1) \\
+(1,1,1,2,1,1;\bar{4},1,1)+(1,1,2,1,1,1;1,1,4)+(1,1,1,2,1,1;1,1,4) \\
+(1,1,1,1,2,1;1,\bar{4},1)+(1,1,1,1,1,2;1,\bar{4},1)+(1,1,1,1,2,1;4,1,1) \\
+(1,1,1,1,1,2;4,1,1)+(2,1,1,1,1,1;1,1,\bar{4})+(1,2,1,1,1,1;1,1,\bar{4}) \\
+2[(1,1,1,1,1,1;\bar{4},4,1)+
(1,1,1,1,1,1;1,\bar{4},4)+(1,1,1,1,1,1;4,1,\bar{4})]$. 
Breaking $SU^{6}(2)$ to a single diagonal $SU(2)$with all
six (2$_{i}$,2$_{j}$) type VEVs of $SU_{i}(2)\times SU_{j}(2)$, and
then
further VEVs of the type (2;4,1,1), (2;1,4,1), and (2;1,1,4) 
to break the $SU(4)$s to $SU(3)$s leads to
the
set of remaining chiral fermions:

2[$(3,\bar{3},1)+(1,3,\bar{3})+(\bar{3},1,3)].$ 

So this route leads to two families.

\bigskip
\bigskip

\noindent If instead we seek a Pati-Salam model, there are several spontaneous
symmetry breaking routes we need to investigate. If we break with
(1,1,1,1,1,1;$\bar{4},4,1)$ scalars to $SU^{6}(2)\times SU_{D}(4)\times
SU_{3}(4)$ we find the fermions remaining chiral are 

$%
(2,1,1,1,1,1;4,1)+(1,2,1,1,1,1;4,1)+ (1,1,2,1,1,1;1,4)+(1,1,1,2,1,1;1,4) \\
+(1,1,2,1,1,1;\bar{4},1)
+(1,1,1,2,1,1;\bar{4},1) +(2,1,1,1,1,1;1,\bar{4})+(1,2,1,1,1,1;1,\bar{4}).$

Now breaking with a 
($4_1,\bar{4}_{3})$ or ($4_2,\bar{4}_{3})$ VEV would render the model vectorlike, so we avoid
this and insted give
VEVs to (2$_{5}$,4$_{1}$) and (2$_{6}$,4$_{1}$) to break $SU_{D}(4)$ to
$SU^{\prime }(2)$. However, this yields at most two families.

We must try another route. If we avoid ($\bar{4},4)$ type VEVs and give
VEVs
only to ($2,4)$ type scalars, we can proceed as follows:
\mbox{$<$}%
2$_{1}$,4$_{2}>,$
\mbox{$<$}%
2$_{2}$,4$_{2}>,$
\mbox{$<$}%
2$_{3}$,\={4}$_{1}>$ and
\mbox{$<$}%
2$_{4}$,\={4}$_{1}>$ VEVsbreak $SU^{6}(2)\times SU^{3}(4)$ to
$SU_{5}(2)\times %
SU_{6}(2)\times SU^{\prime }(2)\times SU^{\prime ^{{}}\prime }(2)\times
SU(4)
$.
Some fermions remain chiral but they are insufficient to construct
families. We conclude that this model will not provide viable
phenomenology.

\bigskip

\underline{20/5 with ${\bf 4=(1_{2},1_{2},1_{2},1_{2})}$ and $N=3$}

The chiral $SU^{4}(3)$ fermions are 4[$(3,\bar{3},1,1)+$ $(1,3,\bar{3}%
,1)+(1,1,3,\bar{3})+(\bar{3},1,1,3)].$ (The $SU(6)$ does not
participate; it
will be ignored.) The only scalars are in representations $(3,1,\bar{3},1)$
+h.c. and $(1,3,1,\bar{3})$+h.c. A
VEV to, say, the first of these, would break $SU_{1}(3)\times SU_{3}(3)$
to
a diagonal $SU_{D}(3)$, and the fermions would become 4[$(3,\bar{3},1)+$
$(%
\bar{3},3,1)+(3,1,\bar{3})+(\bar{3},1,3)]$ under $SU_{D}(3)\times
SU_{2}(3)%
\times SU_{4}(3)$, which is
vectorlike. Hence any allowed VEVs immediately renders the model
vectorlike.

We get no farther with ${\bf 4=(1_{2},1_{2},1_{2},1_{2})}$ and N = 3,
where $%
{\bf 6}={\bf (1_{3},1_{3},1_{3},1_{1},1_{1},1_{1})}$, since this model
has
only half the chirality content of the previous case, and
again VEVs will render it vectorlike.

\bigskip

\underline{21/2 with ${\bf 4}=({\bf 1}_{2}{\bf ,3}_{1})$ and $N=2$.} Now${\bf
6}=({\bf 3}%
_{1}{\bf ,3}_{2}).$ (Other embeddings of the {\bf 4} with n=0 SUSY are
permutation of the representations of this model and therefore all
equivalent.) The fermions have $\chi =108$ and are $%
(2,1,1;6,1)+(1,2,1;6,1)+(1,1,2;6,1)$
$+(2,1,1;1,\bar{6})+(1,2,1;1,\bar{6}%
)+(1,1,2;1,\bar{6})+(1,1,1;\bar{6},6).$ A VEV for a ($\bar{6},6)$ scalar
renders the model vectorlike. Our only other option is to give (2,6)
type
VEVs. A
\mbox{$<$}%
2,1,1;6,1%
\mbox{$>$}%
breaks the gauge group to $SU_{2}(4)\times SU_{3}(2)\times SU(5)\times
SU(6)$
with chiral fermions 2$(1,1;5,1)+(1,2;5,1)+(2,1;5,1)$ $+(1,1;1,\bar{6}%
)+(2,1;1,\bar{6})+(1,2;1,\bar{6})+(1,1,1;\bar{5},6).$ There is
insufficient
fermion content for a three family Pati-Salam model if we identify
$SU_{2}(4)%
\times SU_{3}(2)$ with $SU_{L}(4)\times SU_{R}(2).$ Our only other
choice is
to get one of these $SU(2)$s from $SU(5)\times SU(6).$ For instance a
\mbox{$<$}%
2$_{2}$,5%
\mbox{$>$}%
VEV breaks the gauge group to $SU_{3}(2)\times SU(4)\times SU(6)$ but
the
remaing chiral fermions are 4$(1,4,1)+(2,4,1)$
$+3(1,1,\bar{6})+(2,1,\bar{6}
)+(1,2;1,\bar{6})+(1,1,6)+(1,\bar{4},6).$ We can not identify $SU(4)$
with $%
SU_{PS}(4)$, so this group can only be in $SU(6).$ Breaking $SU(6)$
with an
adjoint to $SU(2)\times SU(4)$ leaves us with $SU(2)\times SU(4)\times
SU(2)%
\times SU(4)$ fermions that are again insufficient for a three family
Pati-Salam model.

\bigskip

\bigskip

\underline{24/7 with ${\bf 4}=({\bf 1}\alpha ,{\bf 1}^{\prime },{\bf 2}\alpha )$ for $N=2$}

This model, the only successful one in the present broad search, has been discussed in detail in \cite{FK2} but for completeness we repeat the derivation here.

The original gauge group at the conformality scale is $SU(4)^3 \times SU(2)^{12}$ with
chiral fermions as given in Section IV and complex scalars as stated in Section V above.

If we break the three $SU(4)$s to a single diagonal $SU(4)$ subgroup, chirality is
lost. To avoid this we break $SU(4)_1$ completely and then break
$SU(4)_{\alpha} \times SU(4)_{\alpha^2}$ to its diagonal
subgroup $SU(4)_D$. The appropriate VEVs are available as $[(4_1, 2_b \alpha^k) + h.c.]$
with $b$ arbitrary but $k=1$ or $k=2$. The second step requires an $SU(4)_D$ singlet VEV
from $(\bar{4}_{\alpha}, 4_{\alpha^2})$ and/or $(4_{\alpha}, \bar{4}_{\alpha^2})$.
Once a choice is made for $b$ (we take $b=4$), the remaining fermions are, in an intuitive notation,:
\begin{equation}
\sum_{\alpha = 1}^{\alpha = 3} [ (2_{\alpha}\alpha, 1, 4_D) + (1, 2_{\alpha}\alpha^{-1}, \bar{4}_D)]
\label{24/7}
\end{equation}
which has the same content as a three family Pati-Salam model, though
with a separate $SU(2)_L \times SU(2)_R$ per family.

To further reduce the symmetry we must arrange to break to a single $SU(2)_L$
and a single $SU(2)_R$.
This is achieved by modifying step one where $SU(4)_1$ was broken. Consider
the block diagonal decomposition of $SU(4)_1$ into $SU(2)_{1L} \times SU(2)_{1R}$.
The representations $(2_{\alpha}\alpha, 4_1)$ and $(2_{\alpha}\alpha^{-1}, 4_1)$ 
decompose as $(2_{\alpha}\alpha, 4_1) \rightarrow (2_{\alpha}\alpha, 2, 1) + (2_{\alpha}\alpha,
1, 2)$ and $(2_{\alpha}\alpha^{-1}, 4_1) \rightarrow (2_{\alpha}\alpha^{-1}, 2, 1) + (2_{\alpha}^{-1},
1, 2)$.
Now if we give VEVs of equal magnitude to the
$(2_a\alpha, 2, 1), a = 1, 2, 3$
and equal magnitude VEVs to the
$(2_a\alpha^{-1}, 1, 2), a = 1, 2, 3$,
we break 
$SU(2)_{1L} \times \Pi_{a=1}^{a=3}SU(2_a\alpha)$
to a single
$SU(2)_L$
and we break
$SU(2)_{1R} \times \Pi_{a=1}^{a=3}SU(2_a\alpha^{-1})$
to a single
$SU(2)_R$.
Finally, VEVs for $(2_4\alpha, 2, 1)$
and $(2_4\alpha, 1, 2)$ as well as
$(2_4\alpha^{-1}, 2, 1)$ and $(2_4\alpha^{-1}, 1, 2)$
ensure that both $SU(2_4\alpha)$ and $SU(2_4\alpha^{-1})$
are broken and that only three families remain chiral.
The final set of chiral fermions is then
$3[(2, 1, 4) + (1, 2, \bar{4})]$ with gauge symmetry
$SU(2)_L \times SU(2)_R \times SU(4)_D$.

To achieve the final reduction to the standard model, an adjoint VEV
from $(\bar{4}_{\alpha}, 4_{\alpha^2})$ and/or $(4_{\alpha}, \bar{4}_{\alpha^2})$
is used to break $SU(4)_D$ to 
$SU(3) \times U(1)$, and
a right-handed doublet is used to break $SU(2)_R$.

\bigskip
\bigskip  

\underline{24/9 with ${\bf 4}=({\bf 1}_{1}\alpha ,{\bf 1}_{2}\alpha^3 ,{\bf 2}\alpha^2
)$ for $N=2$}

The original gauge group at the conformality scale is $SU(4)^4 \times SU(2)^{8}$ with
chiral fermions as given in Section IV and complex scalars as stated in Section V above.

Achievement of chiral families under the Pati-Salam subgroup
$SU(4) \times SU(2)_L \times SU(2)_R$
requires the identifications
$SU(2)_{1_1} = SU(2)_{1_2} = SU(2)_{1_1\alpha} = SU(2)_{1_2\alpha} = SU(2)_L$;
$SU(2)_{1_1\alpha^3} = SU(2)_{1_2\alpha^2} = SU(2)_{1_1\alpha^3} = SU(2)_{1_2\alpha^3} = SU(2)_R$;
while, for example,
$SU(4)_{2} = SU(4)_{2\alpha} = {\bf \bar{4}}$ of $SU(4)$;
$SU(4)_{2\alpha^2} = SU(4)_{2\alpha^3} = {\bf 4}$ of $SU(4)$
where by this simplified notation we imply diagonal subgroups.

But the scalars tabulated for this case in Section V
are insufficient to allow this pattern of spontaneous symmetry breaking,
and hence no interesting model emerges.

\bigskip
\bigskip

\underline{24/9 with ${\bf 4}=({\bf 1}_{1}\alpha ,{\bf 1}_{2}\alpha ,{\bf 2}\alpha
)$ for $N=2$}

The original gauge group at the conformality scale is $SU(4)^4 \times SU(2)^{8}$ with
chiral fermions as given in Section IV and complex scalars as stated in Section V above.

Achievement of chiral families under the Pati-Salam subgroup
$SU(4) \times SU(2)_L \times SU(2)_R$
requires the identifications
$SU(2)_{1_1} = SU(2)_{1_2} = SU(2)_{1_1\alpha} = SU(2)_{1_2\alpha} = SU(2)_L$;
$SU(2)_{1_1\alpha^3} = SU(2)_{1_2\alpha^2} = SU(2)_{1_1\alpha^3} = SU(2)_{1_2\alpha^3} = SU(2)_R$;
while, for example,
$SU(4)_{2} = SU(4)_{2\alpha^3} = {\bf \bar{4}}$ of $SU(4)$;
$SU(4)_{2\alpha} = SU(4)_{2\alpha^2} = {\bf 4}$ of $SU(4)$
where by this simplified notation we imply diagonal subgroups.

But the scalars tabulated for this case in Section V
are insufficient to allow this pattern of spontaneous symmetry breaking,
and hence no interesting model emerges.

\bigskip

\bigskip 

\bigskip

\underline{24/9 with ${\bf 4}=({\bf 1}_{1}\alpha^2 ,{\bf 1}_{2}, {\bf 2}\alpha
)$ for $N=2$}

The original gauge group at the conformality scale is $SU(4)^4 \times SU(2)^{8}$ with
chiral fermions as given in Section IV and complex scalars as stated in Section V above.

Achievement of chiral families under the Pati-Salam subgroup
$SU(4) \times SU(2)_L \times SU(2)_R$
requires the identifications
$SU(2)_{1_1} = SU(2)_{1_2} = SU(2)_{1_1\alpha} = SU(2)_{1_2\alpha} = SU(2)_L$;
$SU(2)_{1_1\alpha^3} = SU(2)_{1_2\alpha^2} = SU(2)_{1_1\alpha^3} = SU(2)_{1_2\alpha^3} = SU(2)_R$;
while, for example,
$SU(4)_{2} = SU(4)_{2\alpha^3} = {\bf \bar{4}}$ of $SU(4)$;
$SU(4)_{2\alpha} = SU(4)_{2\alpha^2} = {\bf 4}$ of $SU(4)$
where by this simplified notation we imply diagonal subgroups.

But the scalars tabulated for this case in Section V
are insufficient to allow this pattern of spontaneous symmetry breaking,
and hence no interesting model emerges.

\bigskip
\bigskip

\newpage

\bigskip

\underline{24/9 with ${\bf 4}=({\bf 2}\alpha ,{\bf 2}\alpha )$ for $N=2$}

The original gauge group at the conformality scale is $SU(4)^4 \times SU(2)^{8}$ with
chiral fermions as given in Section IV and complex scalars as stated in Section V above.

Achievement of chiral families under the Pati-Salam subgroup
$SU(4) \times SU(2)_L \times SU(2)_R$
requires the identifications
$SU(2)_{1_1} = SU(2)_{1_1\alpha} = SU(2)_{1_1\alpha^2} = SU(2)_{1_1\alpha^3} = SU(2)_L$;
$SU(2)_{1_2\alpha} = SU(2)_{1_2\alpha} = SU(2)_{1_2\alpha^2} = SU(2)_{1_2\alpha^3} = SU(2)_R$;
while, for example,
$SU(4)_{2\alpha} = SU(4)_{2\alpha^3} = {\bf 4}$ of $SU(4)$
where by this simplified notation we imply diagonal subgroups, and
$SU(4)_{2}$ and $SU(4)_{2\alpha^2}$ are broken.

But the scalars tabulated for this case in Section V
are insufficient to allow this pattern of spontaneous symmetry breaking,
and hence no interesting model emerges.

\bigskip
\bigskip

\underline{24/13 with ${\bf 4}=({\bf 2}_{1},{\bf 2}_{2})$ for $N=2$}

The original gauge group at the conformality scale is $SU(6) \times SU(4)^3 \times SU(2)^3$ with
chiral fermions as given in Section IV and complex scalars as stated in Section V above.

\bigskip

According to the analysis in Section IV this orbifold permits only two chiral
families and is therefore not of phenomenological interest.

\bigskip
\bigskip

\underline{24/14 with ${\bf 4}=({\bf 2}_{1},{\bf 2}_{2})$ for $N=2$}

The original gauge group at the conformality scale is $SU(4)^4 \times SU(2)^8$ with
chiral fermions as given in Section IV and complex scalars as stated in Section V above.

Achievement of chiral families under the Pati-Salam subgroup
$SU(4) \times SU(2)_L \times SU(2)_R$
requires the identifications
$SU(2)_{1_1} = SU(2)_{1_2} = SU(2)_{1_5} = SU(2)_{1_6} = SU(2)_L$;
$SU(2)_{1_3} = SU(2)_{1_4} = SU(2)_{1_5} = SU(2)_{1_6} = SU(2)_R$;
while, for example,
$SU(4)_{2_2} = SU(4)_{2_3} = {\bf 4}$ of $SU(4)$;
$SU(4)_{2_1} = SU(4)_{2_4} = {\bf \bar{4}}$ of $SU(4)$
where by this simplified notation we imply diagonal subgroups.

But the scalars tabulated for this case in Section V
are insufficient to allow this pattern of spontaneous symmetry breaking,
and hence no interesting model emerges.


\bigskip
\bigskip

\underline{24/15 with ${\bf 4}=({\bf 1}_{2},{\bf 1}_{3},{\bf 2}_{3})$ for $N=2$}

The original gauge group at the conformality scale is $SU(4)^5 \times SU(2)^4$ with
chiral fermions as given in Section IV and complex scalars as stated in Section V above.

Achievement of chiral families under the Pati-Salam subgroup
$SU(4) \times SU(2)_L \times SU(2)_R$
requires the identifications
$SU(2)_{1_1} = SU(2)_{1_3} = SU(2)_L$;
$SU(2)_{1_2} = SU(2)_{1_4} = SU(2)_R$;
while, for example,
$SU(4)_{2_3} = SU(4)_{2_4} = {\bf 4}$ of $SU(4)$,
where by this simplified notation we imply diagonal subgroups.

But the scalars tabulated for this case in Section V
are insufficient to allow this pattern of spontaneous symmetry breaking,
and hence no interesting model emerges.

\bigskip
\bigskip

\underline{24/15 with ${\bf 4}=({\bf 2}_{3},{\bf 2}_{5})$ for $N=2$}

The original gauge group at the conformality scale is $SU(4)^5 \times SU(2)^4$ with
chiral fermions as given in Section IV and complex scalars as stated in Section V above.

\bigskip

According to the analysis in Section IV this orbifold permits only two chiral
families and is hence not phenomenologically interesting.

\bigskip
\bigskip

\underline{24/15 with ${\bf 4}=({\bf 2}_{3},{\bf 2}_{3})$ for $N=2$}

The original gauge group at the conformality scale is $SU(4)^5 \times SU(2)^4$ with
chiral fermions as given in Section IV and complex scalars as stated in Section V above.

Achievement of chiral families under the Pati-Salam subgroup
$SU(4) \times SU(2)_L \times SU(2)_R$
requires the identifications
$SU(2)_{1_1} = SU(2)_{1_3} = SU(2)_L$;
$SU(2)_{1_2} = SU(2)_{1_4} = SU(2)_R$;
while, for example,
$SU(4)_{2_3} = SU(4)_{2_4} = {\bf 4}$ of $SU(4)$
where by this simplified notation we imply diagonal subgroups.

But the scalars tabulated for this case in Section V
are insufficient to allow this pattern of spontaneous symmetry breaking,
and hence no interesting model emerges.

\bigskip
\bigskip

\bigskip

\bigskip

\underline{27/4 with ${\bf 4}=({\bf 1}_{2}{\bf ,3}_{1})$ with $N=2.$}

Here ${\bf 6}=({\bf
3}_{1}{\bf %
,3}_{2})$ and the chiral fermions are given by Equation 29 and all scalars are of type
$%
(2_{i},\bar{6}_{1}),(2_{i},6_{2})$ or $(6_{1},\bar{6}_{2})\ $for
$i=1,2,...,9
$.. A VEV for $(6_{1},\bar{6}_{2})+h.c.$ scalar breaks
$SU_{1}(6)\times SU_{2}(6)$ to $SU_{D}(6)$, and the model becomes
vectorlike. Hence we must break only with (2,6) type scalars if there is
any
hope of a viable model. We give VEVs to $(2_{i},6_{1})$ scalars for $%
i=1,2,...,5$  to break $SU_{1}(6)$ completely, and VEVs to
$(2_{j},6_{2})$
for $j=6,7$ to break $SU_{2}(6)$ to $SU(4)$. Then the remaining unbroken
gauge group is $SU_{8}(2)\times SU_{9}(2)\times SU(4)$ with  
fermions
$(2,1,4)+(1,2,4)+4(1,1,\bar{4})$, which are chiral but not of the correct
form. 

A more successful variation is obtained with $(2_{i},6_{1})$
scalars VEVs for $i=1,2,3$ and $4$ to break the gauge group to
$SU_{5}(2)\times %
SU_{6}(2)\times SU_{7}(2)\times SU_{8}(2)\times SU_{9}(2)\times
SU^{\prime
}(2)\times SU(6)$ and thenVEVs for $(2_{5},6_{2})$ and $(2_{6},6_{2})$
to
break to $SU_{7}(2)\times SU_{8}(2)\times SU_{9}(2)\times SU^{\prime
}(2)%
\times SU(4)$ which has chiral fermions
(2,1,1,1,4)+(1,2,1,1,4)+(1,1,2,1,4)+3(1,1,1,2,\={4}). If we could break
$%
SU_{7}(2)\times SU_{8}(2)\times SU_{9}(2)$ to a diagonal
$SU(2)$ subgroup, we
would have a three-family Pati-Salam model. However, the scalars to
accomplish this are not in the spectrum. If we could give VEVs to $%
(2_{i},6_{1})$ scalars for $i=7,8,9$ to break $SU_{7}(2)\times
SU_{8}(2)%
\times SU_{9}(2)$ to a $U_{Y}(1)$ without disturbing the $SU^{\prime
}(2)$
subgroup of $SU_{1}(6),$ and a further $(2_{j},6_{2})$ VEV, say $%
(2_{1},6_{2}),$ to break $SU(4)$ to $SU_{C}(3),$ then we would have a true
three family standard (i.e., $U_{Y}(1)\times SU_{EW}(2)\times $
$SU_{C}(3))$%
model upon identifying $SU^{\prime }(2)$ with $SU_{EW}(2)$.

\bigskip

\underline{30/2 with\bigskip ${\bf 4=(1}_{{\bf 1}}
{\bf \alpha ,1_{2}},{\bf 2\alpha)}$
and $N=2$}.

\bigskip
\bigskip

Here ${\bf 6}=({\bf 1}_{2}{\bf \alpha ,1}_{2}{\bf \alpha ^{2}},{\bf
2\alpha },%
{\bf 2\alpha ^{2}})$, and the gauge group is $SU^{6}(2)\times
SU^{6}(4).$ This
group has chiral fermions:

 $(2,1,1,1,1,1;1,1,4,1,1,1)
+(1,2,1,1,1,1;1,1,4,1,1,1) \\
+(1,1,2,1,1,1;\bar{4},1,1,1,1,1)
+(1,1,1,2,1,1;\bar{4},1,1,1,1,1)$

$+(1,1,1,1,1,1;1,\bar{4},4,1,1,1)
+(1,1,1,1,1,1;\bar{4},1,4,1,1,1) \\
+2(1,1,1,1,1,1;1,\bar{4},1,4,1,1)
+(1,1,1,1,1,1;\ \bar{4},1,1,4,1,1)$

+$($1$,1,2,1,1,1;1,1,1,1,4,1)+(1,1,1,2,1,1;1,1,1,1,4,1) \\
+(1,1,1,1,2,1;1,1,\bar{4},1,1,1)
+(1,1,1,1,1,2;1,1,\bar{4},1,1,1)$

$+(1,1,1,1,1,1;1,1,1,\bar{4},4,1)
+(1,1,1,1,1,1;1,1,\bar{4},1,4,1) \\
+2(1,1,1,1,1,1;1,1,1,\bar{4},1,4)+(1,1,1,1,1,1;1,1,\bar{4},1,1,4)$

+$(1,1,1,1,2,1;4,1,1,1,1,1)+(1,1,1,1,1,2;4,1,1,1,1,1) \\
+(2,1,1,1,1,1;1,1,1,1,\bar{4},1)
+(1,2,1,1,1,1;1,1,1,1,\bar{4},1)$
$+(1,1,1,1,1,1;4,1,1,1,\bar{4},1)+(1,1,1,1,1,1;4,1,1,1,1,\bar{4}) \\
+2(1,1,1,1,1,1;1,4,1,1,1,\bar{4})+(1,1,1,1,1,1;1,4,1,1,\bar{4},1)$

The spontaneous symmetry breaking analysis for this model is quite
unwieldy,
but for the most part can be carried out systematically.  For example,
breaking with $(1,1,1,1,1,1;1,\bar{4},1,4,1,1),$ $(1,1,1,1,1,1;\\bar{4}%
,1,1,4,1,1),$ $(1,1,1,1,1,1;4,1,1,1,\bar{4},1)$ and
$(1,1,1,1,1,1;1,1,\bar{4}%
,1,1,4)$ VEVs reduces $SU^{6}(4)$ to $SU_{1}(4)\times SU_{D}(4),$ with
fermions remaining chiral in representations: 

$%
(2,1,1,1,1,1;1,4)+(1,2,1,1,1,1;1,4)
+(1,1,2,1,1,1;\bar{4},1)+(1,1,1,2,1,1;\bar{4},1)$

$+($1$,1,2,1,1,1;1,4)+(1,1,1,2,1,1;1,4)
+(1,1,1,1,2,1;1,\bar{4})+(1,1,1,1,1,2;1,\bar{4})$

+$(1,1,1,1,2,1;4,1)+(1,1,1,1,1,2;4,1)+(2,1,1,1,1,1;1,\bar{4})
+(1,2,1,1,1,1;1,\bar{4}).$ 

Now (1,1,1,1,2,1;4,1) and (1,1,1,1,1,2;4,1) VEVs break
$SU_{5}(2)\times SU_{6}(2)\times SU_{1}(4)$ 
to $SU^{\prime }(2)$ with fermions
remaining chiral in the representations:

$(2,1,1,1;4)+(1,2,1,1;4)+($1$,1,2,1;4)
+(1,1,1,2;4) \\
+2(1,1,1,1;\bar{4})+2(1,1,1,1;\bar{4})+(2,1,1,1;\bar{4})
+(1,2,1,1;\bar{4})$ 

\noindent which is already insufficient to provide three
normal
families. Other analyses of spontaneous symmetry breaking toward
constructing a Pati-Salam model starting with this 30/2 model are
similarly
unsucessful. 

An alternative is to seek a trinification model. To
this
end, consider only the $SU^{6}(4)$ fermion
sector

$+(1,\bar{4},4,1,1,1)+(\bar{4},1,4,1,1,1)
+2(1,\bar{4},1,4,1,1) \\
+(\bar{4},1,1,4,1,1)+(1,1,1,\bar{4},4,1)
+(1,1,\bar{4},1,4,1)$

$+2(1,1,1,\bar{4},1,4)+(1,1,\bar{4},1,1,4)
+(4,1,1,1,\bar{4},1) \\
+2(4,1,1,1,1,\bar{4})
+2(1,4,1,1,1,\bar{4})+(1,4,1,1,\bar{4},1)$

Identifying $SU_{1}(4)$ with $SU_{2}(4),$ $SU_{3}(4)$ with $SU_{4}(4)$
and $SU_{5}(4)$ with $SU_{6}(4)$ would lead to five families of the form
$5[(\bar{4},4,1)+(1,\bar{4},4)+(4,1,\bar{4})],$ however 
there are no scalars of
the
type needed to carry this out.

This analysis is not exhaustive and there may be models where
$SU_{L}(2),$ $SU_{R}(2)$ or both are contained in $SU^{6}(4)$. 
Since we are starting
with
a group of rank 24, and seek the standard model of rank 4 or a unified
model
thereof of rank 5 or 6, and since there are 66 Higgs representations in
the
theory, the spontaneous symmetry breaking possibilities are rather
complex.
The $N=3$ case is obviously even more complicated, with initial rank 42,
and
one could try to automate the search for phenomenological models,
although
we have not attempted to do so.

\bigskip

\underline{30/3 with ${\bf 4=(1}_{{\bf 1}}{\bf \alpha ,1_{2}},{\bf 2\alpha }^{{\bf
2}}%
{\bf )}$ and $N=2$.} We now have ${\bf 6}=({\bf 1}_{2}{\bf \alpha ,1}_{2}{\bf \alpha
^{4}},{\bf %
2\alpha }^{{\bf 3}},{\bf 2\alpha ^{2}})$ where ${\bf \alpha ^{5}=1.}$

The chiral  $SU^{10}(2)\times SU^{5}(4)$fermions are

(1$^{10}$;$\bar{4},4,1,1,1)+$(1$^{10}$;$\bar{4},1,4,1,1)+
$(1$^{4},2,1^{5}$;$\bar{4},1,1,1,1)
+$(1$^{5},2,1^{4}$;$\bar{4},1,1,1,1)$$ \\
$+(1$^{10}$;1,$\bar{4},4,1,1)+
$(1$^{10}$;1,$\bar{4},1,4,1)+$(1$^{6},2,1^{3}$;$1,\bar{4},1,1,1)
+$(1$^{7},2,1^{2}$;$1,\bar{4},1,1,1)$$ \\
$+(1$^{10}$;$1,1,\bar{4},4,1)
+$(1$^{10}$;1,1,$\bar{4},1,4)+$(1$^{8},2,1^{1}$;
$1,1,\bar{4},1,1)+$(1$^{9},2$;$1,1,\bar{4},1,1)$$ \\
$+(1$^{10}$;$1,1,1,\bar{4},4)+$(1$^{10}$;$4,1,1,\bar{4},1)
+$(2,1$^{9}$;$1,1,1,\bar{4},1)
+$(1$^{1},2,1^{8}$;$1,1,1,\bar{4},1)$$ \\
$+(1$^{10}$;$4,1,1,1,\bar{4})+$(1$^{10}$;$1,4,1,1,\bar{4})+$($1^{2},2,1^{7}$;
$ 1,1,1,1,\bar{4})+$(1$^{3},2,1^{6}$;$1,1,1,1,\bar{4})$
\bigskip
Consider the bifundamentals only. VEVs for 
($1,1,1,\bar{4},4)\ $and (1,$\bar{4},4,1,1)$ scalars reduce the chiral fermion
sector to $2[(\bar{4},4,1)+(1,\bar{4},4)+(4,1,\bar{4})]$ which provides
at
most a two family model.

If instead we try to construct a Pati-Salam model, and note that there
are
20 (2;4) type fermions, and that we need six appropriate ones of these
for
three families, we must take care in the spontaneous symmetry breaking
to
preserve this much chirality. If we (i), break  $SU_{2}(4)\times $
$SU_{4}(4)%
\times SU_{5}(4)$ completely and (ii) $SU_{1}(4)\times SU_{3}(4)$ to $%
SU_{PS}(4)$ while (iii) equating $SU_{5}(2),$ $SU_{6}(2),$ $SU_{9}(2)\
$and
(iv) equating $SU_{10}(2)$ with $SU_{L}(2)$, and $SU_{1}(2),$
$SU_{2}(2),$ $%
SU_{7}(2)\ $and $SU_{8}(2)$ with $SU_{R}(2),$ and (v) breaking
$SU_{3}(2)%
\times SU_{4}(2)$ completely, we would be left with a 4 family
Pati--Salam
Fmodel. Can we do this?
(ii) is accomplished with (a) (1$^{10}$;$\bar{4},1,4,1,1),$ then (i)
requires (b) (1$^{10}$;1,$\bar{4},1,4,1)$ and (c)
(1$^{10}$;1,$\bar{4},1,1,4)
$to get a $SU_{D}(4)$. Breaking this to nothing, assuming VEVs (a) and
(b)
allow no freedom to rotate the (c) VEV to diagonal form. Now, at this
point,
we are stymied, as there are insufficient (2$_{i}$,2$_{j}$)
representations
of $SU_{i}(2)\times $ $SU_{j}(2)$ to accomplish (v).

Finally, one can imagine that there exist models with either $SU_{L}(2)$
or $%
SU_{R}(2)$ or both coming from $SU^{5}(4)$, but we see not obvious way
to
cary this out, while on the other hand since there are 60 Higgs
representations we are unable to categorically eliminate this
possibility.

\newpage

\bigskip
\bigskip
\bigskip

\section{Summary}

We have shown how $AdS/CFT$ duality leads to a large
class of models which can provide interesting extensions of the standard model of
particle phenomenology. The naturally occurring ${\cal N} = 4$ extended
supersymmetry was completely broken to
${\cal N} = 0$ by choice of orbifolds $S^5/\Gamma$ such that 
$\Gamma \not\subset SU(3)$.

\bigskip

In the present work, we studied systematically all such non-abelian $\Gamma$
with order $g \leq 31$. We have seen how chiral fermions require that the
embedding of $\Gamma$ be neither real nor pseudoreal. This reduces dramatically
the number of possibilities to obtain chiral fermions.
Nevertheless, many candidates for models which contain the
chiral fermions of the three-family standard model were found.

\bigskip
\bigskip

However, the requirement that the spontaneous symmetry breaking down to the correct gauge symmetry
of the standard model be permitted by the prescribed scalar representations
eliminates most of the surviving models. We found only one allowed model
based on the $\Gamma = 24/7$ orbifold. We had initially expected to find more examples in our search.
The moral for model-building is interesting. Without the rigid framework of string duality 
the scalar sector would be arbitrarily chosen to 
permit the required spontaneous symmetry breaking. This is the normal practice in the standard model,
in grand unification, in supersymmetry and so on.
With string duality, the scalar sector is prescribed by the construction
and only in one very special case does it permit the required symmetry breaking.

\bigskip

This leads us to give more credence to the $\Gamma = 24/7$ example that does work
and to encourage its further study to check whether it can have any connection
to the real world.

\newpage

\bigskip
\bigskip
\bigskip

\section*{acknowledgements}

\bigskip
\bigskip
\bigskip

\noindent PHF thanks the Department of Physics and Astronomy, Vanderbilt
University for hospitality while this work was in progress. TWK thanks the Department of Physics 
and Astronomy, at the 
University of North Carolina and the Aspen Center for Physics for hospitality 
while this work was in progress.
This work was supported in part by the US Department of Energy
under Grants No. DE-FG02-97ER-41036 and DE-FG05-85ER-40226.

\bigskip

\bigskip 

\newpage

\section*{Appendix: Multiplication Tables for Non-Abelian Groups with
$g\leq 31$}

The group D$_{3}$=S$_{3}$

\bigskip
\begin{tabular}{|c||c|c|c|}
\hline
$\otimes $ & 1 & 1$^{\prime }\ $ & 2 \\ \hline\hline
1 & 1 & 1$^{\prime }\ $ & 2 \\ \hline
1$^{\prime }\ $ & 1$^{\prime }\ $ & 1 & 2 \\ \hline
2 & 2 & 2 & 1$+$1$^{\prime }+2$ \\ \hline
\end{tabular}

\bigskip

The group D$_{4}$, 8/4

\begin{tabular}{|c||c|c|c|c|c|}
\hline
$\otimes $ & 1$_{1}$ & 1$_{2}$ & 1$_{3}$ & 1$_{4}$ & 2 \\ \hline\hline
1$_{1}$ & 1$_{1}$ & 1$_{2}$ & 1$_{3}$ & 1$_{4}$ & 2 \\ \hline
1$_{2}\ $ & 1$_{2}$ & 1$_{1}$ & 1$_{4}$ & 1$_{3}$ & 2 \\ \hline
1$_{3}$ & 1$_{3}$ & 1$_{4}$ & 1$_{1}$ & 1$_{2}$ & 2 \\ \hline
1$_{4}$ & 1$_{4}$ & 1$_{3}$ & 1$_{2}$ & 1$_{1}$ & 2 \\ \hline
2 & 2 & 2 & 2 & 2 & $1_{1}+1_{2}+1_{3}+1_{4}$ \\ \hline
\end{tabular}

\bigskip

The group Q, 8/5

\begin{tabular}{|c||c|c|c|c|c|}
\hline
$\otimes $ & 1$_{1}$ & 1$_{2}$ & 1$_{3}$ & 1$_{4}$ & 2 \\ \hline\hline
1$_{1}$ & 1$_{1}$ & 1$_{2}$ & 1$_{3}$ & 1$_{4}$ & 2 \\ \hline
1$_{2}\ $ & 1$_{2}$ & 1$_{1}$ & 1$_{4}$ & 1$_{3}$ & 2 \\ \hline
1$_{3}$ & 1$_{3}$ & 1$_{4}$ & 1$_{1}$ & 1$_{2}$ & 2 \\ \hline
1$_{4}$ & 1$_{4}$ & 1$_{3}$ & 1$_{2}$ & 1$_{1}$ & 2 \\ \hline
2 & 2 & 2 & 2 & 2 & $1_{1}+1_{2}+1_{3}+1_{4}$ \\ \hline
\end{tabular}
\bigskip

\bigskip

\newpage

\bigskip The group D$_{5}$, 10/2

\begin{tabular}{|c||c|c|c|c|}
\hline
$\otimes $ & 1 & $1^{\prime }$ & $2$ & 2$^{\prime }$ \\ \hline\hline
1 & 1 & $1^{\prime }$ & $1^{\prime \prime }$ & 2$^{\prime }$ \\ \hline
$1^{\prime }$ & $1^{\prime }$ & $1$ & 1 & 2$^{\prime }$ \\ \hline
$2$ & $2$ & 2 & $1+1^{\prime }+2^{\prime }$ & $2+2^{\prime }$ \\ \hline
2$^{\prime }$ & 2$^{\prime }$ & 2$^{\prime }$ & $2+2^{\prime }$ & $%
1+1^{\prime }+2$ \\ \hline
\end{tabular}

\bigskip

The group T, 12/4

\begin{tabular}{|c||c|c|c|c|}
\hline
$\otimes $ & 1 & $1^{\prime }$ & $1^{\prime \prime }$ & 3 \\
\hline\hline
1 & 1 & $1^{\prime }$ & $1^{\prime \prime }$ & 3 \\ \hline
$1^{\prime }$ & $1^{\prime }$ & $1^{\prime \prime }$ & 1 & 3 \\ \hline
$1^{\prime \prime }$ & $1^{\prime \prime }$ & 1 & $1^{\prime }$ & 3 \\
\hline
3 & 3 & 3 & 3 & $1+1^{\prime }+1^{\prime \prime }+3+3$ \\ \hline
\end{tabular}

\bigskip

\bigskip The group $D_{6},12/3$

\begin{tabular}{|c||c|c|c|c|c|c|}
\hline
$\otimes $ & 1$_{1}$ & 1$_{2}$ & 1$_{3}$ & 1$_{4}$ & 2 & 2$^{\prime }$
\\
\hline\hline
1$_{1}$ & 1$_{1}$ & 1$_{2}$ & 1$_{3}$ & 1$_{4}$ & 2 & 2$^{\prime }$ \\
\hline
$1_{2}$ & 1$_{2}$ & 1$_{1}$ & 1$_{4}$ & 1$_{3}$ & 2$^{\prime }$ & 2 \\
\hline
1$_{3}$ & 1$_{3}$ & 1$_{4}$ & 1$_{1}$ & 1$_{2}$ & $2^{\prime }$ & 2 \\
\hline
1$_{4}$ & 1$_{4}$ & 1$_{3}$ & 1$_{2}$ & $1_{1}$ & 2 & $2^{\prime }$ \\
\hline
2 & 2 & 2$^{\prime }$ & 2$^{\prime }$ & 2 & $1_{1}+1_{4}+2^{\prime }$ &
$%
1_{2}+1_{3}+2$ \\ \hline
2$^{\prime }$ & 2$^{\prime }$ & 2 & 2 & $2^{\prime }$ & $1_{2}+1_{3}+2$
& $
1_{1}+1_{4}+2^{\prime }$ \\ \hline
\end{tabular}
\bigskip

\bigskip

\newpage

\bigskip The group $Q_{6},12/5$

\begin{tabular}{|c||c|c|c|c|c|c|}
\hline
$\otimes $ & 1$_{1}$ & 1$_{2}$ & 1$_{3}$ & 1$_{4}$ & 2 & 2$^{\prime }$
\\
\hline\hline
1$_{1}$ & 1$_{1}$ & 1$_{2}$ & 1$_{3}$ & 1$_{4}$ & 2 & 2$^{\prime }$ \\
\hline
$1_{2}$ & 1$_{2}$ & 1$_{3}$ & 1$_{4}$ & 1$_{1}$ & 2$^{\prime }$ & 2 \\
\hline
1$_{3}$ & 1$_{3}$ & 1$_{4}$ & 1$_{1}$ & 1$_{2}$ & $2$ & 2$^{\prime }$ \\
\hline
1$_{4}$ & 1$_{4}$ & 1$_{1}$ & 1$_{2}$ & $1_{3}$ & 2$^{\prime }$ & $2$ \\
\hline
2 & 2 & 2$^{\prime }$ & 2 & 2$^{\prime }$ & $1_{1}+1_{3}+2^{\prime }$ &
$%
1_{2}+1_{4}+2$ \\ \hline
2$^{\prime }$ & 2$^{\prime }$ & 2 & 2$^{\prime }$ & $2$ &
$1_{2}+1_{4}+2$ & $%
1_{1}+1_{2}+2^{\prime }$ \\ \hline
\end{tabular}

\bigskip

\bigskip \bigskip The group $D_{7},14/2$

$
\begin{tabular}{|c||c|c|c|c|c|}
\hline
$\otimes $ & 1 & 1$^{\prime }$ & 2$_{1}$ & 2$_{2}$ & 2$_{3}$ \\
\hline\hline
1 & 1$_{1}$ & 1$^{\prime }$ & 2$_{1}$ & 2$_{2}$ & 2$_{3}$ \\ \hline
1$^{\prime }\ $ & 1$^{\prime }$ & 1 & 2$_{1}$ & 2$_{2}$ & 2$_{3}$ \\
\hline
2$_{1}$ & 2$_{1}$ & 2$_{1}$ & $1+1^{\prime }+2_{2}$ & $2_{1}+2_{3}$ & $%

2_{2}+2_{3}$ \\ \hline
2$_{2}$ & 2$_{2}$ & 2$_{2}$ & $2_{1}+2_{3}$ & $1+1^{\prime }+2_{3}$ & $%

2_{1}+2_{2}$ \\ \hline
2$_{3}$ & 2$_{3}$ & 2$_{3}$ & $2_{2}+2_{3}$ & 2$_{1}+$2$_{2}$ &
$1+1^{\prime
}+2_{1}$ \\ \hline
\end{tabular}
$

\bigskip

\newpage

The group $(Z_{4}\times Z_{2})\widetilde{\times }Z_{2},16/8$

\begin{tabular}{|c||c|c|c|c|c|c|c|c|c|c|}
\hline
$\otimes $ & 1$_{1}$ & 1$_{2}$ & 1$_{3}$ & 1$_{4}$ & 1$_{5}$ & 1$_{6}$ &
1$%
_{7}$ & 1$_{8}$ & 2 & 2$^{\prime }$ \\ \hline\hline
1$_{1}$ & 1$_{1}$ & 1$_{2}$ & 1$_{3}$ & 1$_{4}$ & 1$_{5}$ & 1$_{6}$ &
1$_{7}$
& 1$_{8}$ & 2 & 2$^{\prime }$ \\ \hline
1$_{2}$ & 1$_{2}$ & 1$_{1}$ & 1$_{4}$ & 1$_{3}$ & 1$_{6}$ & 1$_{5}$ &
1$_{8}$
& 1$_{7}$ & 2 & 2$^{\prime }$ \\ \hline
1$_{3}$ & 1$_{3}$ & 1$_{4}$ & 1$_{1}$ & 1$_{2}$ & 1$_{7}$ & 1$_{8}$ &
1$_{5}$
& 1$_{6}$ & 2 & 2$^{\prime }$ \\ \hline
1$_{4}$ & 1$_{4}$ & 1$_{3}$ & 1$_{2}$ & 1$_{1}$ & 1$_{8}$ & 1$_{7}$ &
1$_{6}$
& 1$_{5}$ & 2 & 2$^{\prime }$ \\ \hline
1$_{5}$ & 1$_{5}$ & 1$_{6}$ & 1$_{7}$ & 1$_{8}$ & 1$_{1}$ & 1$_{2}$ &
1$_{3}$
& 1$_{4}$ & 2$^{\prime }$ & 2 \\ \hline
1$_{6}$ & 1$_{6}$ & 1$_{5}$ & 1$_{8}$ & 1$_{7}$ & 1$_{2}$ & 1$_{1}$ &
1$_{4}$
& 1$_{3}$ & 2$^{\prime }$ & 2 \\ \hline
1$_{7}$ & 1$_{7}$ & 1$_{8}$ & 1$_{5}$ & 1$_{6}$ & 1$_{3}$ & 1$_{4}$ &
1$_{1}$
& 1$_{2}$ & 2$^{\prime }$ & 2 \\ \hline
1$_{8}$ & 1$_{8}$ & 1$_{7}$ & 1$_{6}$ & 1$_{5}$ & 1$_{4}$ & 1$_{3}$ &
1$_{2}$
& 1$_{1}$ & 2$^{\prime }$ & 2 \\ \hline
2 & 2 & 2 & 2 & 2 & 2$^{\prime }$ & 2$^{\prime }$ & 2$^{\prime }$ & 2$%
^{\prime }$ & $1_{5}+1_{6\ }+1_{7}+1_{8}$ & $1_{1}+1_{2}+1_{3}+1_{4}$ \\
\hline
2$^{\prime }$ & 2$^{\prime }$ & 2$^{\prime }$ & 2$^{\prime }$ &
2$^{\prime }$
& 2 & 2 & 2 & 2 & $1_{1}+1_{2}+1_{3}+1_{4}$ & $1_{5}+1_{6\
}+1_{7}+1_{8}$ \\
\hline
\end{tabular}

\bigskip

\newpage

The group $Z_{4}\ \widetilde{\times }Z_{4},16/10$

\begin{tabular}{|c||c|c|c|c|c|c|c|c|c|c|}
\hline
$\otimes $ & 1$_{1}$ & 1$_{2}$ & 1$_{3}$ & 1$_{4}$ & 1$_{5}$ & 1$_{6}$ &
1$%
_{7}$ & 1$_{8}$ & 2 & 2$^{\prime }$ \\ \hline\hline
1$_{1}$ & 1$_{1}$ & 1$_{2}$ & 1$_{3}$ & 1$_{4}$ & 1$_{5}$ & 1$_{6}$ &
1$_{7}$
& 1$_{8}$ & 2 & 2$^{\prime }$ \\ \hline
1$_{2}$ & 1$_{2}$ & 1$_{3}$ & 1$_{4}$ & 1$_{1}$ & 1$_{6}$ & 1$_{7}$ &
1$_{8}$
& 1$_{5}$ & 2$^{\prime }$ & 2 \\ \hline
1$_{3}$ & 1$_{3}$ & 1$_{4}$ & 1$_{1}$ & 1$_{2}$ & 1$_{7}$ & 1$_{8}$ &
1$_{5}$
& 1$_{6}$ & 2 & 2$^{\prime }$ \\ \hline
1$_{4}$ & 1$_{4}$ & 1$_{1}$ & 1$_{2}$ & 1$_{3}$ & 1$_{8}$ & 1$_{5}$ &
1$_{6}$
& 1$_{7}$ & 2$^{\prime }$ & 2 \\ \hline
1$_{5}$ & 1$_{5}$ & 1$_{6}$ & 1$_{7}$ & 1$_{8}$ & 1$_{1}$ & 1$_{2}$ &
1$_{3}$
& 1$_{4}$ & 2 & 2$^{\prime }$ \\ \hline
1$_{6}$ & 1$_{6}$ & 1$_{7}$ & 1$_{8}$ & 1$_{5}$ & 1$_{2}$ & 1$_{3}$ &
1$_{4}$
& 1$_{1}$ & 2$^{\prime }$ & 2 \\ \hline
1$_{7}$ & 1$_{7}$ & 1$_{8}$ & 1$_{5}$ & 1$_{6}$ & 1$_{3}$ & 1$_{4}$ &
1$_{1}$
& 1$_{2}$ & 2 & 2$^{\prime }$ \\ \hline
1$_{8}$ & 1$_{8}$ & 1$_{5}$ & 1$_{6}$ & 1$_{7}$ & 1$_{4}$ & 1$_{1}$ &
1$_{2}$
& 1$_{3}$ & 2$^{\prime }$ & 2 \\ \hline
2 & 2 & 2$^{\prime }$ & 2 & 2$^{\prime }$ & 2 & 2$^{\prime }$ & 2 & 2$%
^{\prime }$ & $1_{1}+1_{3\ }+1_{5}+1_{7}$ & $1_{2}+1_{4}+1_{6}+1_{8}$ \\
\hline
2$^{\prime }$ & 2$^{\prime }$ & 2 & 2$^{\prime }$ & 2 & 2$^{\prime }$ &
2 & 2%
$^{\prime }$ & 2 & $1_{2}+1_{4}+1_{6}+1_{8}$ & $1_{1}+1_{3\
}+1_{5}+1_{7}$
\\ \hline
\end{tabular}

\bigskip

\bigskip 
\newpage
The group $Z_{8}\ \widetilde{\times }Z_{2},16/11$

\begin{tabular}{|c||c|c|c|c|c|c|c|c|c|c|}
\hline
$\otimes $ & 1$_{1}$ & 1$_{2}$ & 1$_{3}$ & 1$_{4}$ & 1$_{5}$ & 1$_{6}$ &
1$%
_{7}$ & 1$_{8}$ & 2 & 2$^{\prime }$ \\ \hline\hline
1$_{1}$ & 1$_{1}$ & 1$_{2}$ & 1$_{3}$ & 1$_{4}$ & 1$_{5}$ & 1$_{6}$ &
1$_{7}$
& 1$_{8}$ & 2 & 2$^{\prime }$ \\ \hline
1$_{2}$ & 1$_{2}$ & 1$_{3}$ & 1$_{4}$ & 1$_{1}$ & 1$_{6}$ & 1$_{7}$ &
1$_{8}$
& 1$_{5}$ & 2$^{\prime }$ & 2 \\ \hline
1$_{3}$ & 1$_{3}$ & 1$_{4}$ & 1$_{1}$ & 1$_{2}$ & 1$_{7}$ & 1$_{8}$ &
1$_{5}$
& 1$_{6}$ & 2 & 2$^{\prime }$ \\ \hline
1$_{4}$ & 1$_{4}$ & 1$_{1}$ & 1$_{2}$ & 1$_{3}$ & 1$_{8}$ & 1$_{5}$ &
1$_{6}$
& 1$_{7}$ & 2$^{\prime }$ & 2 \\ \hline
1$_{5}$ & 1$_{5}$ & 1$_{6}$ & 1$_{7}$ & 1$_{8}$ & 1$_{1}$ & 1$_{2}$ &
1$_{3}$
& 1$_{4}$ & 2 & 2$^{\prime }$ \\ \hline
1$_{6}$ & 1$_{6}$ & 1$_{7}$ & 1$_{8}$ & 1$_{5}$ & 1$_{2}$ & 1$_{3}$ &
1$_{4}$
& 1$_{1}$ & 2$^{\prime }$ & 2 \\ \hline
1$_{7}$ & 1$_{7}$ & 1$_{8}$ & 1$_{5}$ & 1$_{6}$ & 1$_{3}$ & 1$_{4}$ &
1$_{1}$
& 1$_{2}$ & 2 & 2$^{\prime }$ \\ \hline
1$_{8}$ & 1$_{8}$ & 1$_{5}$ & 1$_{6}$ & 1$_{7}$ & 1$_{4}$ & 1$_{1}$ &
1$_{2}$
& 1$_{3}$ & 2$^{\prime }$ & 2 \\ \hline
2 & 2 & 2$^{\prime }$ & 2 & 2$^{\prime }$ & 2 & 2$^{\prime }$ & 2 & 2$%
^{\prime }$ & $1_{2}+1_{4}+1_{6}+1_{8}$ & $1_{1}+1_{3\ }+1_{5}+1_{7}$ \\
\hline
2$^{\prime }$ & 2$^{\prime }$ & 2 & 2$^{\prime }$ & 2 & 2$^{\prime }$ &
2 & 2%
$^{\prime }$ & 2 & $1_{1}+1_{3\ }+1_{5}+1_{7}$ &
$1_{2}+1_{4}+1_{6}+1_{8}$
\\ \hline
\end{tabular}

\bigskip

\newpage

The group $D_{8}$, ($Z_{8}\ \widetilde{\times }Z_{2})^{\prime },16/12$
($%
Q_{8}$ , 16/14, has the same table.)

\bigskip
\begin{tabular}{|c||c|c|c|c|c|c|c|}
\hline
$\otimes $ & 1$_{1}$ & 1$_{2}$ & 1$_{3}$ & 1$_{4}$ & 2$_{1}$ & 2$_{2}$ &
2$%
_{3}$ \\ \hline\hline
1$_{1}$ & 1$_{1}$ & 1$_{2}$ & 1$_{3}$ & 1$_{4}$ & 2$_{1}$ & 2$_{2}$ &
2$_{3}$
\\ \hline
1$_{2}$ & 1$_{2}$ & 1$_{1}$ & 1$_{4}$ & 1$_{3}$ & 2$_{3}$ & 2$_{2}$ &
2$_{1}$
\\ \hline
1$_{3}$ & 1$_{3}$ & 1$_{4}$ & 1$_{1}$ & 1$_{2}$ & 2$_{1}$ & 2$_{2}$ &
2$_{3}$
\\ \hline
1$_{4}$ & 1$_{4}$ & 1$_{3}$ & 1$_{2}$ & 1$_{1}$ & $2_{3}$ & 2$_{2}$ &
2$_{1}$
\\ \hline
2$_{1}$ & 2$_{1}$ & 2$_{3}$ & 2$_{1}$ & 2$_{3}$ & $1_{1}+1_{3\ }+2_{2}$
& $%
2_{1}+2_{3}$ & $1_{2}+1_{4}+2_{2}$ \\ \hline
2$_{2}$ & 2$_{2}$ & 2$_{2}$ & 2$_{2}$ & 2$_{2}$ & $2_{1}+2_{3}$ & $%
1_{1}+1_{2}+1_{3}+1_{4}$ & $2_{1}+2_{3}$ \\ \hline
2$_{3}$ & 2$_{3}$ & 2$_{1}$ & 2$_{3}$ & 2$_{1}$ & $1_{2}+1_{4}+2_{2}$ &
$%
2_{1}+2_{3}$ & $1_{1}+1_{3\ }+2_{2}$ \\ \hline
\end{tabular}

\bigskip

\bigskip The group ($Z_{8}\ \widetilde{\times }Z_{2})^{\prime \prime
},16/13$

\bigskip
\begin{tabular}{|c||c|c|c|c|c|c|c|}
\hline
$\otimes $ & 1$_{1}$ & 1$_{2}$ & 1$_{3}$ & 1$_{4}$ & 2$_{1}$ & 2$_{2}$ &
2$%
_{3}$ \\ \hline\hline
1$_{1}$ & 1$_{1}$ & 1$_{2}$ & 1$_{3}$ & 1$_{4}$ & 2$_{1}$ & 2$_{2}$ &
2$_{3}$
\\ \hline
1$_{2}$ & 1$_{2}$ & 1$_{1}$ & 1$_{4}$ & 1$_{3}$ & 2$_{3}$ & 2$_{2}$ &
2$_{1}$
\\ \hline
1$_{3}$ & 1$_{3}$ & 1$_{4}$ & 1$_{1}$ & 1$_{2}$ & 2$_{3}$ & 2$_{2}$ &
2$_{1}$
\\ \hline
1$_{4}$ & 1$_{4}$ & 1$_{3}$ & 1$_{2}$ & 1$_{1}$ & $2_{1}$ & 2$_{2}$ &
2$_{3}$
\\ \hline
2$_{1}$ & 2$_{1}$ & 2$_{3}$ & 2$_{3}$ & 2$_{1}$ & $1_{2}+1_{3\ }+2_{2}$
& $%
2_{1}+2_{3}$ & $1_{1}+1_{4}+2_{2}$ \\ \hline
2$_{2}$ & 2$_{2}$ & 2$_{2}$ & 2$_{2}$ & 2$_{2}$ & $2_{1}+2_{3}$ & $%
1_{1}+1_{2}+1_{3}+1_{4}$ & $2_{1\ }+2_{3}$ \\ \hline
2$_{3}$ & 2$_{3}$ & 2$_{1}$ & 2$_{1}$ & 2$_{3}$ & $1_{1}+1_{4}+2_{2}$ &
$%
2_{1}+2_{3}$ & $1_{2}+1_{3\ }+2_{2}$ \\ \hline
\end{tabular}

\bigskip
\newpage
\bigskip The group $D_{9},18/4$

\begin{tabular}{|c||c|c|c|c|c|c|}
\hline
$\otimes $ & 1 & $1^{\prime }$ & 2$_{1}$ & 2$_{2}$ & 2$_{3}$ & 2$_{4}$
\\
\hline\hline
1 & 1 & $1^{\prime }$ & 2$_{1}$ & 2$_{2}$ & 2$_{3}$ & 2$_{4}$ \\ \hline
$1^{\prime }$ & $1^{\prime }$ & 1 & 2$_{1}$ & 2$_{2}$ & 2$_{3}$ &
2$_{4}$ \\
\hline
2$_{1}$ & 2$_{1}$ & 2$_{1}$ & $1+1^{\prime }+2_{2}$ & $2_{1}+2_{3}$ & $%

2_{2}+2_{4}$ & $2_{3}+2_{4}$ \\ \hline
2$_{2}$ & 2$_{2}$ & 2$_{2}$ & $2_{1}+2_{3}$ & $1+1^{\prime }+2_{4}$ & $%

2_{1}+2_{4}$ & $2_{2}+2_{3}$ \\ \hline
2$_{3}$ & 2$_{3}$ & 2$_{3}$ & $2_{2}+2_{4}$ & $2_{1}+2_{4}$ &
$1+1^{\prime
}+2_{3}$ & $2_{1}+2_{2}$ \\ \hline
2$_{4}$ & 2$_{4}$ & 2$_{4}$ & $2_{3}+2_{4}$ & $2_{2}+2_{3}$ &
$2_{1}+2_{2}$
& $1+1^{\prime }+2_{1}$ \\ \hline
\end{tabular}
\bigskip

\bigskip

\bigskip The group $(Z_{3}\times Z_{3})\widetilde{\times }Z_{2},18/5$

\begin{tabular}{|c||c|c|c|c|c|c|}
\hline
$\otimes $ & 1 & $1^{\prime }$ & 2$_{1}$ & 2$_{2}$ & 2$_{3}$ & 2$_{4}$
\\
\hline\hline
1 & 1 & $1^{\prime }$ & 2$_{1}$ & 2$_{2}$ & 2$_{3}$ & 2$_{4}$ \\ \hline
$1^{\prime }$ & $1^{\prime }$ & 1 & 2$_{1}$ & 2$_{2}$ & 2$_{3}$ &
2$_{4}$ \\
\hline
2$_{1}$ & 2$_{1}$ & 2$_{1}$ & $1+1^{\prime }+2_{1}$ & $2_{3}+2_{4}$ & $%

2_{2}+2_{4}$ & $2_{2}+2_{3}$ \\ \hline
2$_{2}$ & 2$_{2}$ & 2$_{2}$ & $2_{3}+2_{4}$ & $1+1^{\prime }+2_{2}$ & $%

2_{1}+2_{4}$ & $2_{1}+2_{3}$ \\ \hline
2$_{3}$ & 2$_{3}$ & 2$_{3}$ & $2_{2}+2_{4}$ & $2_{1}+2_{4}$ &
$1+1^{\prime
}+2_{3}$ & $2_{1}+2_{2}$ \\ \hline
2$_{4}$ & 2$_{4}$ & 2$_{4}$ & $2_{2}+2_{3}$ & $2_{1}+2_{3}$ &
$2_{1}+2_{2}$
& $1+1^{\prime }+2_{4}$ \\ \hline
\end{tabular}

\bigskip

\newpage
The group $D_{10}$, 20$/3$

\bigskip
\begin{tabular}{|c||c|c|c|c|c|c|c|c|}
\hline
$\otimes $ & 1$_{1}$ & 1$_{2}$ & 1$_{3}$ & 1$_{4}$ & 2$_{1}$ & 2$_{2}$ &
2$%
_{3}$ & 2$_{4}$ \\ \hline\hline
1$_{1}$ & 1$_{1}$ & 1$_{2}$ & 1$_{3}$ & 1$_{4}$ & 2$_{1}$ & 2$_{2}$ &
2$_{3}$
& 2$_{4}$ \\ \hline
1$_{2}$ & 1$_{2}$ & 1$_{1}$ & 1$_{4}$ & 1$_{3}$ & 2$_{4}$ & 2$_{3}$ &
2$_{1}$
& 2$_{1}$ \\ \hline
1$_{3}$ & 1$_{3}$ & 1$_{4}$ & 1$_{1}$ & 1$_{2}$ & 2$_{1}$ & 2$_{2}$ &
2$_{3}$
& 2$_{4}$ \\ \hline
1$_{4}$ & 1$_{4}$ & 1$_{3}$ & 1$_{2}$ & 1$_{1}$ & $2_{4}$ & 2$_{3}$ &
2$_{1}$
& 2$_{1}$ \\ \hline
2$_{1}$ & 2$_{1}$ & 2$_{4}$ & 2$_{1}$ & 2$_{4}$ & $1_{1}+1_{3\ }+2_{2}$
& $%
2_{1}+2_{3}$ & $2_{2}+2_{4}$ & $1_{2}+1_{4}+2_{3}$ \\ \hline
2$_{2}$ & 2$_{2}$ & 2$_{3}$ & 2$_{2}$ & 2$_{3}$ & $2_{1}+2_{3}$ & $%
1_{1}+1_{3\ }+2_{4}$ & $2_{1}+2_{3}$ & $2_{2}+2_{4}$ \\ \hline
2$_{3}$ & 2$_{3}$ & 2$_{2}$ & 2$_{3}$ & 2$_{2}$ & $2_{2}+2_{4}$ & $%
2_{1}+2_{3}$ & $1_{1}+1_{3\ }+2_{4}$ & $2_{1}+2_{3}$ \\ \hline
2$_{4}$ & 2$_{4}$ & 2$_{1}$ & 2$_{4}$ & 2$_{1}$ & $1_{2}+1_{4}+2_{3}$ &
$%
2_{2}+2_{4}$ & $2_{1}+2_{3}$ & $1_{1}+1_{3\ }+2_{2}$ \\ \hline
\end{tabular}

\bigskip

The group $Z_{5}\ \widetilde{\times }Z_{4}$, 20/5

\begin{tabular}{|c||c|c|c|c|c|}
\hline
$\otimes $ & 1$_{1}$ & 1$_{2}$ & 1$_{3}$ & 1$_{4}$ & 4 \\ \hline\hline
1$_{1}$ & 1$_{1}$ & 1$_{2}$ & 1$_{3}$ & 1$_{4}$ & 4 \\ \hline
1$_{2}\ $ & 1$_{2}$ & 1$_{3}$ & 1$_{4}$ & 1$_{1}$ & 4 \\ \hline
1$_{3}$ & 1$_{3}$ & 1$_{4}$ & 1$_{1}$ & 1$_{2}$ & 4 \\ \hline
1$_{4}$ & 1$_{4}$ & 1$_{1}$ & 1$_{2}$ & 1$_{3}$ & 4 \\ \hline
4 & 4 & 4 & 4 & 4 & $1_{1}+1_{2}+1_{3}+1_{4}+3\times 4$ \\ \hline
\end{tabular}

\bigskip

\newpage

The group $Z_{7}\ \widetilde{\times }Z_{3}$, 21/2

\begin{tabular}{|c||c|c|c|c|c|}
\hline
$\otimes $ & 1$_{1}$ & 1$_{2}$ & 1$_{3}$ & 3$_{1}$ & 3$_{2}$ \\
\hline\hline
1$_{1}$ & 1$_{1}$ & 1$_{2}$ & 1$_{3}$ & 3$_{1}$ & 3$_{2}$ \\ \hline
1$_{2}\ $ & 1$_{2}$ & 1$_{3}$ & 1$_{1}$ & 3$_{1}$ & 3$_{2}$ \\ \hline
1$_{3}$ & 1$_{3}$ & 1$_{1}$ & 1$_{2}$ & 3$_{1}$ & 3$_{2}$ \\ \hline
3$_{1}$ & 3$_{1}$ & 3$_{1}$ & 3$_{1}$ & $3_{1}+3_{2}+3_{2}$ & $%
1_{1}+1_{2}+1_{3}+3_{1}+3_{2}$ \\ \hline
3$_{2}$ & 3$_{2}$ & 3$_{2}$ & 3$_{2}$ & $1_{1}+1_{2}+1_{3}+3_{1}+3_{2}$
& $%
3_{1}+3_{1}+3_{2}$ \\ \hline
\end{tabular}

\bigskip

\bigskip The group $D_{11},22/2$

\begin{tabular}{|c||c|c|c|c|c|c|c|}
\hline
$\otimes $ & 1 & $1^{\prime }$ & 2$_{1}$ & 2$_{2}$ & 2$_{3}$ & 2$_{4}$ &
2$%
_{5}$ \\ \hline\hline
1 & 1 & $1^{\prime }$ & 2$_{1}$ & 2$_{2}$ & 2$_{3}$ & 2$_{4}$ & 2$_{5}$
\\
\hline
$1^{\prime }$ & $1^{\prime }$ & 1 & 2$_{1}$ & 2$_{2}$ & 2$_{3}$ &
2$_{4}$ & 2%
$_{5}$ \\ \hline
2$_{1}$ & 2$_{1}$ & 2$_{1}$ & $1+1^{\prime }+2_{2}$ & $2_{1}+2_{3}$ & $%

2_{1}+2_{4}$ & $2_{3}+2_{5}$ & $2_{4}+2_{5}$ \\ \hline
2$_{2}$ & 2$_{2}$ & 2$_{2}$ & $2_{1}+2_{3}$ & $1+1^{\prime }+2_{4}$ & $
2_{1}+2_{5}$ & $2_{2}+2_{5}$ & $2_{3}+2_{4}$ \\ \hline
2$_{3}$ & 2$_{3}$ & 2$_{3}$ & $2_{1}+2_{4}$ & $2_{1}+2_{5}$ &
$1+1^{\prime
}+2_{5}$ & $2_{1}+2_{4}$ & $2_{2}+2_{3}\ $ \\ \hline
2$_{4}$ & 2$_{4}$ & 2$_{4}$ & $2_{3}+2_{5}$ & $2_{2}+2_{5}$ &
$2_{1}+2_{4}$
& $1+1^{\prime }+2_{3}$ & $2_{1}+2_{2}$ \\ \hline
2$_{5}$ & 2$_{5}$ & 2$_{5}$ & $2_{4}+2_{5}$ & $2_{3}+2_{4}$ &
$2_{2}+2_{3}$
& $2_{1}+2_{2}$ & $1+1^{\prime }+2_{1}$ \\ \hline
\end{tabular}

\bigskip

\newpage

\bigskip The group $D_{12}$, 24$/10$

\bigskip
\begin{tabular}{|c||c|c|c|c|c|c|c|c|c|}
\hline
$\otimes $ & 1$_{1}$ & 1$_{2}$ & 1$_{3}$ & 1$_{4}$ & 2$_{1}$ & 2$_{2}$ &
2$%
_{3}$ & 2$_{4}$ & 2$_{5}$ \\ \hline\hline
1$_{1}$ & 1$_{1}$ & 1$_{2}$ & 1$_{3}$ & 1$_{4}$ & 2$_{1}$ & 2$_{2}$ &
2$_{3}$
& 2$_{4}$ & 2$_{5}$ \\ \hline
1$_{2}$ & 1$_{2}$ & 1$_{1}$ & 1$_{4}$ & 1$_{3}$ & 2$_{1}$ & 2$_{2}$ &
2$_{3}$
& 2$_{4}$ & 2$_{5}$ \\ \hline
1$_{3}$ & 1$_{3}$ & 1$_{4}$ & 1$_{1}$ & 1$_{2}$ & 2$_{5}$ & 2$_{4}$ &
2$_{3}$
& 2$_{2}$ & 2$_{1}$ \\ \hline
1$_{4}$ & 1$_{4}$ & 1$_{3}$ & 1$_{2}$ & 1$_{1}$ & $2_{5}$ & 2$_{4}$ &
2$_{3}$
& 2$_{2}$ & 2$_{1}$ \\ \hline
2$_{1}$ & 2$_{1}$ & 2$_{1}$ & 2$_{5}$ & 2$_{5}$ & $1_{1}+1_{2\ }+2_{2}$
& $%
2_{1}+2_{3}$ & $2_{2}+2_{4}$ & $2_{3}+2_{5}$ & $2_{3}+2_{5}$ \\ \hline
2$_{2}$ & 2$_{2}$ & 2$_{2}$ & 2$_{4}$ & 2$_{4}$ & $2_{1}+2_{3}$ & $%
1_{1}+1_{2\ }+2_{4}$ & $2_{1}+2_{5}$ & $1_{3\ }+1_{4}+2_{4}$ &
$2_{3}+2_{5}$
\\ \hline
2$_{3}$ & 2$_{3}$ & 2$_{3}$ & 2$_{3}$ & 2$_{3}$ & $2_{2}+2_{4}$ & $%
2_{1}+2_{5}$ & $1_{1}+1_{2\ }+1_{3\ }+1_{4}$ & $2_{1}+2_{5}$ &
$2_{2}+2_{4}$
\\ \hline
2$_{4}$ & 2$_{4}$ & 2$_{4}$ & 2$_{2}$ & 2$_{2}$ & $2_{3}+2_{5}$ & $1_{3\

}+1_{4}+2_{4}$ & $2_{1}+2_{5}$ & $1_{1}+1_{2\ }+2_{4}$ & $2_{1}+2_{3}$
\\
\hline
2$_{5}$ & 2$_{5}$ & 2$_{5}$ & 2$_{1}$ & 2$_{1}$ & $2_{3}+2_{5}$ & $%
2_{3}+2_{5}$ & $2_{2}+2_{4}$ & $2_{1}+2_{3}$ & $1_{1}+1_{2\ }+2_{2}$ \\
\hline
\end{tabular}

\bigskip

The group S$_{4}$, 24/12

\begin{tabular}{|c||c|c|c|c|c|}
\hline
$\otimes $ & 1 & 1$^{\prime }$ & 2 & 3 & 3$^{\prime }$ \\ \hline\hline
1 & 1 & 1$^{\prime }$ & 2 & 3 & 3$^{\prime }$ \\ \hline
1$^{\prime }\ $ & 1$^{\prime }$ & 1 & 2 & 3$^{\prime }$ & 3 \\ \hline
2 & 2 & 2 & $1+1^{\prime }+2$ & $3+3^{\prime }$ & $3+3^{\prime }$ \\
\hline
3 & 3 & 3$^{\prime }$ & $3+3^{\prime }$ & $1+2+3+3^{\prime }$ &
$1^{\prime
}+2+3+3^{\prime }$ \\ \hline
3$^{\prime }$ & 3$^{\prime }$ & 3$_{2}$ & $3+3^{\prime }$ & $1^{\prime
}+2+3+3^{\prime }$ & $1+2+3+3^{\prime }$ \\ \hline
\end{tabular}

\bigskip

\newpage

The group $SL_{2}(F_{3})$, $Q\ \widetilde{\times }Z_{3},24/13$

\bigskip
\begin{tabular}{|c||c|c|c|c|c|c|c|}
\hline
$\otimes $ & 1$_{1}$ & 1$_{2}$ & 1$_{3}$ & 2$_{1}$ & 2$_{2}$ & 2$_{3}$ &
3
\\ \hline\hline
1$_{1}$ & 1$_{1}$ & 1$_{2}$ & 1$_{3}$ & 2$_{1}$ & 2$_{2}$ & 2$_{3}$ & 3
\\
\hline
1$_{2}$ & 1$_{2}$ & 1$_{3}$ & 1$_{1}$ & 2$_{2}$ & 2$_{3}$ & 2$_{1}$ & 3
\\
\hline
1$_{3}$ & 1$_{3}$ & 1$_{1}$ & 1$_{2}$ & 2$_{3}$ & 2$_{1}$ & 2$_{2}$ & 3
\\
\hline
2$_{1}$ & 2$_{1}$ & 2$_{2}$ & 2$_{3}$ & $1+3$ & $1^{\prime }+3$ &
$1^{\prime
\prime }+3$ & $2_{1}+2_{2}+2_{3}$ \\ \hline
2$_{2}$ & 2$_{2}$ & 2$_{3}$ & 2$_{1}$ & $1^{\prime }+3$ & $1^{\prime
\prime
}+3$ & $1+3$ & $2_{1}+2_{2}+2_{3}$ \\ \hline
2$_{3}$ & 2$_{3}$ & 2$_{1}$ & 2$_{2}$ & $1^{\prime \prime }+3$ & $1+3$ &
$%
1^{\prime }+3$ & $2_{1}+2_{2}+2_{3}$ \\ \hline
3 & 3 & 3 & 3 & $2_{1}+2_{2}+2_{3}$ & $2_{1}+2_{2}+2_{3}$ & $%
2_{1}+2_{2}+2_{3}$ & 1$_{1}+1_{2}+1_{3\ }+3+3$ \\ \hline
\end{tabular}

\bigskip
\newpage
\bigskip The group $Z_{8}\ \widetilde{\times }Z_{3}$, 24/14

\bigskip
\begin{tabular}{|c||c|c|c|c|c|c|c|c|c|c|c|c|}
\hline
$\otimes $ & 1$_{1}$ & 1$_{2}$ & 1$_{3}$ & 1$_{4}$ & 1$_{5}$ & 1$_{6}$ &
1$%
_{7}$ & 1$_{8}$ & 2$_{1}$ & 2$_{2}$ & 2$_{3}$ & 2$_{4}$ \\ \hline\hline
1$_{1}$ & 1$_{1}$ & 1$_{2}$ & 1$_{3}$ & 1$_{4}$ & 1$_{5}$ & 1$_{6}$ &
1$_{7}$
& 1$_{8}$ & 2$_{1}$ & 2$_{2}$ & 2$_{3}$ & 2$_{4}$ \\ \hline
1$_{2}$ & 1$_{2}$ & 1$_{3}$ & 1$_{4}$ & 1$_{5}$ & 1$_{6}$ & 1$_{7}$ &
1$_{8}$
& 1$_{1}$ & 2$_{2}$ & 2$_{3}$ & 2$_{4}$ & 2$_{1}$ \\ \hline
1$_{3}$ & 1$_{3}$ & 1$_{4}$ & 1$_{5}$ & 1$_{6}$ & 1$_{7}$ & 1$_{8}$ &
1$_{1}$
& 1$_{2}$ & 2$_{3}$ & 2$_{4}$ & 2$_{1}$ & 2$_{2}$ \\ \hline
1$_{4}$ & 1$_{4}$ & 1$_{5}$ & 1$_{6}$ & 1$_{7}$ & 1$_{8}$ & 1$_{1}$ &
1$_{2}$
& 1$_{3}$ & 2$_{4}$ & 2$_{1}$ & 2$_{2}$ & 2$_{3}$ \\ \hline
1$_{5}$ & 1$_{5}$ & 1$_{6}$ & 1$_{7}$ & 1$_{8}$ & 1$_{1}$ & 1$_{2}$ &
1$_{3}$
& 1$_{4}$ & 2$_{1}$ & 2$_{2}$ & 2$_{3}$ & 2$_{4}$ \\ \hline
1$_{6}$ & 1$_{6}$ & 1$_{7}$ & 1$_{8}$ & 1$_{1}$ & 1$_{2}$ & 1$_{3}$ &
1$_{4}$
& 1$_{5}$ & 2$_{2}$ & 2$_{3}$ & 2$_{4}$ & 2$_{1}$ \\ \hline
1$_{7}$ & 1$_{7}$ & 1$_{8}$ & 1$_{1}$ & 1$_{2}$ & 1$_{3}$ & 1$_{4}$ &
1$_{5}$
& 1$_{6}$ & 2$_{3}$ & 2$_{4}$ & 2$_{1}$ & 2$_{2}$ \\
\hline
\end{tabular}

\bigskip

\newpage

\bigskip The group $D_{4}\ \widetilde{\times }Z_{3}$, $24/15$

\begin{tabular}{|c||c|c|c|c|c|c|c|c|c|}
\hline
$\otimes $ & 1$_{1}$ & 1$_{2}$ & 1$_{3}$ & 1$_{4}$ & 2$_{1}$ & 2$_{2}$ &
2$%
_{3}$ & 2$_{4}$ & 2$_{5}$ \\ \hline\hline
1$_{1}$ & 1$_{1}$ & 1$_{2}$ & 1$_{3}$ & 1$_{4}$ & 2$_{1}$ & 2$_{2}$ &
2$_{3}$
& 2$_{4}$ & 2$_{5}$ \\ \hline
1$_{2}$ & 1$_{2}$ & 1$_{1}$ & 1$_{4}$ & 1$_{3}$ & 2$_{2}$ & 2$_{1}$ &
2$_{4}$
& 2$_{3}$ & 2$_{5}$ \\ \hline
1$_{3}$ & 1$_{3}$ & 1$_{4}$ & 1$_{1}$ & 1$_{2}$ & 2$_{1}$ & 2$_{2}$ &
2$_{3}$
& 2$_{4}$ & 2$_{5}$ \\ \hline
1$_{4}$ & 1$_{4}$ & 1$_{3}$ & 1$_{2}$ & 1$_{1}$ & 2$_{2}$ & 2$_{1}$ &
2$_{4}$
& 2$_{3}$ & 2$_{5}$ \\ \hline
2$_{1}$ & 2$_{1}$ & 2$_{2}$ & 2$_{1}$ & 2$_{2}$ & $1_{1}+1_{3}+2_{1}$ &
$%
1_{2}+1_{4}+2_{2}$ & $2_{4}+2_{5}$ & $2_{3}+2_{5}$ & $2_{3}+2_{4}$ \\
\hline
2$_{2}$ & 2$_{2}$ & 2$_{1}$ & 2$_{2}$ & 2$_{1}$ & $1_{2}+1_{4}+2_{2}$ &
$%
1_{1}+1_{3}+2_{1}$ & $2_{3}+2_{5}$ & $2_{4}+2_{5}$ & $2_{3}+2_{4}$ \\
\hline
2$_{3}$ & 2$_{3}$ & 2$_{4}$ & 2$_{3}$ & 2$_{4}$ & $2_{4}+2_{5}$ & $%
2_{3}+2_{5}$ & $1_{2}+1_{4}+2_{1}$ & $1_{1}+1_{3}+2_{2}$ & $2_{1}+2_{2}$
\\
\hline
2$_{4}$ & 2$_{4}$ & 2$_{3}$ & 2$_{4}$ & 2$_{3}$ & $2_{3}+2_{5}$ & $%
2_{4}+2_{5}$ & $1_{1}+1_{3}+2_{2}$ & $1_{2}+1_{4}+2_{1}$ & $2_{1}+2_{2}$
\\
\hline
2$_{5}$ & 2$_{5}$ & 2$_{5}$ & 2$_{5}$ & 2$_{5}$ & $2_{3}+2_{4}$ & $%
2_{3}+2_{4}$ & $2_{1}+2_{2}$ & $2_{1}+2_{2}$ & $1_{1}+1_{2}+1_{3}+1_{4}$
\\
\hline
\end{tabular}

\bigskip

\bigskip \bigskip The group $D_{13},26/2$

\begin{tabular}{|c||c|c|c|c|c|c|c|c|}
\hline
$\otimes $ & 1 & $1^{\prime }$ & 2$_{1}$ & 2$_{2}$ & 2$_{3}$ & 2$_{4}$ &
2$%
_{5}$ & 2$_{6}$ \\ \hline\hline
1 & 1 & $1^{\prime }$ & 2$_{1}$ & 2$_{2}$ & 2$_{3}$ & 2$_{4}$ & 2$_{5}$
& 2$%
_{6}$ \\ \hline
$1^{\prime }$ & $1^{\prime }$ & 1 & 2$_{1}$ & 2$_{2}$ & 2$_{3}$ &
2$_{4}$ & 2%
$_{5}$ & 2$_{6}$ \\ \hline
2$_{1}$ & 2$_{1}$ & 2$_{1}$ & $1+1^{\prime }+2_{2}$ & $2_{1}+2_{3}$ & $%

2_{1}+2_{4}$ & $2_{3}+2_{5}$ & $2_{4}+2_{6}$ & $2_{5}+2_{6}$ \\ \hline
2$_{2}$ & 2$_{2}$ & 2$_{2}$ & $2_{1}+2_{3}$ & $1+1^{\prime }+2_{4}$ & $%

2_{1}+2_{5}$ & $2_{2}+2_{6}$ & $2_{3}+2_{6}$ & $2_{4}+2_{5}$ \\ \hline
2$_{3}$ & 2$_{3}$ & 2$_{3}$ & $2_{1}+2_{4}$ & $2_{1}+2_{5}$ &
$1+1^{\prime
}+2_{6}$ & $2_{1}+2_{6}$ & $2_{2}+2_{5}\ $ & $2_{3}+2_{4}$ \\ \hline
2$_{4}$ & 2$_{4}$ & 2$_{4}$ & $2_{3}+2_{5}$ & $2_{2}+2_{6}$ &
$2_{1}+2_{6}$
& $1+1^{\prime }+2_{5}$ & $2_{1}+2_{4}$ & $2_{2}+2_{3}$ \\ \hline
2$_{5}$ & 2$_{5}$ & 2$_{5}$ & $2_{4}+2_{6}$ & $2_{3}+2_{6}$ &
$2_{2}+2_{5}$
& $2_{1}+2_{4}$ & $1+1^{\prime }+2_{3}$ & $2_{1}+2_{2}$ \\ \hline
2$_{6}$ & 2$_{6}$ & 2$_{6}$ & $2_{5}+2_{6}$ & $2_{4}+2_{5}$ &
$2_{3}+2_{4}$
& $2_{2}+2_{3}$ & $2_{1}+2_{2}$ & $1+1^{\prime }+2_{1}$ \\ \hline
\end{tabular}

\newpage

The group ($Z_{3}\times Z_{3})\widetilde{\times }Z_{3},27/4$

\begin{tabular}{|c||c|c|c|c|c|c|c|c|c|c|c|}
\hline
$\otimes $ & 1$_{1}$ & 1$_{2}$ & 1$_{3}$ & 1$_{4}$ & 1$_{5}$ & 1$_{6}$ &
1$%
_{7}$ & 1$_{8}$ & 1$_{9}$ & 3$_{1}$ & 3$_{2}$ \\ \hline\hline
1$_{1}$ & 1$_{1}$ & 1$_{2}$ & 1$_{3}$ & 1$_{4}$ & 1$_{5}$ & 1$_{6}$ &
1$_{7}$
& 1$_{8}$ & 1$_{9}$ & 3$_{1}$ & 3$_{2}$ \\ \hline
1$_{2}$ & 1$_{2}$ & 1$_{3}$ & 1$_{1}$ & 1$_{5}$ & 1$_{6}$ & 1$_{4}$ &
1$_{8}$
& 1$_{9}$ & 1$_{7}$ & 3$_{1}$ & 3$_{2}$ \\ \hline
1$_{3}$ & 1$_{3}$ & 1$_{1}$ & 1$_{2}$ & 1$_{6}$ & 1$_{4}$ & 1$_{5}$ &
1$_{9}$
& 1$_{7}$ & 1$_{8}$ & 3$_{1}$ & 3$_{2}$ \\ \hline
1$_{4}$ & 1$_{4}$ & 1$_{5}$ & 1$_{6}$ & 1$_{7}$ & 1$_{8}$ & 1$_{9}$ &
1$_{1}$
& 1$_{2}$ & 1$_{3}$ & 3$_{1}$ & 3$_{2}$ \\ \hline
1$_{5}$ & 1$_{5}$ & 1$_{6}$ & 1$_{4}$ & 1$_{8}$ & 1$_{9}$ & 1$_{7}$ &
1$_{2}$
& 1$_{3}$ & 1$_{1}$ & 3$_{1}$ & 3$_{2}$ \\ \hline
1$_{6}$ & 1$_{6}$ & 1$_{4}$ & 1$_{5}$ & 1$_{9}$ & 1$_{7}$ & 1$_{8}$ &
1$_{3}$
& 1$_{1}$ & 1$_{2}$ & 3$_{1}$ & 3$_{2}$ \\ \hline
1$_{7}$ & 1$_{7}$ & 1$_{8}$ & 1$_{9}$ & 1$_{1}$ & 1$_{2}$ & 1$_{3}$ &
1$_{4}$
& 1$_{5}$ & 1$_{6}$ & 3$_{1}$ & 3$_{2}$ \\ \hline
1$_{8}$ & 1$_{8}$ & 1$_{9}$ & 1$_{7}$ & 1$_{2}$ & 1$_{3}$ & 1$_{1}$ &
1$_{5}$
& 1$_{6}$ & 1$_{4}$ & 3$_{1}$ & 3$_{2}$ \\ \hline
1$_{9}$ & 1$_{9}$ & 1$_{7}$ & 1$_{8}$ & 1$_{3}$ & 1$_{1}$ & 1$_{2}$ &
1$_{6}$
& 1$_{4}$ & 1$_{5}$ & 3$_{1}$ & 3$_{2}$ \\ \hline
3$_{1}$ & 3$_{1}$ & 3$_{1}$ & 3$_{1}$ & 3$_{1}$ & 3$_{1}$ & 3$_{1}$ &
3$_{1}$
& 3$_{1}$ & 3$_{1}$ & 3$\times $3$_{2}$ & $\sum_{i=1}^{9}$1$_{i}$ \\
\hline
3$_{2}$ & 3$_{2}$ & 3$_{2}$ & 3$_{2}$ & 3$_{2}$ & 3$_{2}$ & 3$_{2}$ &
3$_{2}$
& 3$_{2}$ & 3$_{2}$ & $\sum_{i=1}^{9}$1$_{i}$ & 3$\times $3$_{1}$ \\
\hline
\end{tabular}

\bigskip
\newpage
The group $Z_{9}\widetilde{\times }Z_{3},27/5$ [Note this table is the
same
as for ($Z_{3}\times Z_{3})\widetilde{\times }Z_{3}.$]

\begin{tabular}{|c||c|c|c|c|c|c|c|c|c|c|c|}
\hline
$\otimes $ & 1$_{1}$ & 1$_{2}$ & 1$_{3}$ & 1$_{4}$ & 1$_{5}$ & 1$_{6}$ &
1$%
_{7}$ & 1$_{8}$ & 1$_{9}$ & 3$_{1}$ & 3$_{2}$ \\ \hline\hline
1$_{1}$ & 1$_{1}$ & 1$_{2}$ & 1$_{3}$ & 1$_{4}$ & 1$_{5}$ & 1$_{6}$ &
1$_{7}$
& 1$_{8}$ & 1$_{9}$ & 3$_{1}$ & 3$_{2}$ \\ \hline
1$_{2}$ & 1$_{2}$ & 1$_{3}$ & 1$_{1}$ & 1$_{5}$ & 1$_{6}$ & 1$_{4}$ &
1$_{8}$
& 1$_{9}$ & 1$_{7}$ & 3$_{1}$ & 3$_{2}$ \\ \hline
1$_{3}$ & 1$_{3}$ & 1$_{1}$ & 1$_{2}$ & 1$_{6}$ & 1$_{4}$ & 1$_{5}$ &
1$_{9}$
& 1$_{7}$ & 1$_{8}$ & 3$_{1}$ & 3$_{2}$ \\ \hline
1$_{4}$ & 1$_{4}$ & 1$_{5}$ & 1$_{6}$ & 1$_{7}$ & 1$_{8}$ & 1$_{9}$ &
1$_{1}$
& 1$_{2}$ & 1$_{3}$ & 3$_{1}$ & 3$_{2}$ \\ \hline
1$_{5}$ & 1$_{5}$ & 1$_{6}$ & 1$_{4}$ & 1$_{8}$ & 1$_{9}$ & 1$_{7}$ &
1$_{2}$
& 1$_{3}$ & 1$_{1}$ & 3$_{1}$ & 3$_{2}$ \\ \hline
1$_{6}$ & 1$_{6}$ & 1$_{4}$ & 1$_{5}$ & 1$_{9}$ & 1$_{7}$ & 1$_{8}$ &
1$_{3}$
& 1$_{1}$ & 1$_{2}$ & 3$_{1}$ & 3$_{2}$ \\ \hline
1$_{7}$ & 1$_{7}$ & 1$_{8}$ & 1$_{9}$ & 1$_{1}$ & 1$_{2}$ & 1$_{3}$ &
1$_{4}$
& 1$_{5}$ & 1$_{6}$ & 3$_{1}$ & 3$_{2}$ \\ \hline
1$_{8}$ & 1$_{8}$ & 1$_{9}$ & 1$_{7}$ & 1$_{2}$ & 1$_{3}$ & 1$_{1}$ &
1$_{5}$
& 1$_{6}$ & 1$_{4}$ & 3$_{1}$ & 3$_{2}$ \\ \hline
1$_{9}$ & 1$_{9}$ & 1$_{7}$ & 1$_{8}$ & 1$_{3}$ & 1$_{1}$ & 1$_{2}$ &
1$_{6}$
& 1$_{4}$ & 1$_{5}$ & 3$_{1}$ & 3$_{2}$ \\ \hline
3$_{1}$ & 3$_{1}$ & 3$_{1}$ & 3$_{1}$ & 3$_{1}$ & 3$_{1}$ & 3$_{1}$ &
3$_{1}$
& 3$_{1}$ & 3$_{1}$ & 3$\times $3$_{2}$ & $\sum_{i=1}^{9}$1$_{i}$ \\
\hline
3$_{2}$ & 3$_{2}$ & 3$_{2}$ & 3$_{2}$ & 3$_{2}$ & 3$_{2}$ & 3$_{2}$ &
3$_{2}$
& 3$_{2}$ & 3$_{2}$ & $\sum_{i=1}^{9}$1$_{i}$ & $3\times 3_{1}$ \\
\hline
\end{tabular}

\bigskip

\newpage

The group $D_{14}$, 28$/3$

\bigskip
\begin{tabular}{|c||c|c|c|c|c|c|c|c|c|c|}
\hline
$\otimes $ & 1$_{1}$ & 1$_{2}$ & 1$_{3}$ & 1$_{4}$ & 2$_{1}$ & 2$_{2}$ &
2$%
_{3}$ & 2$_{4}$ & 2$_{5}$ & 2$_{6}$ \\ \hline\hline
1$_{1}$ & 1$_{1}$ & 1$_{2}$ & 1$_{3}$ & 1$_{4}$ & 2$_{1}$ & 2$_{2}$ &
2$_{3}$
& 2$_{4}$ & 2$_{5}$ & 2$_{6}$ \\ \hline
1$_{2}$ & 1$_{2}$ & 1$_{1}$ & 1$_{4}$ & 1$_{3}$ & 2$_{1}$ & 2$_{2}$ &
2$_{3}$
& 2$_{4}$ & 2$_{5}$ & 2$_{6}$ \\ \hline
1$_{3}$ & 1$_{3}$ & 1$_{4}$ & 1$_{1}$ & 1$_{2}$ & 2$_{6}$ & 2$_{5}$ &
2$_{4}$
& 2$_{3}$ & 2$_{2}$ & 2$_{1}$ \\ \hline
1$_{4}$ & 1$_{4}$ & 1$_{3}$ & 1$_{2}$ & 1$_{1}$ & 2$_{6}$ & 2$_{5}$ &
2$_{4}$
& 2$_{3}$ & 2$_{2}$ & 2$_{1}$ \\ \hline
2$_{1}$ & 2$_{1}$ & 2$_{1}$ & 2$_{6}$ & 2$_{6}$ & $1_{1}+1_{2\ }+2_{2}$
& $%
2_{1}+2_{3}$ & $2_{2}+2_{4}$ & $2_{3}+2_{5}$ & $2_{4}+2_{6}$ &
$1_{3}+1_{4\
}+2_{5}$ \\ \hline
2$_{2}$ & 2$_{2}$ & 2$_{2}$ & 2$_{5}$ & 2$_{5}$ & $2_{1}+2_{3}$ & $%
1_{1}+1_{2\ }+2_{4}$ & $2_{1}+2_{5}$ & $2_{2}+2_{6}$ & $1_{3}+1_{4\
}+2_{3}$
& $2_{4}+2_{6}$ \\ \hline
2$_{3}$ & 2$_{3}$ & 2$_{3}$ & 2$_{4}$ & 2$_{4}$ & $2_{2}+2_{4}$ & $%
2_{1}+2_{5}$ & $1_{1}+1_{2\ }+2_{6}$ & $1_{3}+1_{4\ }+2_{1}$ &
$2_{2}+2_{6}$
& $2_{3}+2_{5}$ \\ \hline
2$_{4}$ & 2$_{4}$ & 2$_{4}$ & 2$_{3}$ & 2$_{3}$ & $2_{3}+2_{5}$ & $
2_{2}+2_{6}$ & $1_{3}+1_{4\ }+2_{1}$ & $1_{1}+1_{2\ }+2_{6}$ &
$2_{1}+2_{5}$
& $2_{2}+2_{4}$ \\ \hline
2$_{5}$ & 2$_{5}$ & 2$_{5}$ & 2$_{2}$ & 2$_{2}$ & $2_{4}+2_{6}$ & $%
1_{3}+1_{4\ }+2_{3}$ & $2_{2}+2_{6}$ & $2_{1}+2_{5}$ & $1_{1}+1_{2\
}+2_{4}$
& $2_{1}+2_{3}$ \\ \hline
2$_{6}$ & 2$_{6}$ & 2$_{6}$ & 2$_{1}$ & 2$_{1}$ & $1_{3}+1_{4\ }+2_{5}$
& $%
2_{4}+2_{6}$ & $2_{3}+2_{5}$ & $2_{2}+2_{4}$ & $2_{1}+2_{3}$ &
$1_{1}+1_{2\
}+2_{2}$ \\ \hline
\end{tabular}
\bigskip

\bigskip

\newpage

The group $D_{5}\times Z_{3}$, 30/2

\bigskip
\begin{tabular}{|c||c|c|c|c|c|c|c|c|c|c|c|c|}
\hline
$\otimes $ & 1$_{1}$ & 1$_{2}$ & 1$_{3}$ & 1$_{4}$ & 1$_{5}$ & 1$_{6}$ &
2$%
_{1}$ & 2$_{2}$ & 2$_{3}$ & 2$_{4}$ & 2$_{5}$ & 2$_{6}$ \\ \hline\hline
1$_{1}$ & 1$_{1}$ & 1$_{2}$ & 1$_{3}$ & 1$_{4}$ & 1$_{5}$ & 1$_{6}$ &
2$_{1}$
& 2$_{2}$ & 2$_{3}$ & 2$_{4}$ & 2$_{5}$ & 2$_{6}$ \\ \hline
1$_{2}$ & 1$_{2}$ & 1$_{3}$ & 1$_{4}$ & 1$_{5}$ & 1$_{6}$ & 1$_{1}$ &
2$_{5}$
& 2$_{6}$ & 2$_{1}$ & 2$_{2}$ & 2$_{3}$ & 2$_{4}$ \\ \hline
1$_{3}$ & 1$_{3}$ & 1$_{4}$ & 1$_{5}$ & 1$_{6}$ & 1$_{1}$ & 1$_{2}$ &
2$_{3}$
& 2$_{4}$ & 2$_{5}$ & 2$_{6}$ & 2$_{1}$ & 2$_{2}$ \\ \hline
1$_{4}$ & 1$_{4}$ & 1$_{5}$ & 1$_{6}$ & 1$_{1}$ & 1$_{2}$ & 1$_{3}$ &
2$_{1}$
& 2$_{2}$ & 2$_{3}$ & 2$_{4}$ & 2$_{5}$ & 2$_{6}$ \\ \hline
1$_{5}$ & 1$_{5}$ & 1$_{6}$ & 1$_{1}$ & 1$_{2}$ & 1$_{3}$ & 1$_{4}$ &
2$_{5}$
& 2$_{6}$ & 2$_{1}$ & 2$_{2}$ & 2$_{3}$ & 2$_{4}$ \\ \hline
1$_{6}$ & 1$_{6}$ & 1$_{1}$ & 1$_{2}$ & 1$_{3}$ & 1$_{4}$ & 1$_{5}$ &
2$_{3}$
& 2$_{4}$ & 2$_{5}$ & 2$_{6}$ & 2$_{1}$ & 2$_{2}$ \\ \hline
2$_{1}$ & 2$_{1}$ & 2$_{5}$ & 2$_{3}$ & 2$_{1}$ & 2$_{5}$ & 2$_{3}$ & $%

1_{1}+1_{4}+2_{1}$ & $2_{1}+2_{2}$ & $1_{3}+1_{6}+2_{4}$ & $2_{3}+2_{4}$
& $%
1_{2}+1_{5}+2_{6}$ & $2_{5}+2_{6}$ \\
\hline
\end{tabular}

\bigskip
\newpage
\bigskip The group $D_{15},30/4$

\begin{tabular}{|c||c|c|c|c|c|c|c|c|c|}
\hline
$\otimes $ & 1 & $1^{\prime }$ & 2$_{1}$ & 2$_{2}$ & 2$_{3}$ & 2$_{4}$ &
2$%
_{5}$ & 2$_{6}$ & 2$_{7}$ \\ \hline\hline
1 & 1 & $1^{\prime }$ & 2$_{1}$ & 2$_{2}$ & 2$_{3}$ & 2$_{4}$ & 2$_{5}$
& 2$%
_{6}$ & 2$_{7}$ \\ \hline
$1^{\prime }$ & $1^{\prime }$ & 1 & 2$_{1}$ & 2$_{2}$ & 2$_{3}$ &
2$_{4}$ & 2%
$_{5}$ & 2$_{6}$ & 2$_{7}$ \\ \hline
2$_{1}$ & 2$_{1}$ & 2$_{1}$ & $1+1^{\prime }+2_{2}$ & $2_{1}+2_{3}$ & $%

2_{1}+2_{4}$ & $2_{3}+2_{5}$ & $2_{4}+2_{6}$ & $2_{5}+2_{7}$ &
$2_{6}+2_{7}$
\\ \hline
2$_{2}$ & 2$_{2}$ & 2$_{2}$ & $2_{1}+2_{3}$ & $1+1^{\prime }+2_{4}$ & $%

2_{1}+2_{5}$ & $2_{2}+2_{6}$ & $2_{3}+2_{7}$ & $2_{4}+2_{7}$ &
$2_{5}+2_{6}$
\\ \hline
2$_{3}$ & 2$_{3}$ & 2$_{3}$ & $2_{1}+2_{4}$ & $2_{1}+2_{5}$ &
$1+1^{\prime
}+2_{6}$ & $2_{1}+2_{7}$ & $2_{2}+2_{7}\ $ & $2_{3}+2_{6}$ &
$2_{4}+2_{5}$
\\ \hline
2$_{4}$ & 2$_{4}$ & 2$_{4}$ & $2_{3}+2_{5}$ & $2_{2}+2_{6}$ &
$2_{1}+2_{7}$
& $1+1^{\prime }+2_{7}$ & $2_{1}+2_{6}$ & $2_{2}+2_{5}$ & $2_{3}+2_{4}$
\\
\hline
2$_{5}$ & 2$_{5}$ & 2$_{5}$ & $2_{4}+2_{6}$ & $2_{3}+2_{7}$ &
$2_{2}+2_{7}$
& $2_{1}+2_{6}$ & $1+1^{\prime }+2_{5}$ & $2_{1}+2_{4}$ & $2_{2}+2_{3}$
\\
\hline
2$_{6}$ & 2$_{6}$ & 2$_{6}$ & $2_{5}+2_{7}$ & $2_{4}+2_{7}$ &
$2_{3}+2_{6}$
& $2_{2}+2_{5}$ & $2_{1}+2_{4}$ & $1+1^{\prime }+2_{3}$ & $2_{1}+2_{2}$
\\
\hline
2$_{7}$ & 2$_{7}$ & 2$_{7}$ & $2_{6}+2_{7}$ & $2_{5}+2_{6}$ &
$2_{4}+2_{5}$
& $2_{3}+2_{4}$ & $2_{2}+2_{3}$ & $2_{1}+2_{2}$ & $1+1^{\prime }+2_{1}$
\\
\hline
\end{tabular}

\bigskip

\newpage

\section*{Figure caption.}

\bigskip
\bigskip

\noindent Quiver diagram for chiral fermions in 24/7 model.

\bigskip


\newpage

\begin{references}
\bibitem{mandelstam}  S. Mandelstam, Nucl. Phys. {\bf B213,} 149 (1983).

\bibitem{KS}  S. Kachru and E. Silverstein, Phys. Rev. Lett. {\bf 80,} 4855
(1998).

\bibitem{maldacena}  J. Maldacena, Adv. Theor. Math. Phys. {\bf 2,} 231
(1998).

\bibitem{F1}  P.H. Frampton, Phys. Rev. {\bf D60,} 041901 (1999).

\bibitem{WS}  P.H. Frampton and W. F. Shively, Phys. Lett. {\bf B454,} 49
(1999).

\bibitem{CV}  P.H. Frampton and C. Vafa, {\tt hep-th/9903226}.

\bibitem{F2}  P.H. Frampton, Phys. Rev. {\bf D60,} 087004 (1999).

\bibitem{F3}  P.H. Frampton, Phys. Rev. {\bf D60,} 107505 (1999).
\bibitem{chiral}
P.H. Frampton and S.L. Glashow, Phys. Lett. {\bf 190B,} 157 (1987);\\
Phys. Rev. Lett. {\bf 58,} 2168 (1987).
\bibitem{331}
P.H. Frampton, Phys. Rev. Lett. {\bf 69,} 2889 (1992).
\bibitem{hexagon}
P.H. Frampton and T.W. Kephart, Phys. Rev. Lett. {\bf 50,} 1343 (1983); {\it ibid} {\bf 50,} 1347 (1983).
\bibitem{books}  Useful sources of information on the finite groups include:%
\newline
D.E. Littlewood, {it The Theory of Group Characters and Matrix
Representations of Groups} (Oxford 1940);\newline
M. Hamermesh, {\it Group Theory and Its Applications to Physical Problems}
(Addison-Wesley, 1962);\newline
J.S. Lomont, {\it Applications of Fimite Groups} (Academic, 1959), reprinted
by Dover (1993);\newline
A.D. Thomas and G.V. Wood, {\it Group Tables} (Shiva, 1980).
\bibitem{guts}
Another (non-minimal) approach to family symmetry
appeared in: 
P.H. Frampton, Phys. Lett. {\bf B88,} 299 (1979).
\bibitem{Pak}
S. Pakvasa and H. Sugawara, Phys. Lett. {\bf B73,} 61 (1978).
\bibitem{FK}  
P.H. Frampton and T.W. Kephart, Int. J. Mod. Phys. {\bf A10,}
4689 (1995).
\bibitem{PS}  J.C. Pati and A. Salam, Phys. Rev. {\bf D10,} 275 (1974).%
\newline
R.N. Mohapatra and J.C. Pati, Phys. Rev. {\bf D11,} 566 (1975).\newline
R.N. Mohapatra and G. Senjanovic, Phys. Rev. {\bf D12,} 1502 (1975)
\bibitem{V}  H. Verlinde, Nucl. Phys. {\bf B580,} 264 (2000).
\bibitem{RS}  L. Randall and R. Sundrum, Phys. Rev. Lett. {\bf 83} 3370 (1999);
{\it ibid} {\bf 83,} 4690 (1999).\newline
J. Lykken and L. Randall, JHEP {\bf 0006,} 014 (2000).
\bibitem{GW}  
W.D. Goldberger and M. B. Wise, Phys. Rev. {\bf D60,} 107505 (1999);
Phys. Rev. Lett. {\bf 83,} 4922 (1999); Phys. Lett. {\bf B475,} 275 (2000).
\bibitem{antoniadis}  
I. Antoniadis, Phys. Lett. {\bf B246} (1990) 377; \newline
I. Antoniadis and K. Benakli, Phys. Lett. {\bf B326} (1994) 69;\newline
I. Antoniadis, K. Benakli and M. Quiros, Phys. Lett. {\bf B331} (1994) 313.%
\newline
J. D. Lykken, Phys. Rev. {\bf D54} (1996) 3693.\newline
I. Antoniadis {\it et al}, Phys. Lett. {\bf B436,} 257 (1998).\newline
I. Antoniadis, S. Dimopoulos, A. Pomarol and M. Quiros, Nucl. Phys. {\bf %
B544,} 503 (1999).\newline
K. Dienes, E. Dudas and T. Gherghetta, Phys. Lett. {\bf B436} (1998) 55;
Nucl. Phys. {\bf 537,} 47 (1999); Nucl. Phys. {\bf B543,} 387 (1999).\newline
P.H. Frampton and A. Rasin, Phys. Lett. {\bf 460B,} 313 (1999).\newline
D. Ghilencea and G.G. Ross, Phys. Lett. {\bf B442} (1998) 165.\newline
C.D. Carone, Phys. Lett. {\bf B454,} 70 (1999).\\
C. Bachas, JHEP {\bf 9811,} 023 (1998). \\
G. Shiu and S.H.H. Tye, Phys. Rev. {\bf D58} (1998) 106007. \newline
Z. Kakushadze and S.H.H. Tye, Nucl. Phys. {\bf B548,} 180 (1999). \newline
Z. Kakushadze, Nucl. Phys. {\bf B551,} 549 (1999).\newline
N. Arkani-Hamed, S. Dimopoulos and G. Dvali, Phys. Lett {\bf 429} (1998)506.
\bibitem{FK2}  
P.H. Frampton and T.W. Kephart, Phys. Lett. {\bf B485,} 403 (2000).
\bibitem{DM}
M. Douglas and G. Moore. {\tt hep-th/9603167}

\end{references}
\end{document}